\documentclass[pra,twocolumn,letterpaper,tighten,float,superscriptaddress]{revtex4}
\usepackage{graphicx}
\usepackage{latexsym}
\usepackage{amsmath,amssymb}
\usepackage{bm} 
\usepackage{color}
\usepackage{epsfig}

\DeclareMathOperator{\sgn}{sgn}

\definecolor{myblue}{rgb}{.93, .93, 1}

\newcommand{\bsub}{\begin{subequations}}
\newcommand{\esub}{\end{subequations}}

\begin{document}
	
\title{Finite-temperature spectroscopy of dirty helical Luttinger
  liquids}
	
\author{Tzu-Chi~Hsieh}\email{tzuchi.hsieh@colorado.edu}
\affiliation{Department of Physics and Center for Theory of Quantum
  Matter, University of Colorado Boulder, Boulder, Colorado 80309,
  USA} \author{Yang-Zhi~Chou}\email{yzchou@umd.edu}
\affiliation{Department of Physics, Condensed Matter theory center and
  the Joint Quantum Institute, University of Maryland, College Park,
  Maryland 20742, USA} \affiliation{Department of Physics and Center
  for Theory of Quantum Matter, University of Colorado Boulder,
  Boulder, Colorado 80309, USA} \author{Leo
  Radzihovsky}\email{radzihov@colorado.edu} \affiliation{Department of
  Physics and Center for Theory of Quantum Matter, University of
  Colorado Boulder, Boulder, Colorado 80309, USA}
	
\date{\today}
	
\begin{abstract}
  We develop a theory of finite-temperature momentum-resolved
  tunneling spectroscopy (MRTS) for disordered, interacting
  two-dimensional topological-insulator edges. The MRTS complements
  conventional electrical transport measurement in characterizing the
  properties of the helical Luttinger liquid edges.  Using standard
  bosonization technique, we study low-energy spectral function and
  the MRTS tunneling current, providing a detailed description
  controlled by disorder, interaction, and temperature, taking into
  account Rashba spin orbit coupling, interedge interaction and
  distinct edge velocities.  Our theory provides a systematic
  description of the spectroscopic signals in the MRTS measurement and
  we hope will stimulate future experimental studies on the
  two-dimensional time-reversal invariant topological insulator.
\end{abstract}
	
\maketitle
	
\section{Introduction}
	
Topology has become an important component of and has revolutionized
modern condensed matter physics over the past few decades. Strikingly,
topological condensed matter phenomena are robust to local
heterogeneities (disorder), sample geometry, and other low-energy
microscopic details. A paradigmatic example is the chiral edge state
of the integer quantum Hall effect, which gives a quantized $e^2/h$
Hall conductance per channel, robust to local perturbations. Another
significant advance is the prediction of a time-reversal (TR)
symmetric topological insulators (TI)
\cite{Kane2005_1,Kane2005_2,Bernevig2006,Hasan2010_RMP,Qi2011_RMP} and more generally
symmetry-protected TIs \cite{SenthilARCMP}, that stimulated numerous
theoretical
\cite{Xu2006,Wu2006,Teo2009,Maciejko2009,Schmidt2012,Vayrynen2013} and
experimental investigations
\cite{Konig2007,Knez2011,Suzuki2013,Du2015,Li2015,Qu2015,Ma2015,Nichele2016,Nguyen2016,Couedo2016,Fei2017,Du2017,Li2017,Tang2017,Wu2018,Chen2018QSH_WSe2,Ugeda2018WSe2,Reis2017}
(also see reviews and references therein,
\cite{Hasan2010_RMP,Qi2011_RMP,Dolcetto2015,SenthilARCMP,Rachel2018,Lunczer2019}).
	
A 2D time-reversal symmetric TI
\cite{Kane2005_1,Kane2005_2,Bernevig2006} (of class AII
\cite{Hasan2010_RMP}) is a fully gapped bulk insulator with its edge
hosting counter-propagating Kramers pairs of electrons. The
time-reversal symmetric disorder cannot backscatter in the absence of
interactions (though it can for an interacting edge, e.g., via a
two-particle backscattering \cite{Wu2006,Xu2006,Chou2018EdgeGlass})
with edge electrons propagating ballistically, thus avoiding Anderson
localization. Such ideal topologically protected helical Luttinger
liquid (hLL) edge \cite{Wu2006,Xu2006} is expected to exhibit a
quantized $e^2/h$ zero-temperature conductance, controls the
low-energy properties of the TI, and provides a new platform for
studying and testing the low-energy Luttinger liquid (LL) theory of
interacting one-dimensional electronic systems.
	
\begin{figure}[t!]
		\includegraphics[width=0.375\textwidth]{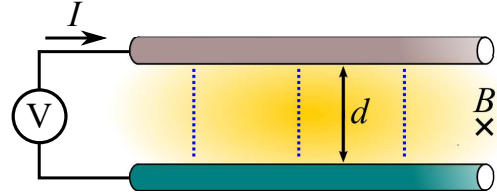}
		\caption{A schematic of an experimental setup for
                  momentum-resolved tunneling spectroscopy, with
                  dashed lines indicating the single particle
                  tunneling between two separated topological
                  insulator edges, and momentum transfer tuned with an
                  out-of-plane magnetic field applied to the yellow
                  shaded region.}
		\label{Fig:setup}
\end{figure}
	
In contrast to the quantum Hall edges, a transport in 2D TR symmetric
TI edges is sensitive to a set of microscopic details. At the simplest
level a hLL is predicted to exhibit interaction strength-dependent
power-laws in frequency, voltage and temperature
\cite{Maciejko2009,Tanaka2011,Lezmy2012,Kainaris2014,Chou2015}.  In a
more detailed analysis, the primary finite-temperature conductance
correction is believed to come from charge puddles near the edge
\cite{Vayrynen2013,Vayrynen2014}. The charge puddles can behave like
Kondo impurities \cite{Vayrynen2014,Maciejko2009,Tanaka2011} and can
generate insulator-like finite-temperature conductivity
\cite{Vayrynen2016}.  External noise \cite{Vayrynen2018} and intraedge
inelastic
interaction \cite{Maciejko2009,Schmidt2012,Kainaris2014,Chou2015} are
also predicted to give nontrivial conductance corrections.  To our
knowledge, however, the existing experiments have not systematically
demonstrated the finite-temperature conductivity predicted by any of
the above theories.  Among various other potential explanations (see
e.g.,
\cite{Pikulin2014,Hu2016,Li2017_Hidden_edge,Skolasinski2018,Chou2018EdgeGlass,Novelli2019})
is a novel spontaneous symmetry-breaking localization due to an
interplay of TR \textit{symmetric} disorder and interaction
\cite{Chou2018EdgeGlass}, in contrast to Anderson localization due to
a magnetic ordering of an extensive number of the Kondo impurities
\cite{Altshuler2013,Hsu2017}.  Generally, one expects that disorder
with \textit{weak} interactions does not modify the edge state dc
conductance \cite{Kane2005_2,Xie2016}.  In light of above puzzling
transport measurements, an independent experimental probe of the
helical Luttinger liquid (hLL) edges is highly desirable.
		
In the present study, we calculate a spectral function of a
disordered, finite-temperature hLL, and based on it develop a theory
of the finite-temperature momentum-resolved tunneling spectroscopy
(MRTS)
\cite{Auslaender2002,Steinberg2008,Jompol2009,Tsyplyatyev2015,Tsyplyatyev2016,Jin2019}
between two (TR symmetrically) disordered, interacting TI helical
edges. Such MRTS setup thereby provides an independent spectral
characterization of the hLLs, complementary to conventional transport.
In contrast to earlier work \cite{Braunecker2018}, which focused on
clean short zero-temperature hLLs, we study disordered interacting
long TI edges at finite temperatures. In the absence of interedge
interaction, the tunneling current spectroscopy is simply related to a
convolution of two fermionic edge spectral functions, that we compute
in a detailed closed form. An interedge interaction requires a
nonperturbative treatment. Utilizing bosonization, perturbatively in
the tunneling we derive the disorder-averaged, finite temperature MRTS
tunneling current, that depends sensitively on mismatch of edge
velocities.  In contrast to conventional LL
edges \cite{Carpentier2002}, TR symmetric disorder does not
back-scatter helical edge electrons. Thus our low-energy analysis
makes predictions that are nonperturbative in interaction and
disorder, providing a detailed characterization of a hLL that should
be experimentally accessible.
	
Before delving into details of the analysis, we summarize our results
in Sec.~\ref{sec:results}. Then, in Sec.~\ref{sec:Single_Edge},
utilizing bosonization we study the finite-temperature spectral
function of a helical edge of a TR invariant TI in the presence of
symmetry-preserving disorder and interactions. In Sec.~\ref{sec:MRTS},
building on the single-edge analysis we study the interedge tunneling,
showing that it can be used as a momentum-resolved spectroscopic probe
of helical edges, with momentum and frequency tuned by an external
magnetic field and interedge voltage, respectively, as illustrated in
Fig.~\ref{Fig:setup}.  We conclude in Sec.~\ref{sec:conclusion} with a
discussion of using this momentum-resolved tunneling spectroscopy to
unambiguously experimentally identify TI edges, that have resisted
clear identification in a conventional transport measurements. We
relegate much of our somewhat technical analysis to numerous
appendices.
	
\section{Summary of main results}
\label{sec:results}
	
We briefly summarize the key results of our study, detailed in
subsequent sections of the manuscript. Utilizing bosonization we
studied finite temperature spectral properties of an interacting
helical edge of a TR invariant TI in the presence of
symmetry-preserving disorder.  Although a number of similar analyses
have appeared in the literature
\cite{Luther1974,Meden1992,Voit1993,Orgad2001}, to the best of our
knowledge our computation is the most detailed and complete at finite
temperature. Inside the hLL
phase \cite{Wu2006,Xu2006,Chou2018EdgeGlass}, the edge is fully
characterized by a Luttinger parameter $K$ and exponent
$\gamma\equiv\frac{1}{4}\left(K+K^{-1}\right)-\frac{1}{2}$, with $K=1$
($\gamma=0$) in a non-interacting limit and $K<1$ ($\gamma>0$) for
repulsive interaction. 

\begin{figure}[t!]
  \includegraphics[width=0.35\textwidth]{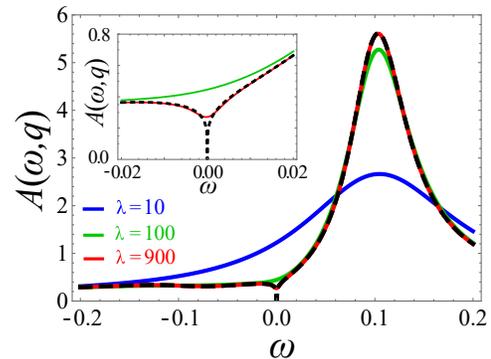}
  \caption{Single helical edge, disorder-averaged spectral function,
    illustrated for a set of temperatures (characterized by a thermal
    length $\lambda=v\beta$), disorder length $\xi=30$, $vq=0.1$, and
    interaction parameter
    $\gamma\equiv\frac{1}{4}\left(K+K^{-1}\right)-\frac{1}{2}$ taken
    to be $0.1$. The black dashed line denotes the zero temperature
    ($\lambda=\infty$) spectral function. The inset shows the details
    of thermal rounding of the zero-bias anomaly. The frequency and the 
    length are in units of $v\alpha^{-1}$ and $\alpha$ respectively.}
\label{fig:Aqw}
\end{figure}

We derive a detailed expression for the disorder-averaged, 
low-temperature spectral function Eq.~(\ref{Eq:G_dis_asym}), that in the
limit of strong disorder $\Delta$ is given by
\begin{align}
  A(\omega,q)\approx T^{2\gamma}\frac{\xi/\pi}{(q\xi)^{2}+1}
  f_{\gamma}\left(\frac{\omega}{T}\right),
\end{align}
where $\xi=2v^2/K^2\Delta$ is a disorder length scale, $v$ is the edge
velocity and $T$ is temperature (with $\beta=T^{-1}$ the inverse temperature). For convenience, we set ${\hbar=k_B=1}$ throughout this paper. Above,
\begin{align}
  f_\gamma\left(x\right)\sim\left\lbrace\begin{array}{cc}
      x^{2\gamma},&\text{ for }x\gg 1\\
      1,&\text{ for }x\ll 1
\end{array}\right.
\end{align}
is a scaling function, with the exact form given by the Euler Beta
function derived in the main text,
Eq.~(\ref{Eq:G_dis_low_T}). The complete expression for a
right-mover $A(\omega,q)$, characterized by a broad peak at $\omega=v
q$ and a zero-bias anomaly at $\omega = 0$, is illustrated for a set
of temperatures in Fig.~\ref{fig:Aqw}. The broadening of the
quasiparticle peak is described by the full width at half maximum
(FWHM) $4\pi\gamma T+2v\xi^{-1}$, which suggests that a probe of the
momentum-resolved spectral function can be used to quantify the
interaction and (forward-scattering) disorder strength. 

We note that although generically one expects sample heterogeneity to
smear out sharp features of a clean system, here disorder average of
the finite-momentum spectral function, $A(\omega,q)$ brings out the
sharp zero-bias anomaly that is otherwise absent at finite
momentum. This counter-intuitive effect arises due to impurities
providing the momentum needed to shift the $q=0$ zero-frequency
anomaly to a finite momentum $q$, as shown in Fig.~\ref{fig:Aqw}. All figures in this paper are plotted in the units of $v\alpha^{-1}$ and $\alpha$ for frequency and length respectively, where $\alpha$ is the ultraviolet cutoff length scale in LL theory.
	
\begin{figure}[t!]
\includegraphics[width=0.35\textwidth]{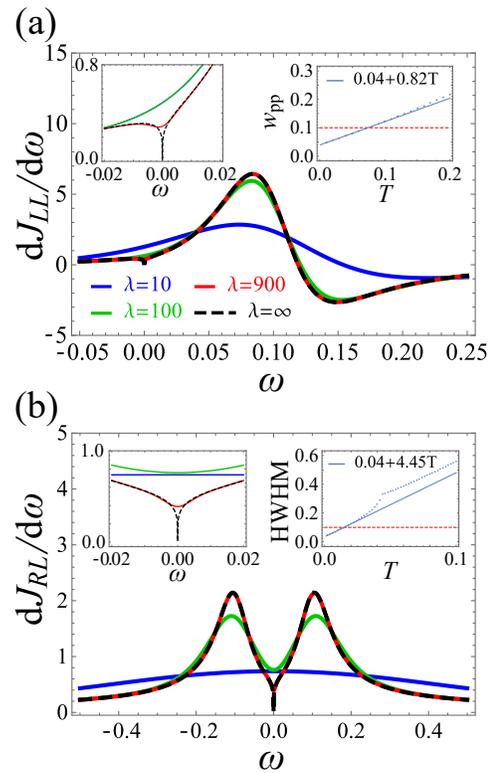}
\caption{\textit{Finite-temperature} differential tunneling
  conductance $dJ(\omega=eV/\hbar,Q=2\pi Bd/\phi_0)/d\omega$ in the
  presence of \textit{forward-scattering disorders} (with disorder
  length $\xi=20$) around (a) two left Fermi points and (b) right and
  left Fermi points, illustrated for a range of temperatures
  characterized by the thermal length $\lambda=v\beta$. Velocity and interaction are taken to be identical 
  for the two edges,
  with the more generic expression given in the main text. The
  interaction parameter $\gamma\equiv\frac{1}{4}\left(K+K^{-1}
  \right)-\frac{1}{2}$ is taken to be $0.05$ and $vq=-0.1$. The left
  inset is the magnification of the zero-bias anomaly. The right insets show 
  linear temperature dependence of (a) the distance (in $\omega$) between the $\text{left-positive}$ and 
  $\text{right-negative}$ peaks ($w_{pp}$) and (b) the half width at half maximum (HWHM) respectively for 
  $T<|vq|$ (red dashed lines). The frequency and the length are in units of $v\alpha^{-1}$ and $\alpha$ respectively.}
\label{fig:currentJqw}
\end{figure}

Our second key prediction is that of the finite-temperature
momentum-resolved interedge tunneling current
$J(\omega=eV/\hbar,Q=2\pi Bd/\phi_0)$ in the presence of disorder and
interaction, and tunable by an external magnetic field $B$ and voltage
bias $V$, as illustrated in a schematic setup of Fig.~\ref{Fig:setup}.
In the above, $d$ denotes the distance between two edges and $\phi_0=h/e$ is
the magnetic flux quantum. Importantly, $Q$ is the momentum shift between the energy bands of the two edges controlled by the external magnetic field. The representative predictions for the tunneling current, computed perturbatively in the tunneling are given by the following analytical
expressions. For the vertical geometry (Fig.~\ref{Fig:vertical setup}) with identical edges (same velocity and interaction but different Fermi wavevectors $k_{F,1}\neq k_{F,2}$), the tunneling
current in the absence of disorder and interedge interaction is
well approximated by
$J_{RR}(\omega,Q+k_{F,1}-k_{F,2})+J_{LL}(\omega,Q-k_{F,1}+k_{F,2})$,
where
\begin{align}
  J_{LL}(\omega,q)&=-2et_0^2\left(\frac{2\pi\alpha}{\beta
      v}\right)^{4\gamma}\frac{1}{4\pi^2v}\sin(2\pi\gamma)
  \nonumber\\
  &\quad\,\times \text{Im}\left\lbrace
    B\left[\frac{\beta(-i\omega+ivq)}{4\pi}+\gamma+1,-1-2\gamma\right]\right.
  \nonumber\\
  &\quad\quad\,\,\times
  \left. B\left[\frac{\beta(-i\omega-ivq)}{4\pi}+\gamma,1-2\gamma\right]
  \right\rbrace,\label{Eq:J_LL_equalv}
\end{align}
and $J_{RR}(\omega,q)=J_{LL}(\omega,-q)$. For the horizontal geometry
(Fig.~\ref{Fig:horizontal setup}) with identical edges,
the tunneling current is given by
$J_{RL}(\omega,Q+k_{F,1}+k_{F,2})+J_{LR}(\omega,Q-k_{F,1}-k_{F,2})$
where
\begin{align}
  J_{RL/LR}(\omega,q)&=-2et_0^2\left(\frac{2\pi\alpha}{\beta
      v}\right)^{4\gamma}\frac{1}{4\pi^2v}\sin(2\pi\gamma)
  \nonumber\\
  &\quad\,\times \text{Im}\left\lbrace
    B\left[\frac{\beta(-i\omega+ivq)}{4\pi}+\gamma+\frac{1}{2},-2\gamma\right]\right.
 \nonumber\\
 &\quad\quad\,\,\times
  \left. B\left[\frac{\beta(-i\omega-ivq)}{4\pi}+\gamma+\frac{1}{2},-2\gamma\right]\right\rbrace.\label{Eq:J_RL_equalv_hori}
\end{align}
The effects of forward-scattering disorder can be included through a
convolution with a Lorentzian (with width $\xi^{-1}$, where $\xi$ is
the disorder length). With Eqs.~(\ref{Eq:J_LL_equalv}) and (\ref{Eq:J_RL_equalv_hori}), the differential tunneling conductance can be derived. The differential tunneling conductance for both vertical and horizontal
geometries are plotted in Fig.~\ref{fig:currentJqw}. We discuss
the more generic case (e.g., including interedge interaction, distinct
edge velocities, etc) in Sec~\ref{sec:MRTS}.

\begin{figure}[t!]
\includegraphics[width=0.45\textwidth]{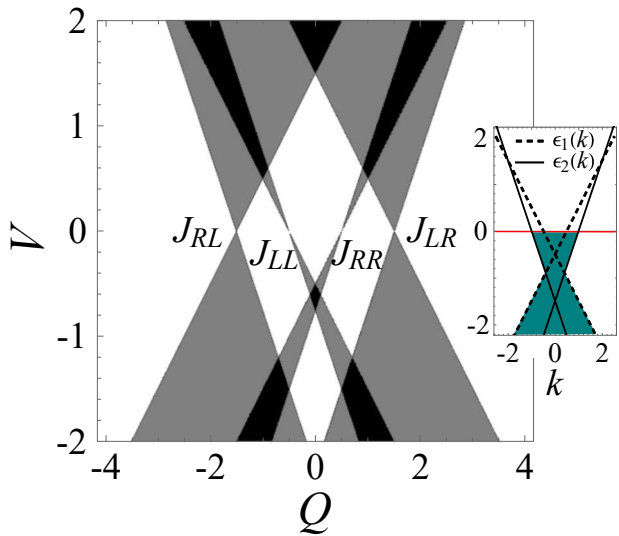}
\caption{Schematic diagram of momentum-resolved tunneling current for
  two non-interacting TI edges. There are four tunneling regions (gray
  color, labeled by $J_{\alpha\alpha'}$) for low bias $V\approx 0$,
  corresponding to tunneling between $\alpha$ (of edge 1) and
  $\alpha'$ (of edge 2) Fermi points. The current flows from edge 1 to
  2 (positive current) and from edge 2 to 1 (negative current) for
  positive and negative bias voltages respectively. The black regions
  indicate that tunneling happens between two pairs of Fermi
  points. Inset: energy bands of edge 1 (dashed line) and edge 2
  (solid line) used for generating the main figure. The red line
  denotes the Fermi energy at equilibrium ($V=0$).}
\label{fig:kinematics}
\end{figure}

A map of tunneling current can be constructed by tuning $B$ and $V$
independently.  In the absence of the interaction, the tunneling
currents are nonzero only in the kinematically allowed regions
\cite{Carpentier2002} illustrated in Fig.~\ref{fig:kinematics}. The
interactions modify the kinematically allowed region as we discuss
in the main text.	
	
We now turn to the detailed analysis that leads to the above results,
as well as exploration of a number of different parameters and
experimental geometries.
	
\section{Single edge: Model and spectral function}
\label{sec:Single_Edge}
	
The edge states of a two-dimensional time-reversal symmetric
topological insulator exhibit counterpropagating fermion Kramers
pair. In contrast to a conventional Luttinger liquid, the TR symmetry
on the edge constrains relevant interactions and disorder perturbatons
to be forward-scattering only. The absence of Anderson localization is
the manifestation of the topological protection of the TI edges.  The
gapless insulating localized states can still appear through
spontaneous TR symmetry breaking for $K<3/8$ \cite{Wu2006,Xu2006} due
to an interplay of interaction and disorder
\cite{Chou2018EdgeGlass}. In this work, we exclusively focus on the
$K>3/8$ hLL phase.  We next introduce the minimal model for such
disordered hLL and then study its finite-temperature spectral function
using bosonization \cite{Shankar_Book,Giamarchi_Book}.
	
\subsection{Weakly interacting generic hLL}
A helical edge is characterized by a Kramers pair of right-moving,
$c_{+}(k)$ and left-moving, $c_{-}(k)$ fermions at each quasi-momentum
$k$. Under antiunitary TR operation $\mathcal{T}$, the TR
symmetric partners are related to each other by
$\mathcal{T}c_{\pm}(k)\mathcal{T}^{-1}=\pm c_{\mp}(-k)$
($\mathcal{T}^2=-1$). To describe low energy physics around
Fermi points $\pm k_F$, the field operators can be expressed in terms
of the slowly-varying fermionic degrees of freedom $R$ and $L$ near
$k_F$ and $-k_F$, respectively,
\begin{align}
  c_{+}(x)=&\int \frac{dk}{2\pi}\,e^{ikx}c_{+}(k)\approx
  e^{ik_{F}x}R(x)
  \nonumber\\
  c_{-}(x)=&\int \frac{dk}{2\pi}\,e^{ikx}c_{-}(k)\approx
  e^{-ik_{F}x}L(x).
\label{Eq:field_op_band_basis}
\end{align}
	
In TI samples without mirror symmetry, the Rashba spin-orbit coupling
(RSOC) is generically present. The primary effect of the RSOC is
to induce momentum-dependent spin rotation \cite{Schmidt2012,Rod2015}. As a result, the
field operators with a definite spin projection $\uparrow$ and
$\downarrow$ are a linear combination of chiral fields \cite{Xie2016} (also see
Appendix~\ref{App:chiral decomposition} for a derivation),
\begin{align}
  \label{Eq:field_op_spin_basis}
  c_{\uparrow}(x)\approx &e^{ik_{F}x}R(x)-i\zeta
  e^{-ik_{F}x}\partial_{x}L(x)
  \nonumber\\
  c_{\downarrow}(x)\approx &e^{-ik_{F}x}L(x)-i\zeta^*
  e^{ik_{F}x}\partial_{x}R(x),
\end{align}
where length $\zeta$ encodes the degree of ``spin rotation texture''.
In a simple model discussed in \cite{Schmidt2012},
$\zeta=2k_{F}/k_0^2$, where $k_0$ characterizes the strength of
RSOC. In Eq.~(\ref{Eq:field_op_spin_basis}), the spin quantization
axis is chosen such that $\uparrow$ and $\downarrow$ match the spins
at Fermi points $\pm k_F$ respectively. Next, we construct the
low-energy Hamiltonian for the helical edge. 

The kinetic part of the Hamiltonian is given by
\begin{align}\label{Eq:H_hLL_kinetic}
  H_{\text{0}}&= v_F\int dx\left[R^{\dagger}\left(-i\partial_x
      R\right)-L^{\dagger}\left(-i\partial_x L\right)\right]
\end{align}
where $v_F$ is the Fermi velocity. The interaction and disorder parts
of the Hamiltonian couple to the electron density, given by
\begin{align}
\rho(x)=& R^{\dagger}R+L^{\dagger}L
\nonumber\\
\label{Eq:rho_RSOC}&-\left\lbrace i\zeta
  e^{-i2k_{F}x}\left[R^{\dagger}(\partial_{x}L)-(\partial_{x}R^{\dagger})L\right]+\text{H.c.}\right\rbrace,
\end{align}
where only terms up to $O(\zeta)$ are kept.  The low-energy expansion
of the electron density contains a slowly-varying (low momentum
transfer) and a fast-varying ($2k_F$ momentum transfer)
contributions. It is important to note that Eq.~(\ref{Eq:rho_RSOC}) is
invariant under TR operation ($R\rightarrow L$, $L\rightarrow -R$, and
$i\rightarrow -i$).
	
It is instructive to consider a chemical potential shift coupled to
the density, $\rho(x)$ given by (\ref{Eq:rho_RSOC}) in the presence of
RSOC. The key observation is that the shifted Hamiltonian can be
brought back to the original gapless form (\ref{Eq:H_hLL_kinetic})
\begin{align}
  H_0'\equiv& H_0-\delta\mu\int dx\,\rho(x)\\
  \label{Eq:H_uni_chem}=&v_F'\int
  dx\left[R'^{\dagger}\left(-i\partial_x
      R'\right)-L'^{\dagger}\left(-i\partial_x L'\right)\right],
\end{align}
with $k_F$-dependent rotation of the quantization axis of the helical
fermions,
\begin{align}\label{Eq:rotation_RSOC_uni}
  \left[\begin{array}{c}
      R'(x)\\
      L'(x)
\end{array}
  \right]=e^{-i\hat{\sigma}_zk_F'
    x}e^{-i\hat{\sigma}_y\theta/2}e^{i\hat{\sigma}_zk_Fx}\left[\begin{array}{c}
      R(x)\\
      L(x)
\end{array}
	\right],
\end{align}
characterized by $\theta=\tan^{-1}\left(2\delta\mu\,\zeta/v_F\right)$,
$k_F'=\frac{v_Fk_F+\delta\mu}{v_F'}$, and
$v_F'=\sqrt{v_F^2+\left(2\delta\mu\, \zeta\right)^2}$ (to simplify the
expression we have taken $\zeta$ to be real).  The gapless helical
edge remains topologically protected against uniform RSOC as long as
the bulk gap is finite \cite{Kane2005_1}.
	
The key qualitative distinguishing feature of hLL is that TR
invariance forbids Anderson localization of the edge Kramers pairs by
{\em nonmagnetic} impurities.  In the absence of RSOC this is manifest
as the density operator, $R^{\dagger}R+L^{\dagger}L$ is only
forward-scattering.  In the presence of both RSOC and the TR symmetric
disorder, a position-dependent rotation can again map the theory to
the 1D massless Dirac Hamiltonian in a fixed realization of disorder
\cite{Xie2016}. Thus, low-energy effects of TR invariant disorder on
the helical edges of a TI are qualitatively captured by random forward
scattering perturbation,
\begin{align}\label{Eq:H_dis}
  H_{\text{dis}}=&\int dx
  \,V(x)\left[R^{\dagger}R+L^{\dagger}L\right].
\end{align}
Without loss of generality, we take the random potential $V(x)$ to
have zero-mean and Gaussian statistics characterized by disorder
average
\begin{align}
  \overline{V(x)V(y)}=\Delta \delta(x-y),
  \label{VVcorr}
\end{align}
with variance amplitude, $\Delta$.
	
Within the stable hLL phase, the interaction is dominated by
forward-scattering, given by
\begin{align}\label{Eq:H_hLL_int}
  H_{\text{int}}=
  &\int\limits_x:\left[UR^{\dagger}R(x+\alpha)R^{\dagger}R(x)\right.
  \nonumber\\
  +&U'\left.R^{\dagger}R(x+\alpha)L^{\dagger}L(x)+(R\to L):\right]
\end{align}
where $U$ and $U'$ are the screened short-range components of Coulomb
interaction and $\alpha$ is the ultraviolet cutoff length scale. We
neglect the backscattering components (in the presence of RSOC)
\cite{Schmidt2012,Kainaris2014,Chou2015} since they are subdominant in
the regime studied in this work.
	
The Hamiltonian
$H_{\text{hLL}}=H_{\text{0}}+H_{\text{int}}+H_{\text{dis}}$ given by
Eqs.~(\ref{Eq:H_hLL_kinetic}),~(\ref{Eq:H_dis}),
and~(\ref{Eq:H_hLL_int}) is the minimal model of the interacting,
dirty helical edge of a topological insulator protected by TR
symmetry. As we will see next, the model is exactly solvable by
bosonization, allowing a nonperturbative description of TI's helical
edge.
	
\subsection{Bosonization}
\label{Sec:single_edge_bosonization}
	
To treat Luttinger interaction and disorder,
$H_{\text{int}}+H_{\text{dis}}$ nonperturbatively we utilize a
standard bosonization analysis \cite{Shankar_Book,Giamarchi_Book},
summarized in Appendix~\ref{App:bosonization_convention}. Using the imaginary-time path-integral
formalism, the disordered helical Luttinger liquid is characterized by
the imaginary-time action,
$\mathcal{S}=\mathcal{S}_{\text{hLL}}+\mathcal{S}_{\text{dis}}$, where
\begin{align}
  \label{Eq:S_hLL}
  \mathcal{S}_{\text{hLL}}=&\int\limits_{\tau,x}\left\{\frac{i}{\pi}\left(\partial_x\theta\right)\left(\partial_{\tau}\phi\right)+\frac{v}{2\pi}\left[K\left(\partial_x\phi\right)^2+\frac{1}{K}\left(\partial_x\theta\right)^2\right]\right\},\\
  \mathcal{S}_{\text{dis}}=&\int\limits_{\tau,x}V(x)\frac{1}{\pi}\partial_x\theta,
\end{align}
with $\theta$ the phonon-like boson field and $\phi$ the phase boson
field. The number density and number current operators are given by
$\rho=\frac{1}{\pi}\partial_x\theta$ and
$J=-\frac{1}{\pi}\partial_t\theta$, respectively. Although the action
$S_{\text{hLL}}$ takes the form of a conventional spinless Luttinger
liquid (LL) \cite{Giamarchi_Book}, the physics of this {\em helical}
LL differs significantly because of distinct TR transformations of
$\theta$ and $\phi$ here, due to nontrivial spin content of the
corresponding helical edge fermions (see Appendix~\ref{App:bosonization_convention}). As noted above this latter
property has important physical manifestations, as for example
forbidding potential impurity backscattering in the absence of umklapp
interactions.
	
We note that the forward-only scattering disorder, can be fully
non-perturbatively taken into account by shifting $V(x)$ from the
action via a linear transformation on $\theta$,
$\mathcal{S}_{\text{hLL}}[\theta,\phi]+\mathcal{S}_{\text{dis}}[\theta]\rightarrow\mathcal{S}_{\text{hLL}}[\tilde{\theta},\phi]+\text{constant}$,
where
\begin{align}
  \tilde{\theta}(\tau,x)=\theta(\tau,x)+\frac{K}{v}\int_{-\infty}^{x}V(y)dy.
\end{align}
	
Under this shift, the correlation functions of $\theta$ transform
covariantly. For instance,
\begin{align}
  \left\langle e^{-in\theta(\tau,x)}e^{in\theta(0,0)}\right\rangle
  =e^{-i\frac{nK}{v}\int_{0}^{x}V(y)dy}\left\langle
    e^{-in\tilde\theta(\tau,x)}e^{in\tilde\theta(0,0)}\right\rangle,
\end{align}
shifts by a $V(x)$-dependent phase factor, that now allows for an
exact disorder average of the correlation function. In the above, $n$ is a constant controlling the scaling dimension of the operator. Gaussian
statistics of $V(x)$, with variance (\ref{VVcorr}) then gives
\begin{align}\label{Eq:dis_ave_corr}
  \overline{\left\langle
      e^{-in\theta(\tau,x)}e^{in\theta(0,0)}\right\rangle}
  =e^{-\frac{n^2K^2\Delta}{2v^2}|x|}\left\langle
    e^{-in\tilde\theta(\tau,x)}e^{in\tilde\theta(0,0)}\right\rangle.
\end{align}
Forward-scattering disorder thus suppresses power-law Luttinger liquid
correlations, cutting them off exponentially beyond a correlation
length $\xi = 2v^2/(n^2 K^2\Delta)$, that in momentum space
corresponds to smearing the disorder-free power-law peak via a
convolution with a Lorentzian, with width set by $1/\xi$.
	
\subsection{Spectral function}
	
\begin{figure}[t!]
\includegraphics[width=0.35\textwidth]{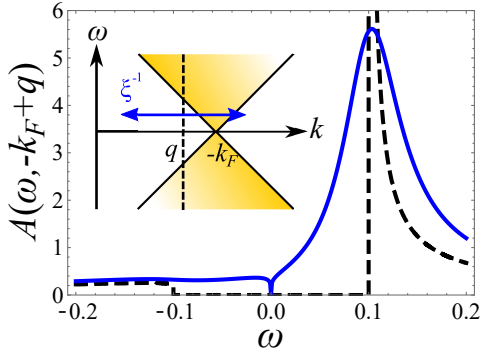}
\caption{$\textit{Zero-temperature}$ spectral function along the cut
  through $k=-k_{F}+q$ indicated by a dashed line in the inset with full
  (dashed) curve for disordered (clean) case. The interaction
  parameter $\gamma\equiv\frac{1}{4}\left(K+K^{-1}\right)-\frac{1}{2}$
  is set to 0.1. Inset: The spectral function in the
  vicinity of the left Fermi point. The yellow shaded region indicates
  finite weight of the clean spectral function. The width of the blue
  double arrow is the inverse length scale ($\xi^{-1}$) set by the
  strength of forward-scattering disorder.}
\label{Fig:Spectral_fcn_dis_vs_clean}
\end{figure}
	
\begin{figure}[t!]
  \includegraphics[width=0.35\textwidth]{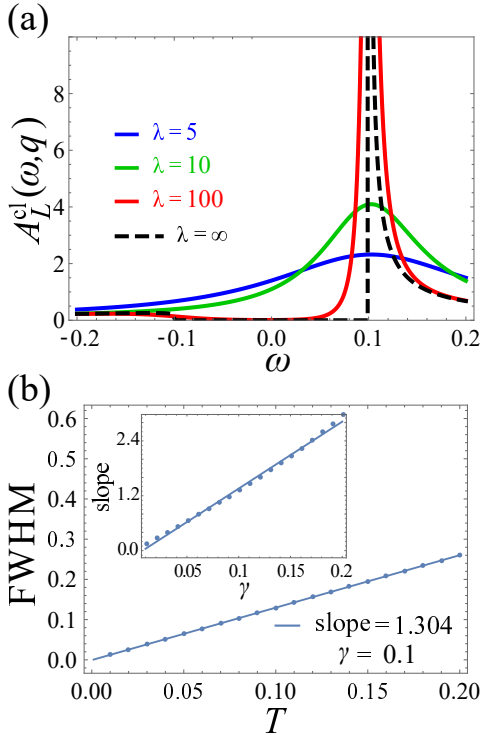}
  \caption{$\textit{Finite-temperature}$ clean (disorder-free) spectral function for a
    set of temperatures characterized by thermal de Broglie length,
    $\lambda=v\beta$. (a) The interaction parameter $\gamma=0.1$ and
    $vq=-0.1$. (b) Quasiparticle peak width as a function of
    temperature $T$ and $\gamma$. The linear dependence on $T$ and
    $\gamma$ shows $\text{FWHM}\approx 4\pi\gamma T$. The frequency and the 
    length are in units of $v\alpha^{-1}$ and $\alpha$ respectively.}
  \label{Fig:Spectral_fcn_cl}
\end{figure}
	
\begin{figure}[t!]
  \includegraphics[width=0.35\textwidth]{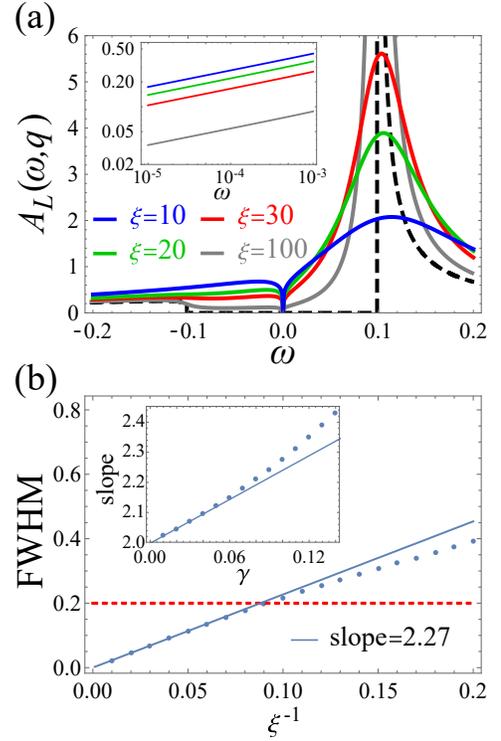}
  \caption{$\textit{Zero-temperature}$ spectral function with
    $\textit{forward-scattering disorder}$. The interaction parameter
    $\gamma=0.1$ and $vq=-0.1$. (a) Zero bias anomaly (ZBA) appears at
    $\omega=0$ for all disorder strengths $\xi^{-1}$, with the same
    exponent $2\gamma$ (inset), where
    $\xi=\frac{2v^2}{K^2\Delta}$. (b) The quasiparticle peak is
    broadened by disorders with a (half) width $~\xi^{-1}$ for
    $\text{FWHM}<|vq|$ (dashed red line), beyond which ZBA modifies the
    linear dependence. Inset: The slope $=2$ in the noninteracting
    limit and the disorder-strength dependence becomes more sensitive
    for stronger interaction. The frequency and the 
    length are in units of $v\alpha^{-1}$ and $\alpha$ respectively.}
  \label{Fig:Spectral_fcn_dis_T0}
\end{figure}
	
\begin{figure}[t!]
  \includegraphics[width=0.35\textwidth]{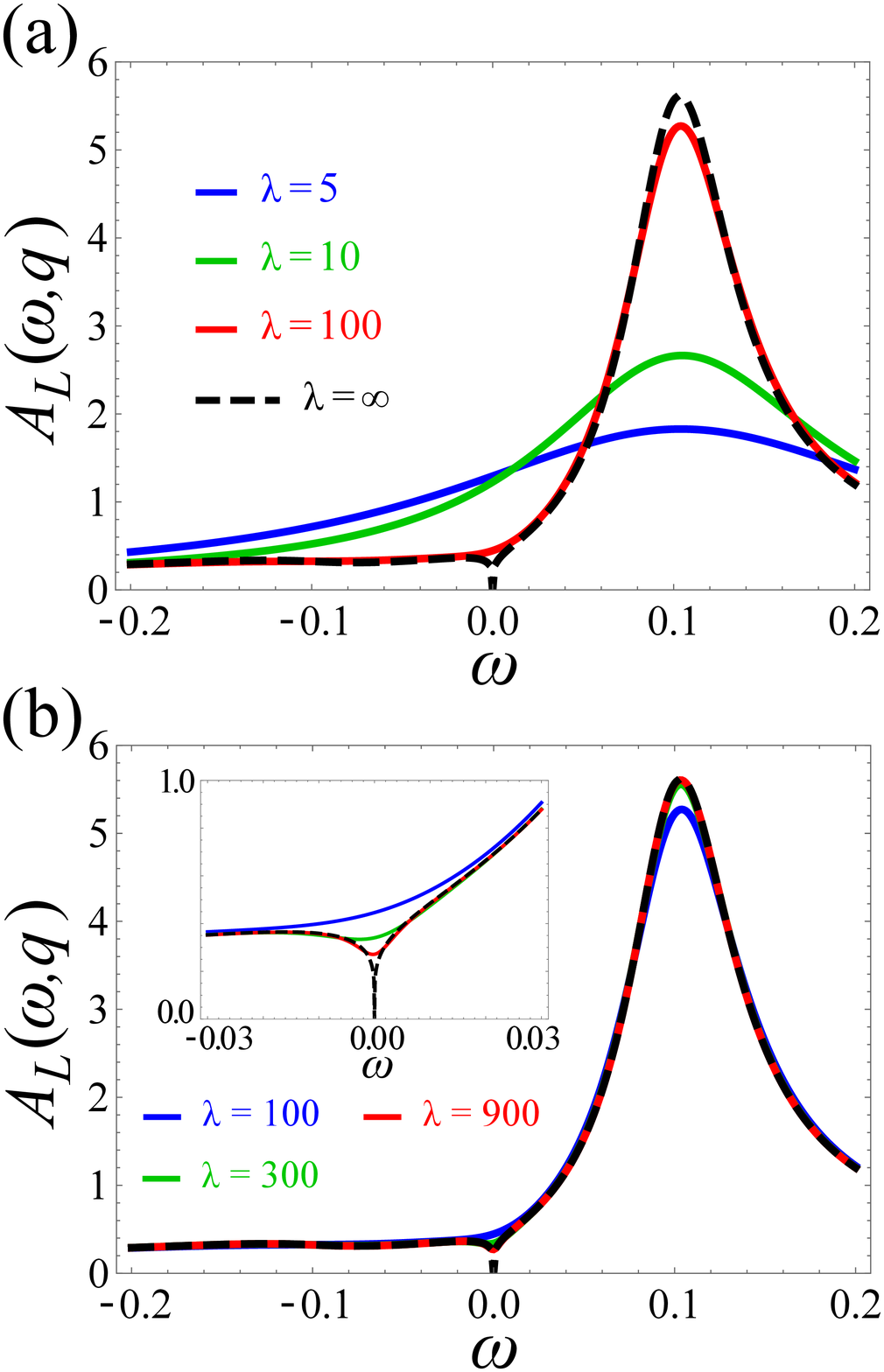}
  \caption{$\textit{Finite-temperature}$ spectral function with
    $\textit{forward-scattering disorder}$. (a) High temperature
    regime $\lambda\ll\xi$. (b) Low temperature regime
    $\lambda\gg\xi$. The disorder length $\xi=30$, the interaction
    parameter $\gamma=0.1$ and $vq=-0.1$. The frequency and the 
    length are in units of $v\alpha^{-1}$ and $\alpha$ respectively.}
  \label{Fig:Spectral_fcn_dis}
\end{figure}
	
\subsubsection{Clean spectral function}
	
In the clean limit, the imaginary time-ordered, single particle
space-time Green function at finite temperature is well-known for a
spinless LL \cite{Giamarchi_Book}. Although physically hLL and LL are
quite distinct, because the actions of the two systems are identical
at a leading order, we find that the single-edge spectral function for
a hLL is identical to that of a spinless LL. The calculation can be
carried out at zero temperature followed by a conformal mapping [a
mapping from a $(\tau,x)$ 2D plane to a cylinder in the
space-imaginary time domain] to get the finite temperature
expression. The finite temperature Green function can be also obtained
directly through the Matsubara technique. We provide a complemented
derivation using the latter approach in Appendix~\ref{App:Der_space_time_G}. Both analyses consistently give the single
particle imaginary time-ordered Green function for the right and left
movers,
\begin{align}
  &\mathcal{G}_{R}(\tau,x) =-\langle
  \hat{T}_{\tau}R(\tau,x)R^{\dag}(0,0)\rangle
  \nonumber\\
  &=-\frac{i}{2\pi\alpha}\frac{(\frac{\pi\alpha}{\beta
      v})^{2\gamma+1}}{\left[\sinh\left(\frac{\pi(x+iv\tau)}{\beta
          v}\right)\right]^{\gamma+1}\left[\sinh\left(\frac{\pi(x-iv\tau)}{\beta
          v}\right)\right]^{\gamma}}
  \\
  &\mathcal{G}_{L}(\tau,x) =-\langle
  \hat{T}_{\tau}L(\tau,x)L^{\dag}(0,0)\rangle
  \nonumber\\
  &=\frac{i}{2\pi\alpha}\frac{(\frac{\pi\alpha}{\beta
      v})^{2\gamma+1}}{\left[\sinh\left(\frac{\pi(x+iv\tau)}{\beta
          v}\right)\right]^{\gamma}\left[\sinh\left(\frac{\pi(x-iv\tau)}{\beta
          v}\right)\right]^{\gamma+1}},
  \label{Eq:G in space-time}
\end{align}
where $\alpha$ is the ultraviolet cutoff length scale,
$\gamma=\frac{1}{4}(K+K^{-1})-\frac{1}{2}$ and $\hat{T}_{\tau}$ denotes imaginary-time ordering. The spectral function can
be computed in the standard way by Fourier transforming the imaginary
time-ordered Green function $\mathcal{G}_{R/L}(\tau,x)$ and then
analytically continuing to real frequencies
$i\omega_n\to\omega+i\eta$, where $\eta\rightarrow 0^+$. The disorder-free (``clean'') spectral
function $A^{\text{cl}}(\omega,q)$ is then given by
\begin{equation}\label{Eq:SpectralF_formula}
  A^{\text{cl}}_{R/L}(\omega,q)=-\frac{1}{\pi}\text{Im}[G^{\text{ret}}_{R/L}(\omega,q)],
\end{equation}
where the retarded Green function $G^{\text{ret}}_{R/L}(\omega,q)$ is
computed using standard analysis, detailed in Appendix~\ref{App:Der_ret_G},
\begin{align}
  \label{Eq:G_clean}
  G^{\text{ret}}_{R/L}(\omega,q)=&i\frac{\beta(\frac{2\pi\alpha}{\beta
      v})^{2\gamma}}{4\pi^2}\sin\left(\pi\gamma\right)
  \nonumber\\
  &\times B\left[-i\frac{\beta(\omega_{\eta}\mp
      vq)}{4\pi}+\frac{\gamma}{2},1-\gamma\right]
  \nonumber\\
  &\times B\left[-i\frac{\beta(\omega_{\eta}\pm
      vq)}{4\pi}+\frac{\gamma+1}{2},-\gamma\right]
\end{align}
with $q=k\mp k_F$ for the right (subscript $R$) and left (subscript
$L$) movers, respectively. In Eq.~(\ref{Eq:G_clean}), $B$ is the Euler
Beta function and $\omega_{\eta}\equiv \omega+i0^+$. To the best of
our knowledge, the full expression of $G^{\text{ret}}_{R/L}(\omega,q)$
has not appeared in the literature, with only the imaginary part (or
the greater/lesser Green functions) given in Ref.~\cite{Orgad2001}. A
few remarks of our results: (i) In doing Fourier transformation, we
consider the approximate space-time Green function valid for
$v\tau,x>\alpha$, (ii) The zero temperature limit of
Eq.~(\ref{Eq:G_clean}) is in good agreement with the result in
Ref.~\cite{Meden1992} (both real and imaginary parts) for $\gamma<0.5$
at low energy $\omega/v,q<1/\alpha$, (iii) The finite temperature
spectral function derived from Eq.~(\ref{Eq:G_clean}) is consistent
with the result in Ref.~\cite{Orgad2001}, (iv) Our expression satisfies the
Kramers-Kronig relation for $\gamma<0.5$.
	
We first discuss the clean spectral function. At zero temperature, the
spectral weight is constrained within the ``light cone'' [the yellow
shaded region in the inset of
Fig.~\ref{Fig:Spectral_fcn_dis_vs_clean}]. The quasiparticle peak is a
power-law singularity located at $\omega=-vq$ for the left mover, and
at $\omega=vq$ for the right mover, with the exponent
$\gamma=\frac{1}{4}(K+K^{-1})-\frac{1}{2}$, as illustrated in
Fig.~\ref{Fig:Spectral_fcn_dis_vs_clean}
\cite{Luther1974,Meden1992,Voit1993,Voit1995}. We plot in
Fig.~\ref{Fig:Spectral_fcn_cl}(a) the non-zero temperature,
disorder-free (left) spectral function for different values of thermal
length $\lambda=v\beta$ and $vq=-0.1$, illustrating thermal broadening of the ``light cone''
constraint. Throughout this paper, we plot the spectral function (and the momentum-resolved tunneling spectroscopy in the next section) in the low energy regime $\omega/v,q<\alpha^{-1}$ (note $\alpha=1$), where our low-energy Hamiltonian is valid.
	
The power-law threshold singularity is smeared at finite temperature, displaying
low temperature $\lambda^{-1}\ll |q|$ (quantum) and high temperature
$\lambda^{-1}\gg |q|$ (classical) regimes. For the former, the
quasiparticle peak remains asymmetric, while for the latter, the
smeared peak approaches a Lorentzian at high temperature. The
broadening of the peak is nicely captured by a $2\pi\gamma T$
inelastic rate as discussed by Le Hur \cite{Le_Hur2002,Le_Hur2006}. We
note that the linear in $T$ and $\gamma$ broadening is very robust
starting from low temperature until $T$ becomes comparable to the
ultraviolet cutoff, as illustrated in
Fig.~\ref{Fig:Spectral_fcn_cl}(b).
	
The features discussed above can be understood in the following. The
Beta functions in the exact expression (\ref{Eq:G_clean}) can be
expressed through an integral identity,
\begin{align}
  B\left(-i\frac{\kappa}{2}+\frac{C}{2},1-C\right)&=2\int_{0}^{\infty}d\xi
  e^{i\kappa\xi}\left(2\sinh\xi\right)^{-C},
  \label{Eq:Beta_fn_integral_form}
\end{align}
which gives the retarded Green function expressed as integrals over
the light-cone coordinates $\xi_{\pm}=vt\pm x$ $(\tau\sim it)$
\begin{align}
  G^{\text{ret}}_{R/L}(\omega,q)=&\frac{i}{2\beta
    v^2}\sin\left(\pi\gamma\right)(\frac{\pi\alpha}{\beta
    v})^{2\gamma}
  \nonumber\\
  &\times\int_{0}^{\infty}d\xi_{\pm}e^{i(\omega\mp
    vq)\xi_{\pm}/2v}\sinh\left(\frac{\pi\xi_{\pm}}{\beta
      v}\right)^{-\gamma}
  \nonumber\\
  &\times\int_{0}^{\infty}d\xi_{\mp}e^{i(\omega\pm
    vq)\xi_{\mp}/2v}\sinh\left(\frac{\pi\xi_{\mp}}{\beta
      v}\right)^{-\gamma-1}.
  \label{Eq:G_clean_integral_form}
\end{align}
The low-temperature ($\omega\mp vq\gg\frac{1}{\beta}$) power-law and
high-temperature ($\omega\mp vq\ll\frac{1}{\beta}$) Lorentzian forms
of the quasiparticle peak respectively correspond to the two different
limits of integral representation in
Eq.~(\ref{Eq:Beta_fn_integral_form}): $\sinh(x)\approx x$ for $|x|\ll
1$ and $\sinh(x)\approx \sgn(x)e^{-|x|}/2$ for $|x|\gg 1$.
	
\subsubsection{Disorder-averaged spectral function}
	
In the presence of disorder, the momentum is no longer a good quantum
number. However, generic spectroscopic experiments probe the
disorder-averaged spectral function, analysis of which we discuss
next. As emphasized in Sec.~\ref{sec:Single_Edge}, TR invariance
constrains heterogeneities to nonmagnetic impurities that can only
forward-scatter. The resulting disorder can thus be treated exactly
and in real space is given by Eq.~(\ref{Eq:dis_ave_corr}). In momentum
space, disorder thus smears the disorder-free spectral function
through its convolution with a Lorentzian, and is given by,
\begin{align}\label{Eq:G_dis}
  \overline{A_{R/L}(\omega,q)}=&\int_{-\infty}^{\infty}dk\frac{\xi^{-1}/\pi}{(k-q)^2+\xi^{-2}}
  A^{\text{cl}}_{R/L}(\omega,k),
\end{align}
illustrated in Fig.~\ref{Fig:Spectral_fcn_dis_vs_clean}, where
$\xi=\frac{2v^2}{K^2\Delta}$ is the mean-free path set by the
forward-scattering disorder. Despite this expected smearing of sharp
features by disorder, we observe that disorder-averaged spectral
function, $\overline{A_{R/L}(\omega,q)}$, illustrated in
Fig.~\ref{Fig:Spectral_fcn_dis_T0}(a) in fact exhibits (even at finite
momentum $q$) a disorder-induced zero-bias anomaly (ZBA),
$\overline{A_{R/L}(\omega,q)}\propto \mathcal{C}|\omega|^{2\gamma}$
\cite{Giamarchi_Book}, with exponent $\gamma$ and amplitude
$\mathcal{C}=\frac{1}{\pi^2
  v}\sin\left(2\pi\gamma\right)|\Gamma(-2\gamma)|(\frac{\alpha}{v})^{2\gamma}\frac{\xi^{-1}}{q^{2}+\xi^{-2}}$,
that is independent of disorder strength.  The origin of this finite
$q$ ZBA is most transparent in the strong disorder limit ($\xi q\ll
1$), where we can approximate the Lorentzian in Eq.~(\ref{Eq:G_dis})
simply by a constant $\xi/\pi$, with the convolution thereby reducing
to an integral over $k$, giving a local density of states, which is
known to exhibit a ZBA \cite{Giamarchi_Book}. Physically, this
counter-intuitive effect is due to impurities providing the momentum
needed to shift the $q=0$ zero-frequency anomaly to a finite momentum
$q$.

In contrast, the power-law peak at $\omega = v q$ is indeed broadened
by disorder, with the width $\propto K^{2}\Delta/v$, decreasing with
stronger repulsive interactions, in contrast to thermal effects in
disorder-free system discussed above [see
Fig.~\ref{Fig:Spectral_fcn_dis_T0}(b)].
	
In the presence of both finite temperature and disorder one expects a
broadening of the disorder-free, zero-temperature spectral
function. Indeed we find that at high temperature, such that
$\lambda\ll\xi$, the broadening of the quasiparticle peak is dominated
by thermal effect, with spectral function reducing to the finite $T$
clean case [see Fig.~\ref{Fig:Spectral_fcn_dis}(a)].  In particular,
the quasiparticle peak approaches a Lorentzian with a (half)
width $\approx 2\pi\gamma T+v\xi^{-1}$, corresponding to temporal exponential decay rate of the momentum-time Green function [read by a replacement $\tau\to it$ and $x\to -vt$ in Eq.~(\ref{Eq:G in space-time})] \cite{Le_Hur2006} at high temperature. As we will show below,
the prediction of the peak width in the high temperature limit works
surprisingly well even at low temperature.
	
Instead, at low temperatures, such that $\lambda\gg\xi$, the spectral
peak broadening is dominated by disorder as is clearly reflected in
Fig.~\ref{Fig:Spectral_fcn_dis}(b). We note the ZBA at $\omega=0$ is
thermally rounded for $\omega\ll\omega^{*}(T)=v/\lambda\approx
T$. This can be understood in the following way: the disorder-induced
exponential decay results in an \text{effective} constraint $|x|=|\xi_{+}-\xi_{-}|/2\approx
0$ in Eq.~(\ref{Eq:G_clean_integral_form}), giving
\begin{align}
  G^{\text{ret}}_{\text{dis},R(L)}(\omega,q)\approx &\frac{i}{\beta
    v}\sin\left(\pi\gamma\right)(\frac{\pi\alpha}{\beta v})^{2\gamma}
  \nonumber\\
  &\times\int_{-\infty}^{\infty}dxe^{-iqx}e^{-\frac{|x|}{\xi}}
  \nonumber\\
  &\times\int_{0}^{\infty}dt e^{i\omega_{\eta} t}\sinh\left(\frac{\pi
      t}{\beta}\right)^{-2\gamma-1},
  \label{Eq:G_dis_low_T_integral_expression}
\end{align}
working in the strong disorder limit, so the integral domain of $x$
may be extended to infinity. Using the definition of Beta function in
Eq.~(\ref{Eq:Beta_fn_integral_form}) then gives
\begin{subequations}\label{Eq:G_dis_low_T}
  \begin{align}
    G^{\text{ret}}_{\text{dis},R(L)}(\omega,q)\approx&\frac{i}{\pi
      v}\sin\left(\pi\gamma\right)(\frac{2\pi\alpha}{\beta
      v})^{2\gamma}\frac{2\xi^{-1}}{q^2+\xi^{-2}}
    \nonumber\\
    &\times
    B\left(-i\frac{\beta\omega_{\eta}}{2\pi}+\frac{2\gamma+1}{2},-2\gamma\right),\\
    \propto&\begin{cases}
      \omega^{2\gamma}, &\text{ for }\omega\gg\omega^*,\\
      T^{2\gamma}, &\text{ for }\omega\ll\omega^*.
    \end{cases}
  \end{align}
\end{subequations}
The full Beta function encodes the crossover between
$\omega^{2\gamma}$ for high frequency $\omega\gg\omega^*$ (low $T$)
and $T^{2\gamma}$ at low frequency $\omega\ll\omega^*$ (high $T$). The
former is precisely the ZBA discussed above; the latter is consistent
with the result previously reported by Le Hur \cite{Le_Hur2006}.
	
\subsubsection{Asymptotic expression}
	
\begin{figure}[t!]
		\includegraphics[width=0.35\textwidth]{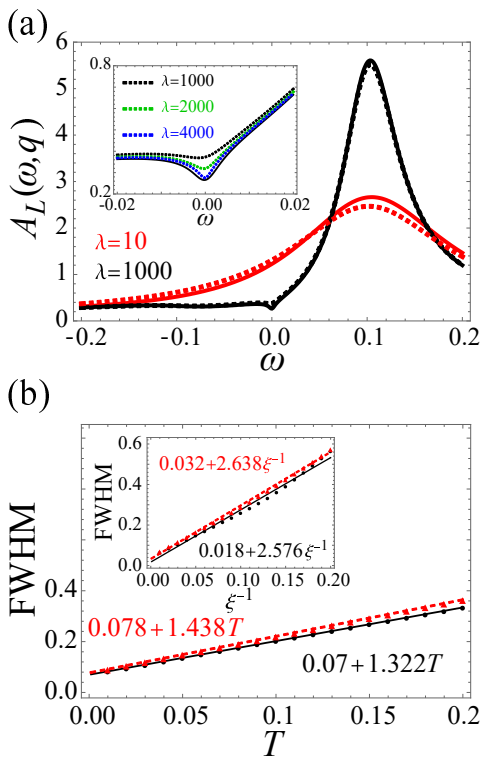}
                \caption{$\textit{Finite-temperature}$ spectral function with
                  $\textit{forward-scattering disorder}$ plotted using the
                  exact (solid line) and the asymptotic (dashed line)
                  Green functions. (a) Spectral function at low
                  ($\lambda=1000$) and high ($\lambda=10$)
                  temperatures.  Inset: asymptotic spectral function
                  for $\lambda=1000,2000,4000$ and the exact spectral
                  function for $\lambda=1000$. (b) Quasiparticle peak
                  width as a function of temperature $T$ and
                  $\xi^{-1}$ (inset). In the above, the interaction
                  parameter $\gamma=0.1$ and $vq=-0.1$. The frequency and the 
                  length are in units of $v\alpha^{-1}$ and $\alpha$ respectively.}
\label{Fig:Spectral_fcn_asym_vs_exact}
\end{figure}
	
In the low-temperature limit, the convolution expression (\ref{Eq:G_dis})
for the spectral function at a finite temperature and disorder, can be
simplified by using the Stirling formula for the single-particle Green
function. We thereby obtain the following asymptotic form
\begin{align}
\label{Eq:G_clean_asym}
G^{\text{ret}}_{L}(\omega,q)\sim&
-i\left(\frac{\alpha}{2v}\right)^{2\gamma}\frac{\Gamma\left(1-\gamma\right)}{\Gamma\left(1+\gamma\right)}\left[-i(\omega+vq)+2\pi\gamma
  T\right]^{\gamma-1}
\nonumber\\
&\times\left[-i(\omega-vq)+2\pi(\gamma+1)T\right]^{\gamma},
\end{align}
that allows us to carry out the convolution in Eq.~(\ref{Eq:G_dis})
and obtain the asymptotic expression for the disorder-averaged
low-temperature Green function (see
Appendix~\ref{App:Der_ret_G_dis}). By choosing a complex contour on
the upper complex plane, the disordered Green function is given by
\begin{align}
  \label{Eq:G_dis_asym}
  G^{\text{ret}}_{\text{dis},L}(\omega,q)\sim
  G^{\text{ret}}_{L}(\omega,q+i\xi^{-1})+G^{\text{ret}}_{2,L}(\omega,q),
\end{align}
where the first term on the right hand side is the residue from the
Lorentzian function and the second term comes from the integral around
the branch cut, evaluated in Appendix~\ref{App:Der_ret_G_dis} with the
result given in Eq.~(\ref{Eq:G_dis_2}). From this asymptotic
expression, we expect the quasiparticle peak to be located at
$\omega=-vq$ with an exponent $\gamma-1$ broadened by thermal and
disorder effects to a width $\approx 2\pi\gamma T+v\xi^{-1}$. The zero
bias anomaly at $\omega=0$ has exponent $2\gamma$ and is rounded only
by the thermal effects with a scale $2\pi(2\gamma+1)T$. As shown in
Fig.~\ref{Fig:Spectral_fcn_asym_vs_exact}, the asymptotic formula
gives a good approximation to the exact spectral function, especially
at low temperature where Stirling formula approximation is valid. At
zero temperature, this asymptotic prediction becomes exact as
Eq.~(\ref{Eq:G_clean_asym}) is exact under such condition. However,
this analytical expression only \textit{asymptotically} captures the
low temperature behavior of the zero-bias anomaly 
[inset of Fig.~\ref{Fig:Spectral_fcn_asym_vs_exact}(a)]. Nevertheless,
the quasiparticle peak is well described by the asymptotic formula,
showing a peak width $\approx 2\pi\gamma T+v\xi^{-1}$ in
Fig.~\ref{Fig:Spectral_fcn_asym_vs_exact}(b).
	
As we have seen above, the spectral function of a single helical edge
reveals the fractionalized properties of the hLL. However, (except for
absence of Anderson localization due to the forbidden disorder elastic
backscattering) it fails to distinguish the helical edge of a TI from
a conventional LL as for example describing a spin-polarized
one-dimensional conductor.
	
To bring out special properties of the hLL, we thus next turn to the
analysis of the momentum and energy resolved inter-helical-edge
tunneling, developing the theory of MRTS.
	
\section{Two edges: Momentum-resolved tunneling}
\label{sec:MRTS}

We study the momentum and energy resolved interedge tunneling
spectroscopy, which, as we will show exhibits distinctive signatures
of the hLL, characterizing an edge of a time-reversal invariant
topological insulator with $K>3/8$
\cite{Wu2006,Xu2006,Chou2018EdgeGlass}. A schematic of a vertical (co-planar)
geometry of an experimental setup that we study is illustrated in
Fig.~\ref{Fig:vertical setup} (Fig.~\ref{Fig:horizontal setup}). This is the TI edge counter-part of
the setup studied for a conventional LL in \cite{Carpentier2002} and
demonstrated experimentally
\cite{Auslaender2002,Steinberg2008,Jompol2009,Tsyplyatyev2015,Tsyplyatyev2016,Jin2019}.
In such a setup, the momentum transfer $Q=2\pi Bd/\phi_0$ and
frequency $\omega = eV/\hbar$ can be independently tuned by a
transverse magnetic field $B$ and interedge source-drain bias $V$. In
the above, $d$ denotes the distance between two edges, $\phi_0=h/e$ is
the magnetic flux quantum and $e>0$ is the elementary charge.
	
In the rest of the section, we first derive the tunneling current from
linear response theory. Then, bosonization is employed to anticipate
both the intraedge and the interedge interactions. We discuss various
situations ranging from the quantum spin Hall limit ($S_z$ spin
conservation) to the generic situations (i.e., Rashba spin orbit
coupling, disorder, distinct edge velocities and interaction
strengths). The analytical expressions for the finite-temperature
tunneling currents (with the same edge velocity) are the main new
results of this work.
	
\subsection{Tunneling current}
	
Following Ref.~\cite{Carpentier2002}, we consider two parallel quantum
edges with a separation that allows weak interedge tunneling
current. The coupled edges Hamiltonian is given by
$H=H_{1}+H_{2}+H_{\text{int}}+H_{\text{tun}}$, where
\begin{align}\label{Eq:H_12}
  H_{a}=&\sum_{\alpha=\pm}\int\limits_k\left[\varepsilon_{a\alpha}(k)-\mu_{a}\right]c^{\dagger}_{a\alpha}(k)c_{a\alpha}(k),\\
  H_{\text{int}}=&U_{12}\int\limits_x\rho_1(x)\rho_2(x)+\sum_{a=1,2}U_a\int\limits_x\rho_a(x)\rho_a(x),\\
  H_{\text{tun}}=&-t_0\sum_{s=\uparrow\downarrow}\int\limits_x
  \left[c^{\dagger}_{1s}(x)c_{2s}(x)+c^{\dagger}_{2s}(x)c_{1s}(x)\right].
\end{align}
In the above expressions,
$\varepsilon_{a\alpha}(k)=\epsilon_{a\alpha}(k)-\epsilon_{a\alpha}(k_{F,a\alpha})$
with $\epsilon_{a\alpha}(k)$ the band dispersion for edge $a=1,2$,
$U_a>0$ ($U_{12}>0$) is the intraedge (interedge) Coulomb interaction
(screened by a gate), $\rho_1$ ($\rho_2$) is the density of edge 1
(edge 2), $t_0$ is the interedge tunneling amplitude, $c_{a\alpha}$ is
the annihilation operator for the chiral fermion with chirality
$\alpha=+/-$ (not to be confused with the ultraviolate length scale)
on the edge $a$, and $c_{as}$ is the annihilation operator for the
physical fermion with spin $s$ on the edge $a$.  We will consider the
electrochemical potentials $\mu_1=eV$ ($e>0$) and $\mu_2=0$ such that
current flows from edge 1 to 2 (2 to 1) for positive (negative)
interedge source-drain bias $V$.  Importantly, $c_{a\alpha}(k)$ and
$c_{as}(k)$ are related to each other via
Eqs.~(\ref{Eq:field_op_band_basis}) and
(\ref{Eq:field_op_spin_basis}), detailed in Appendix~\ref{App:chiral
  decomposition}.
	
In the presence of an external magnetic field applied transversely to
the plane defined by the two edges, tunneling electrons experience a
Lorentz force, included through the Peierls substitution
$c^{\dag}_{2}c_{1}\to c^{\dag}_{2}c_{1}e^{i(-e/\hbar)\int^{d}_{0}dy
  A_{y}(x,y)}$, where $d$ is the interedge $y$ separation. For
magnetic field $\vec{B}=-B\hat{z}$, we choose the Landau gauge
$\vec{A}=-Bx\hat{y}$ in which the associated Berry phase is included
via the replacement $c_{1}(x)\to c_{1}(x)e^{iQx}$, where $Q=2\pi B
d/\phi_0$.  As a result, $H_{1}$, $H_{2}$, $H_{\text{int}}$ remain
unchanged and the tunneling operator, $H_{\text{tun}}$ is replaced by
\begin{align}
  H^{Q}_{\text{tun}}=&-t_0\sum_{s=\uparrow\downarrow}\int\limits_x
  \left[c^{\dagger}_{1s}(x)c_{2s}(x)e^{-iQx}+\text{H.c.}\right],
\end{align}
where H.c. denotes the Hermitian conjugate.
	
We are interested in the tunneling current from edge 1 to edge 2. This
can be derived by computing the time derivative of the charge in edge
1, $\hat{I}_{\text{tun}}=-\frac{1}{i}\left[eN_1,H\right]=\int
dx\hat{J}(x)$, where
\begin{align}
  \hat{J}(x)=
  iet_0\sum_{s}\left[c^{\dagger}_{2s}(x)c_{1s}(x)e^{iQx}-c^{\dagger}_{1s}(x)c_{2s}(x)e^{-iQx}\right]
\end{align}
is the tunneling current density. For the clean case, the expectation
value of the tunneling current density is position independent, and
thus the tunneling current $I_{\text{tun}}$ is proportional to the
length of the tunneling region. For disordered case that we treat
below, we will study disordered averaged current that is again
$x$-independent.
	
To compute the expectation value of the tunneling current density, we
work in interaction representation with respect to perturbation
$H^{Q}_{\text{tun}}$. We select $H_{12}=H_{1}+H_2+H_{\text{int}}$ and
$H_I=H^{Q}_{\text{tun}}$. The expectation value of the tunneling
current density at time $t$ is given by
\begin{align}
  \label{Eq:J_xt_formalism}
  J =\frac{1}{Z}\text{Tr}\left[e^{-\beta H_{12}}
    \hat{U}^{\dagger}(t)\hat{J}(x)\hat{U}(t) \right],
\end{align}
where $\hat{U}(t)=\hat{U}_{12}(t)\hat{U}_I(t)$,
$\hat{U}_{12}(t)=e^{-iH_{12}t}$,
$\hat{U}_I(t)=\hat{T}\exp\left[-i\int_{-\infty}^{t}dt'
  H^I_{\text{tun}}(t')\right]$ ($\hat{T}$ the time-ordering operator),
$H^I_{\text{tun}}(t)\equiv
e^{iH_{12}t}H^{Q}_{\text{tun}}e^{-iH_{12}t}$, and $\beta$ is the
inverse temperature. Importantly, $Z\equiv \text{Tr}[e^{-\beta
  H_{12}}]$ is the ``unperturbed'' partition function, with two edges
in thermal equilibrium at the same temperature (due to interedge
interaction), but kept at the electrochemical potential difference
$\mu_1-\mu_2 = eV$. Equation~(\ref{Eq:J_xt_formalism}) gives the
expectation of the tunneling current density at time $t$ corresponding
to turning on the single-particle tunneling in the infinite past. The
tunneling current $J$ is in the steady state, i.e., $t$ (and $x$)
independent, and clearly vanishes to $O(t_0)$. Relegating the details
to Appendix~\ref{App:Der_J}, standard analysis perturbative in $t_0$
to leading $O(t_0^2)$ order gives,
\begin{align}\label{Eq:J_xt_formalism2}
  J(\omega=eV/\hbar,Q)\approx et_0^2\left[J_{1\to 2}(\omega,Q)-J_{2\to 1}(\omega,Q)\right],
\end{align}
where
\begin{align}
  \nonumber J_{1\to 2}(\omega,Q)=&\sum_{s,s'}\int_{-\infty}^{\infty}\!\!dt'\int_{-\infty}^{\infty}\!\!dx'\,e^{i\omega t'}e^{-iQx'}\\
  &\times\left\langle c^{\dagger}_{1s'}c_{2s'}(t',x')c^{\dagger}_{2s}c_{1s}(0,0)  \right\rangle,\\
  \nonumber J_{2\to 1}(\omega,Q)=&\sum_{s,s'}\int_{-\infty}^{\infty}\!\!dt'\int_{-\infty}^{\infty}\!\!dx'\,e^{i\omega t'}e^{-iQx'}\\
  &\times\left\langle
    c^{\dagger}_{2s}c_{1s}(0,0)c^{\dagger}_{1s'}c_{2s'}(t',x')\right\rangle.
\end{align}
	
We calculate the tunneling current (\ref{Eq:J_xt_formalism}) using
bosonization and utilizing imaginary time and Matsubara analytic
continuation (see Appendix~\ref{App:analytic_con_J}). To this end,
using spectral decomposition, we relate physical current $J$ to the
Matsubara correlator $\mathcal{J}(i\omega_n,Q)$,
\begin{align}\label{Eq:FD_relation}
  J=2et_0^2\text{Im}\left[\mathcal{J}(i\omega_n\to \omega+i\eta,Q)\right],
\end{align}
where $\mathcal{J}(i\omega_n,Q)$ is a Fourier transform of the
imaginary-time ordered correlator defined by
\begin{align}
  \mathcal{J}(i\omega_n,Q)=&\int_{0}^{\beta}d\tau\int_{-\infty}^{\infty}dx
  e^{i(\omega_n\tau-Qx)}\mathcal{J}(\tau,x),
\end{align}
with the space-imaginary time correlation function given by
\begin{align}\label{Eq:current_imaginary_time_correlator}
  \mathcal{J}(\tau,x)=&\sum_{s,s'=\uparrow\downarrow}\left\langle\hat{T}_{\tau}c_{1s'}^{\dag}c_{2s'}(\tau,x)c_{2s}^{\dag}c_{1s}(0,0)\right\rangle.
\end{align}
	
\subsection{Bosonization}
As we have done in Sec.~\ref{Sec:single_edge_bosonization} for the
single-edge, here too we utilize standard bosonization to treat
Luttinger interaction and disorder to compute the interedge tunneling
current. The imaginary-time action of the two-edge setup (without
interedge tunneling) is given by
$\mathcal{S}=\mathcal{S}_{12}+\mathcal{S}_{\text{dis}}$, where
\begin{align}
  \label{Eq:S_12}
  S_{12}=&\sum_{a=1,2}\,\int\limits_{\tau,x}\left\{\frac{v_{a}}{2\pi}\left[K_{a}(\partial_x
      \phi_{a})^{2}+\frac{1}{K_{a}}(\partial_x
      \theta_{a})^{2}\right]\right.
  \nonumber\\
  &+\left.\frac{i}{\pi}\left(\partial_x\theta_a\right)\left(\partial_{\tau}\phi_a\right)\right\}+\frac{U_{12}}{\pi^2}\int\limits_{\tau,x}\left[\partial_x\theta_{1}(x)\right]\left[\partial_x\theta_{2}(x)\right]
  \nonumber\\
  \mathcal{S}_{\text{dis}}=&\int\limits_{\tau,x}\left[V_1(x)\frac{1}{\pi}\partial_x\theta_1+V_2(x)\frac{1}{\pi}\partial_x\theta_2\right],
\end{align}
Because they appear on distinct edges, we take the disorder potentials
$V_a(x)$ to be independent, zero-mean Gaussian fields with
$\overline{V_a(x)V_{a'}(y)}=\Delta_a \delta_{aa'}\delta(x-y)$.  We
ignore interedge backscattering interactions that are only relevant
under certain commensurate conditions \cite{Chou2019_2hLL}.  The
bosonized action $S_{12}$ (\ref{Eq:S_12}) is quadratic and therefore
can be written in diagonalized form. We provide the details of the
explicit transformation in Appendix~\ref{App:Bosonization_two_edge}
analogous to Ref.~\cite{Orignac2011}. After diagonalizing $S_{12}$,
the forward-scattering disorder, $S_{\text{dis}}$ can be taken into
account via a linear transformation on the $\theta_a$ fields. For
instance, in the limit $U_{12}=0$, where the action $S_{12}$ is in its
diagonalized form, the disorder-averaged correlation function is given
by
\begin{align}
  &\overline{\left\langle
      e^{-in_1\theta_1(\tau,x)}e^{in_2\theta_2(\tau,x)}e^{-in_2\theta_2(0,0)}e^{in_1\theta_1(0,0)}\right\rangle}
  \nonumber\\
  \nonumber=&e^{-\sum_{a}\frac{n_a^2K_a^2\Delta_a}{2v_a^2}|x|}\\
  &\times\left\langle
    e^{-in_1\tilde\theta_1(\tau,x)}e^{in_2\tilde\theta_2(\tau,x)}e^{-in_2\tilde\theta_2(0,0)}e^{in_1\tilde\theta_1(0,0)}\right\rangle.
\end{align}
We note that this is a generalized version of
Eq.~(\ref{Eq:dis_ave_corr}). For $U_{12}\neq 0$, one has to first
diagonalize the two-edge problem (see
Appendix~\ref{App:Bosonization_two_edge}), and then average over
disorder to obtain the disorder-averaged correlation function.
	
\subsection{$S_{z}$-conserved edge: quantum spin Hall limit}
For a 2D TI with an out-of-plane reflection symmetry
($z\to -z$), the spin quantization axis of the helical edge is
generally along this $z$-axis due to spin-orbit coupling of the form
$(\vec{p}\times\vec{E})\cdot\vec{\sigma}$, where electrons with
in-plane momentum $\vec{p}$ feels an out-of-plane ($z$-axis directed)
effective magnetic field due to the in-plane electric field (or
crystal field polarization) $\vec{E}$, by symmetry transverse to the
TI edge.  Such $S_z$-conserved topological insulator features
quantized spin-Hall conductance. It is important to note that $S_z$
conservation is not robust as RSOC generically
breaks any spin conservation.  However, it is helpful to first
consider this technically simpler special case.  More generic
non-spin-conserving case can be built from the results derived in this
section.
	
We first consider idealized case of $S_z$-conserved edges in the
absence of disorder or Zeeman field. At low source-drain bias, we
decompose the fermion fields so that the imaginary-time tunneling
current correlator in Eq.~(\ref{Eq:current_imaginary_time_correlator})
is written in terms of tunneling processes between different Fermi
points
\begin{align}
  \label{Eq:S_z_conserved_tunneling_matrix_elemtents}
	\mathcal{J}(\tau,x)=t^{RR}\mathcal{J}_{RR}+t^{LL}\mathcal{J}_{LL}+t^{RL}\mathcal{J}_{RL}+t^{LR}\mathcal{J}_{LR},
\end{align}
where $t^{RR}$, $t^{LL}$, $t^{RL}$, $t^{LR}$ are constants
proportional to the square of the tunneling matrix elements and
\begin{align}
  &\mathcal{J}_{RR}(\tau,x)=e^{-i\delta
    k_Fx}\left\langle\hat{T}_{\tau}R_1^{\dag}R_2(\tau,x)R_2^{\dag}R_1(0,0)\right\rangle,
  \nonumber\\
  &\mathcal{J}_{LL}(\tau,x)=e^{i\delta
    k_Fx}\left\langle\hat{T}_{\tau}L_1^{\dag}L_2(\tau,x)L_2^{\dag}L_1(0,0)\right\rangle,
  \nonumber\\
  &\mathcal{J}_{RL}(\tau,x)=e^{-ik_{F,T}x}\left\langle\hat{T}_{\tau}R_1^{\dag}L_2(\tau,x)L_2^{\dag}R_1(0,0)\right\rangle,
  \nonumber\\
  &\mathcal{J}_{LR}(\tau,x)=e^{ik_{F,T}x}\left\langle\hat{T}_{\tau}L_1^{\dag}R_2(\tau,x)R_2^{\dag}L_1(0,0)\right\rangle.
\end{align}
In the above, $\delta k_F=k_{F,1}-k_{F,2}$ and
$k_{F,T}=k_{F,1}+k_{F,2}$ ($k_{F,a}=k_{F,a\pm}$ as we assume TR
symmetry holds on each edge). The physical tunneling current $J$ can
then be obtained via the standard analytic continuation
(\ref{Eq:FD_relation}).
	
\subsubsection{Vertical geometry}

\begin{figure}[t!]
  \includegraphics[width=0.4\textwidth]{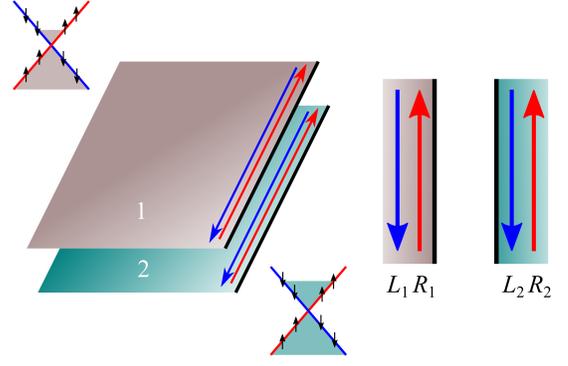}
  \caption{Vertical setup of two topological insulators in the quantum
    spin Hall limit. In both edges, right movers carry up spin and
    left movers carry down spins. The tunneling matrix elements
    between the two edge forbid any mixing of $L^{\dagger}_2R_1$ or
    $R^{\dagger}_2L_1$ (i.e. $t^{RL}=t^{LR}=0$). The tunneling current
    is govern by the momentum transfer of order $|k_{F,1}-k_{F,2}|$.}
  \label{Fig:vertical setup}
\end{figure}

\begin{figure}[t!]
\includegraphics[width=0.35\textwidth]{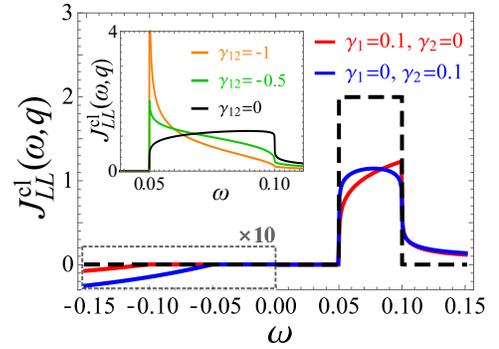}
\caption{$\textit{Zero-temperature}$ clean (disorder-free) tunneling
  current with $\textit{different}$ velocities for a set of
  interaction parameters. The edge velocities and momentum (magnetic
  field) are set to $v_1q=-0.1$ and $v_2q=-0.05$. $et_0^2=1$. The
  black dashed line indicates the non-interacting case
  ($\gamma_1=\gamma_2=0$). The red (blue) curve denotes the case of
  interacting edge 1 (edge 2), where finite current appears for
  $\omega<v_1q$ ($\omega<v_2q$) due to fractionalization in chiral
  degrees of freedom. The inset shows the effects of
  \textit{repulsive} interedge interaction. The interaction parameters
  and edge velocities are set to $\gamma_+=\gamma_-=0.05$, $v_+q=-0.1$
  and $v_-q=-0.05$. The frequency and the 
  length are in units of $v_1\alpha^{-1}$ and $\alpha$ respectively.}
\label{Fig:JLL_zeroT_cl}
\end{figure}
	
\begin{figure}[t!]
\includegraphics[width=0.35\textwidth]{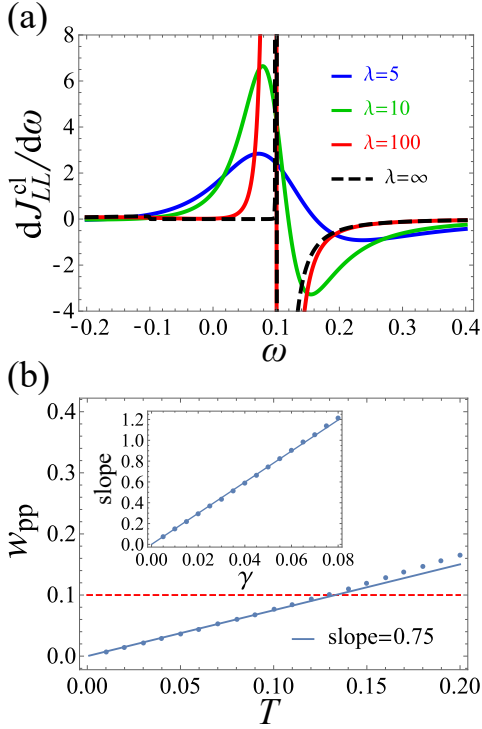}
\caption{$\textit{Finite-temperature}$ clean (disorder-free) differential tunneling
  conductance for \textit{identical} edge velocities with
  $vq=-0.1$. (a) At zero temperature, a positive delta function
  (negative power-law singularity) is located at $\omega=-vq\pm
  0^{+}$.  At finite temperatures, the two peaks are broadened and move
  towards the left and the right respectively. The intraedge
  interaction parameter is set to $2\gamma=\gamma_1+\gamma_2=0.1$.
  (b) Peak-to-peak distance ($w_{pp}$) as a function of $T$ and
  $\gamma$. The linear dependence on $T$ and $\gamma$ shows
  $w_{pp}\approx 7.5\gamma T$ for $w_{pp}<|vq|$ (red dashed line). The frequency and the 
  length are in units of $v\alpha^{-1}$ and $\alpha$ respectively.}
\label{Fig:JLL_finiteT_cl}
\end{figure}
	
\begin{figure}[t!]
\includegraphics[width=0.35\textwidth]{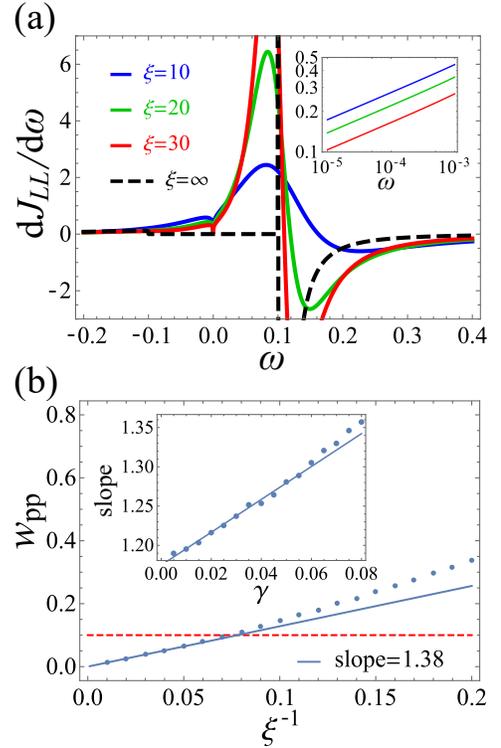}
\caption{$\textit{Zero-temperature}$ differential tunneling
  conductance with \textit{forward-scattering disorders} for
  \textit{identical} edge velocities. The momentum is taken to be
  $vq=-0.1$.  (a) Zero bias anomaly appears at $\omega=0$ for all
  disorder strengths $\xi^{-1}$, with the same exponent $4\gamma$
  (inset), where $\xi^{-1}=\frac{K_1^2\Delta_1}
  {2v_1}+\frac{K_2^2\Delta_2}{2v_2}$. The interaction parameter
  $\gamma=0.05$.  (b) Peak-to-peak distance ($w_{pp}$) as a function
  of $\xi^{-1}$ and $\gamma$. $w_{pp}$ is proportional to $\xi^{-1}$
  for $w_{pp}<|vq|$ (red dashed line). The disorder-strength
  dependence becomes more sensitive for stronger interaction (inset). The frequency and the 
  length are in units of $v\alpha^{-1}$ and $\alpha$ respectively.}
\label{Fig:JLL_zeroT_dis}
\end{figure}

\begin{figure}[t!]
\includegraphics[width=0.35\textwidth]{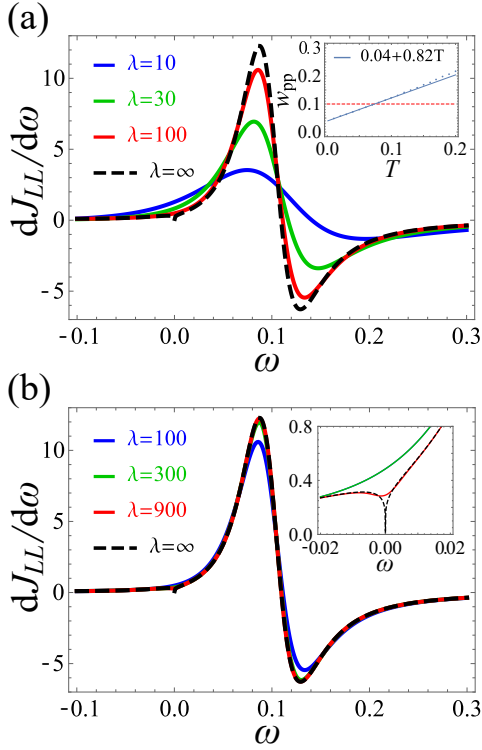}
\caption{$\textit{Finite-temperature}$ differential tunneling
  conductance with \textit{forward-scattering disorders} for
  \textit{identical} edge velocities. (a) High temperature
    regime $\lambda\ll\xi$. Inset: the peak-to-peak distance $w_{pp}$
  depends linearly on T for $w_{pp}<|vq|$ (red dashed line). (b) Low temperature regime
    $\lambda\gg\xi$. The momentum is taken to be $vq=-0.1$.  The intraedge interaction parameter is set to
  $\gamma=0.05$ and disorder length $\xi=20$. The frequency and the 
  length are in units of $v\alpha^{-1}$ and $\alpha$ respectively.}
\label{Fig:JLL_finiteT_dis}
\end{figure}
	
\begin{figure}[t!]
\includegraphics[width=0.35\textwidth]{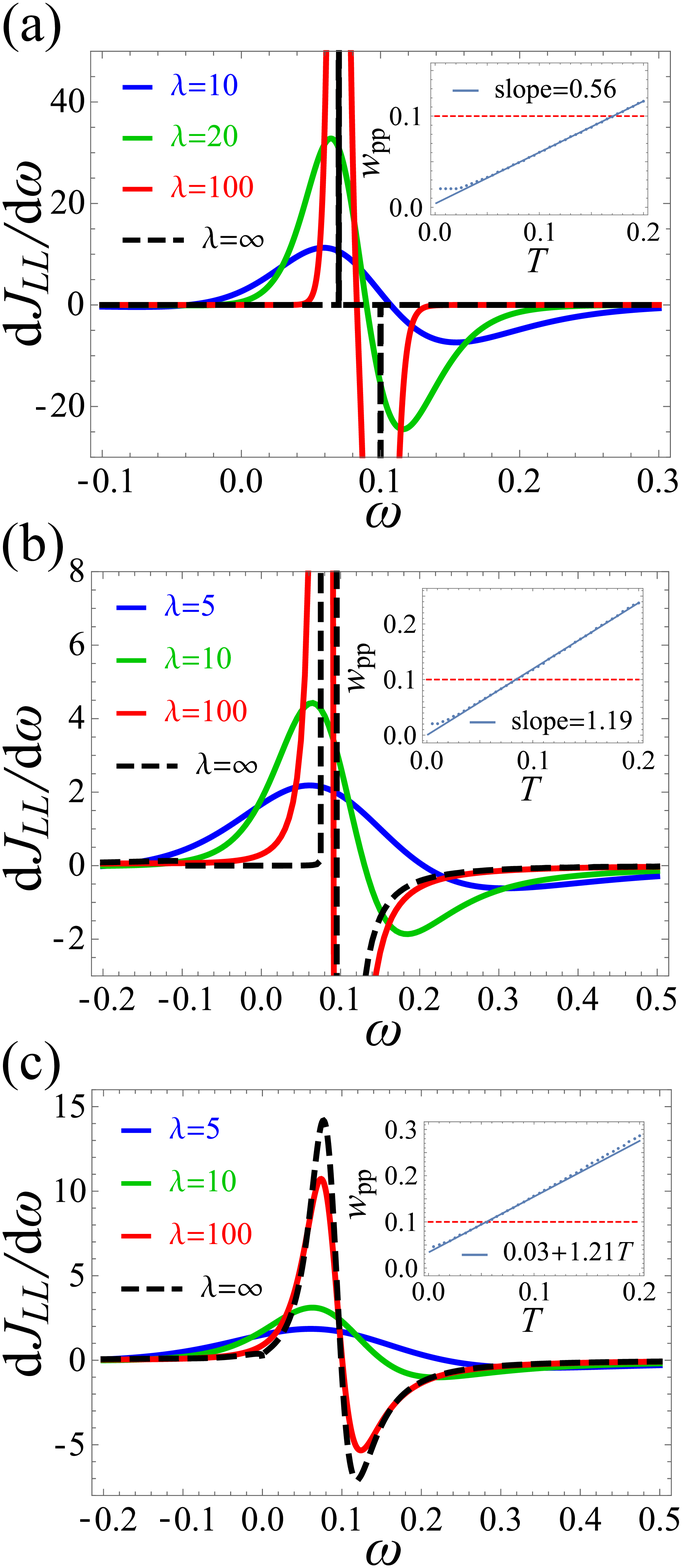}
\caption{Differential tunneling conductance for \textit{distinct} edge
  velocities. The edge velocities and momentum are set to $v_+q=-0.1$
  and $v_-q=-0.08$. (a) Thermal broadening of the
  \textit{non-interacting} clean tunneling peak (b) Thermal broadening
  of the \textit{interacting} ($\gamma_+=0.1$, $\gamma_-=0$,
  $\gamma_{12}=0$) clean tunneling peak (c) Thermal broadening of the
  \textit{interacting} ($\gamma_+=0.1$, $\gamma_-=0$, $\gamma_{12}=0$)
  tunneling peak in the presence of \textit{forward- scattering
    disorders} $\xi=30$. The frequency and the 
length are in units of $v_1\alpha^{-1}$ and $\alpha$ respectively.}
\label{Fig:JLL_2v}
\end{figure}
	
For vertical geometry illustrated in Fig.~\ref{Fig:vertical setup},
two $S_z$ conserved edges have exactly the same spin orientation. The
low-energy expressions of the fermionic $S_z$ eigenstate fields
(Appendix~\ref{App:chiral decomposition} with $k_0\to\infty$) are
given by
\begin{align}
  c_{a\uparrow}\approx
  e^{ik_{F,a}x}R_a(x),\quad\,c_{a\downarrow}\approx
  e^{-ik_{F,a}x}L_a(x).
\end{align}
Plugging the expression above into the imaginary-time correlator in
Eq.~(\ref{Eq:current_imaginary_time_correlator}), we obtain
\begin{align}
  \mathcal{J}_{\text{QSH},\perp}(\tau,x)=\mathcal{J}_{RR}(\tau,x)+\mathcal{J}_{LL}(\tau,x).
\end{align}
Thus, indeed, there is no tunneling current contributions
corresponding to backscattering between the right and left Fermi
points ($t^{RL}=t^{LR}=0$). These are forbidden by the $S_z$
conserving $U(1)$ spin-rotational symmetry, as such contribution
requires a spin flip $\uparrow\leftrightarrow\downarrow$, whose matrix
element identically vanishes in the presence of TR symmetry and in the
absence of Rashba spin-orbit interaction. The momentum-resolved
tunneling current is given by
\begin{align}
  &J(\omega,Q)\approx J_{RR}(\omega,Q+\delta k_F)+J_{LL}(\omega,Q-\delta k_F).
\end{align}
Below we will focus on $J_{LL}(V,Q)$ because the other term can be
obtained via the relation $J_{RR}(\omega,q)=J_{LL}(\omega,-q)$ if we assume TR
symmetry holds independently on each edge.

The space imaginary-time correlator can be calculated for generic
intraedge LL interactions (see
Appendix~\ref{App:Bosonization_two_edge}), given by
\begin{align}
  &\left\langle\hat{T}_{\tau}L_1^{\dag}L_2(\tau,x)L_2^{\dag}L_1(0,0)\right\rangle=-\frac{1}{(2\pi\alpha)^2}\prod_{b=\pm}
  \nonumber\\
  &\times\frac{(\frac{\pi\alpha}{\beta
      v_b})^{2\gamma_{b}+1}}{\left[\sinh\left(\frac{\pi(x+iv_b\tau)}{\beta
          v_b}\right)\right]^{\gamma_b-\frac{b}{2}\gamma_{12}}\left[\sinh\left(\frac{\pi(x-iv_b\tau)}{\beta
          v_b}\right)\right]^{\gamma_b+\frac{b}{2}\gamma_{12}+1}},
  \label{Eq:JLL_xt}
\end{align}
where $v_{\pm}$ encodes the velocity in the diagonal basis, and
$\gamma_{\pm}$, $\gamma_{12}$ are the anomalous exponents. The
explicit forms of $v_{\pm}$, $\gamma_{\pm}$, and $\gamma_{12}$ are
given in Appendix~\ref{App:Bosonization_two_edge}. Notice that
$\gamma_{12}<0$ ($\gamma_{12}>0$) for repulsive (attractive) interedge
interaction. In particular, for $U_{12}=0$, $\mathcal{J}_{LL}$ simply
reduces to a product of two single-particle Green functions with the
parameters given by $\gamma_{12}=0$, $\gamma_{\pm}=\gamma_{1,2}$ and
$v_{\pm}=v_{1,2}$. For identical edges ($v_1=v_2$ and $K_1=K_2$),
$v_+$ ($v_-$) is associated with the velocity of symmetric
(antisymmetric) interedge degrees of freedom.
	
In evaluating $J_{LL}(\omega,q)$, we use two different ways (detailed in
Appendix~\ref{App:Der_J}): (i) evaluate the tunneling current as a
convolution of two spectral functions if $U_{12}=0$, and (ii)
Analytically continue to real time and Fourier transform. In
particular, method (i) works well if we assume one of the edges is
non-interacting and therefore the corresponding spectral function is
just a delta function. On the other hand, method (ii) works well at
$T=0$ since one integral variable can be integrated over analytically
in that situation.
	
\paragraph{Zero-temperature, clean case:}
We start with the simplest case: zero temperature, no disorder and no
interaction. In this case, the tunneling current is simply given by a
box function
\begin{align}
  J^0_{LL}(\omega,q)=\frac{et_0^2\sgn(\omega)}{|v_1-v_2|}\Theta\left[-(\omega+v_2q)(\omega+v_1q)\right].
\end{align}
In the limit $v_1\to v_2\equiv v$, the tunneling current becomes a
delta function $J^0_{LL}(\omega,q)=-et_0^2q\delta(\omega+vq)$. Similar
to the spectral function, the presence of interaction makes the
tunneling peak less sharp and display power-law features as
illustrated in Fig.~\ref{Fig:JLL_zeroT_cl} ($et_0^2=1$ hereafter). In
the presence of interedge interaction, the eigenmodes are
anti-symmetric-like (subscript $-$) and symmetric-like (subscript $+$)
linear combinations of the two edges. As illustrated in the inset of
Fig.~\ref{Fig:JLL_zeroT_cl}, strong \textit{repulsive} interedge
interaction ($U_{12}\gg v_1-v_2$) makes the tunneling current
diverge at $\omega=-v_-q$.
	
\paragraph{Finite-temperature, clean case: } Now, we discuss the
finite-temperature tunneling current in the absent of disorder.  For
the special case $v_1=v_2$ and $U_{12}=0$, we can perform Fourier
transform analytically [using Eq.~(\ref{Eq:F_wn_q})]. The
finite-temperature clean tunneling current is given by
\begin{align}
  J^{\text{cl}}_{LL}(\omega,q)&=-2et_0^2\left(\frac{2\pi\alpha}{\beta
      v}\right)^{4\gamma}\frac{1}{4\pi^2v}\sin(2\pi\gamma)
  \nonumber\\
  &\quad\,\times \text{Im}\left\lbrace
    B\left[\frac{\beta(-i\omega+ivq)}{4\pi}+\gamma+1,-1-2\gamma\right]\right.
  \nonumber\\
  &\quad\quad\,\,\times
  \left. B\left[\frac{\beta(-i\omega-ivq)}{4\pi}+\gamma,1-2\gamma\right]
  \right\rbrace,\label{Eq:J_cl_LL_equalv}
\end{align}
where $\gamma=(\gamma_1+\gamma_2)/2$ is the average interaction
parameter. In the noninteracting limit (i.e., $\gamma=0$ and
$U_{12}=0$), the tunneling current becomes temperature independent as
the strict kinematic constraint of two equal velocity, in contrast to the distinct velocity case discussed below. For identical edges ($K_1=K_2=K$ and $v_1=v_2=v$), we can also obtain analytical
expression for $U_{12}\neq 0$ because the space-time correlator in
Eq.~(\ref{Eq:JLL_xt}) only depends on the velocity of anti-symmetric
mode $v_-$. The resulting clean tunneling current takes the same form
as Eq.~(\ref{Eq:J_cl_LL_equalv}) but with the replacement $v\to
v_-=v\sqrt{1-U_{12}K/2\pi^2v}$ and $\gamma\to (K_-+K_-^{-1})/4-1/2$,
where $K_-=K/\sqrt{1- U_{12}K/2\pi^2v}$. At zero temperature, the
tunneling current exhibits a power singularity at $\omega=-vq$, which
becomes two peaks (or one anti-symmetric peak) for the differential
tunneling conductance as shown in
Fig.~\ref{Fig:JLL_finiteT_cl}. Remarkably, the peak-to-peak distance
is captured by $w_{pp}\approx 7.5\gamma T$ for $w_{pp}<|vq|$. For
$w_{pp}>|vq|$, the broadening of the left (positive-valued) peak is
dominated by the thermal excitation around the Fermi point and the
linear dependence breaks down.

\paragraph{Disordered case: }
Now we discuss the effects of forward-scattering disorder (evaluated
through a convolution with a Lorentzian characterized by a disorder
strength
$\xi^{-1}=\frac{K_1^2\Delta_1}{2v_1}+\frac{K_2^2\Delta_2}{2v_2}$). At
zero temperature, exact analytical expression is derived in
Eq.~(\ref{Eq:J_dis_T0}). Similar to spectral function in a single
edge, the differential tunneling conductance features a
disorder-induced ZBA in a power-law form
$dJ_{LL}/d\omega\propto|\omega|^{4\gamma}$, independent of disorder
strength, as shown in Fig.~\ref{Fig:JLL_zeroT_dis}(a). The
peak-to-peak distance exhibits a linear dependence on $\xi^{-1}$ for
$w_{pp}<|vq|$ as illustrated in
Fig.~\ref{Fig:JLL_zeroT_dis}(b). Different from thermal broadening,
the disorder can smear out the peak even at zero temperature (see the
inset). For $w_{pp}>|vq|$, tunneling weights from opposite momentum
(i.e. having different sign of q) will start to contribute, which
gives opposite currents, and the linear dependence fails. At finite
temperature, $w_{pp}$ still depends linearly on $T$ despite the
presence of finite disorder [see Fig.~\ref{Fig:JLL_finiteT_dis}(a)],
which suggests that the disorder strength $\xi^{-1}$ and interaction
strength $\gamma$ can both be quantified through a temperature
dependence measure on $w_{pp}$. Figure~\ref{Fig:JLL_finiteT_dis}(b)
shows that ZBA gets rounded at finite temperature. The thermal
rounding takes the similar form as Eq.~(\ref{Eq:G_dis_low_T}).
	
\paragraph{Distinct velocity: }
When $U_{12}> 0$, the system is in general characterized by two
distinct velocities $v_{\pm}$ with $v_-<v_+$ (even for identical
edges) and an exponent $\gamma_{12}$ [given by
Eq.~(\ref{Eq:gamma_12})], encoding the correction due to the
  interaction between the two edges ($U_{12}$). The
interaction-driven inequality of velocities, $v_-<v_+$ has
qualitatively important effects on the tunneling current. This is in
contrast to nonvanishing $\gamma_{12}$, that does not modify the
tunneling current qualitatively. We therefore take $\gamma_{12}=0$ for
simplicity. With such an approximate, the effects of interedge
interaction $U_{12}$ still enter by modifying $v_\pm$ and
$\gamma_\pm$. The $\gamma_{12}=0$ approximation affects the analytical
form of the tunneling peak in the clean limit (see the inset of
Fig.~\ref{Fig:JLL_zeroT_cl}) but does not change the thermal
broadening rate because the exponential decay factor at large time
does not depend on $\gamma_{12}$ [see Eq.~(\ref{Eq:JLL_xt})]. Also, in
the presence of the disorders, by power counting in
Eq.~(\ref{Eq:JLL_xt}) we expect that the ZBA of the differential
tunneling conductance to be characterized by a power-law exponent
$2\gamma_++2\gamma_-$, which is also independent of $\gamma_{12}$ (but
does dependent on $U_{12}$). At zero temperature, the clean
differential tunneling conductance is featured by two singularities
located at $\omega=-v_+q$ and $\omega=-v_-q$. One prominent effect of
$v_+\neq v_-$, as illustrated in Fig.~\ref{Fig:JLL_2v}(a), is on
thermal broadening of the tunneling peak, even in the non-interacting
limit. The absence of thermal broadening in the same velocity case is
due to the strict kinematic constraint which is fine-tuned.
Remarkably, the thermal broadening (due to distinct velocities) is
linear in $T$ at high temperature [see inset of
Fig.~\ref{Fig:JLL_2v}(a)]. The temperature dependence should also be
proportional to the velocity difference, i.e. $\propto (v_+-v_-)T$, if
$v_+-v_-\ll v_++v_-$.  In the presence of interaction, with or without
disorders, the peak-to-peak distance still exhibits a considerable
linear in $T$ regime [see Fig.~\ref{Fig:JLL_2v}(b) and (c)]. However,
the zero temperature peak width (or $w_{pp}$) is now determined by
both the disorder strength $\xi^{-1}$ and $(v_+-v_-)q$. In evaluating
Fig.~\ref{Fig:JLL_2v}(b) and (c), we set one of the interaction
parameter to zero $\gamma_-=0$ and use the asymptotic expression in Eq.~(\ref{Eq:G_dis_asym}) for the symmetric-like branch for
computational convenience. More generally, we expect the interaction
facilitated thermal broadening rate is determined by
$\gamma_++\gamma_-$.

\subsubsection{Horizontal geometry}
	
\begin{figure}[t!]
  \includegraphics[width=0.4\textwidth]{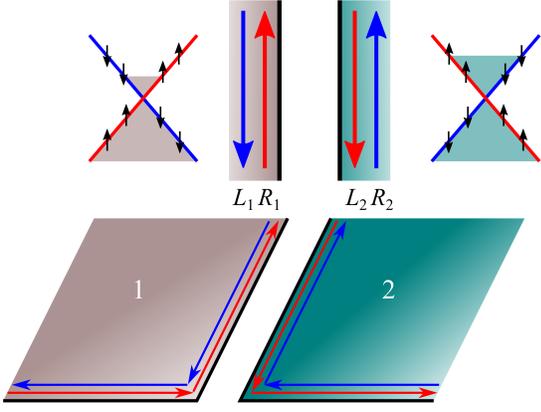}
  \caption{Horizontal setup of two topological insulators in the
    quantum spin Hall limit. The two co-moving edges have opposite
    spin orientations. As a result, the tunneling matrix element
    forbids the mixing of $R^{\dagger}_2R_1$ and $L^{\dagger}_2L_1$
    (i.e. $t^{RR}=t^{LL}=0$). The tunneling current is govern by the
    momentum transfer of order $|k_{F,1}+k_{F,2}|$.  }
  \label{Fig:horizontal setup}
\end{figure}
	
\begin{figure}[t!]
  \includegraphics[width=0.35\textwidth]{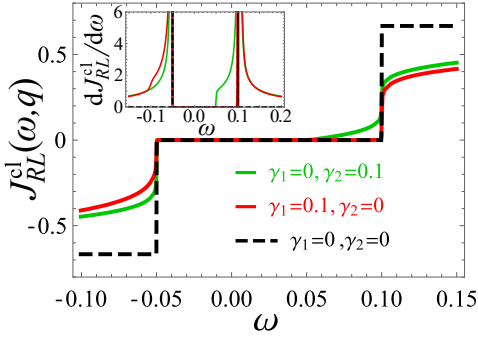}
  \caption{\textit{Zero-temperature} clean (disorder-free) tunneling current
    for a set of intraedge interaction parameters. The interedge
    interaction is ignored in this plot. The edge velocities and
    momentum are set to $v_1q=-0.1$ and $v_2q=-0.05$. The black dashed
    line indicates the non-interacting case
    ($\gamma_1=\gamma_2=0$). The red (green) line denotes the case
    that edge 1 (edge 2) becomes interacting, where finite current
    appears for $\omega<v_1q$ ($\omega>-v_2q$) due to
    fractionalization in chiral degrees of freedom. The inset shows
    the corresponding differential tunneling conductance. The frequency and the 
    length are in units of $v_1\alpha^{-1}$ and $\alpha$ respectively.}
  \label{Fig:JRL_zeroT_cl}
\end{figure}
	
\begin{figure}[t!]
  \includegraphics[width=0.35\textwidth]{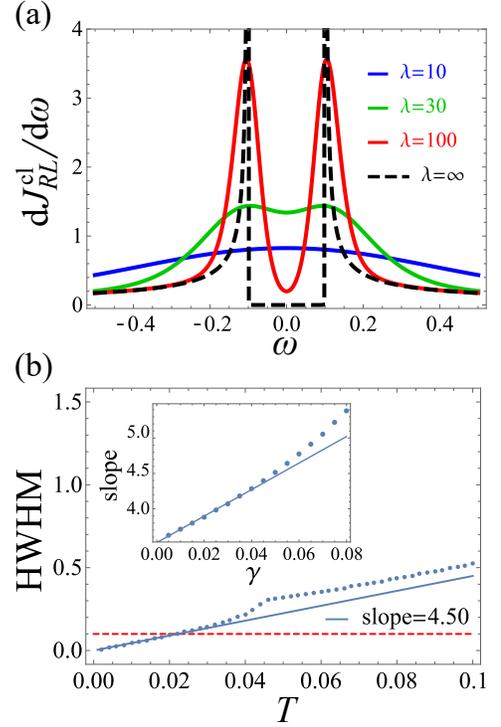}
  \caption{$\textit{Finite-temperature}$ clean (disorder-free) differential tunneling
    conductance for \textit{identical} edge velocities with
    $vq=-0.1$. (a) At zero temperature, two power-law singularities
    are located at $\omega=\pm vq$. At finite temperature, the two
    peaks get broadened and merge into a single peak at $\omega=0$
    with increasing temperature. The intraedge interaction parameter
    is set to $2\gamma=\gamma_1+\gamma_2=0.1$. (b) Half width at half
    maximum versus temperature. The linear dependence on $T$ holds for
    $\text{HWHM}<|vq|$ (red dashed line) and the slope has an
    $\text{offset}\approx 3.5$ in the non-interacting limit (inset). The frequency and the 
    length are in units of $v\alpha^{-1}$ and $\alpha$ respectively.}
  \label{Fig:JRL_finiteT_cl}
\end{figure}
	
\begin{figure}[t!]
  \includegraphics[width=0.35\textwidth]{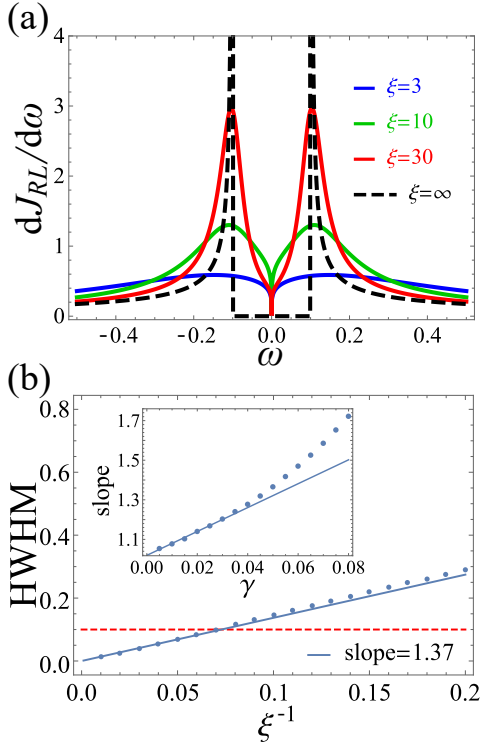}
  \caption{$\textit{Zero-temperature}$ differential tunneling
    conductance with \textit{forward-scattering disorders} for
    \textit{identical} edge velocities. The momentum is taken to be
    $vq=-0.1$.  (a) Zero bias anomaly appears at $\omega=0$ for all
    $\xi^{-1}$, with the same exponent $4\gamma$, where
    $\xi^{-1}=\frac{K_1^2\Delta_1} {2v_1}+\frac{K_2^2\Delta_2}{2v_2}$
    characterizing the strength of disorder. The interaction parameter
    $\gamma=0.05$.  (b) Half width at half maximum versus
    temperature. $\text{HWHM}$ is proportional to $\xi^{-1}$ for
    $\text{HWHM}<|vq|$ (red dashed line). Inset: The slope $\approx 1$
    in the noninteracting limit and the disorder-strength dependence
    becomes more sensitive for stronger interaction. The frequency and the 
    length are in units of $v\alpha^{-1}$ and $\alpha$ respectively.}
  \label{Fig:JRL_zeroT_dis}
\end{figure}
	
\begin{figure}[t!]
  \includegraphics[width=0.35\textwidth]{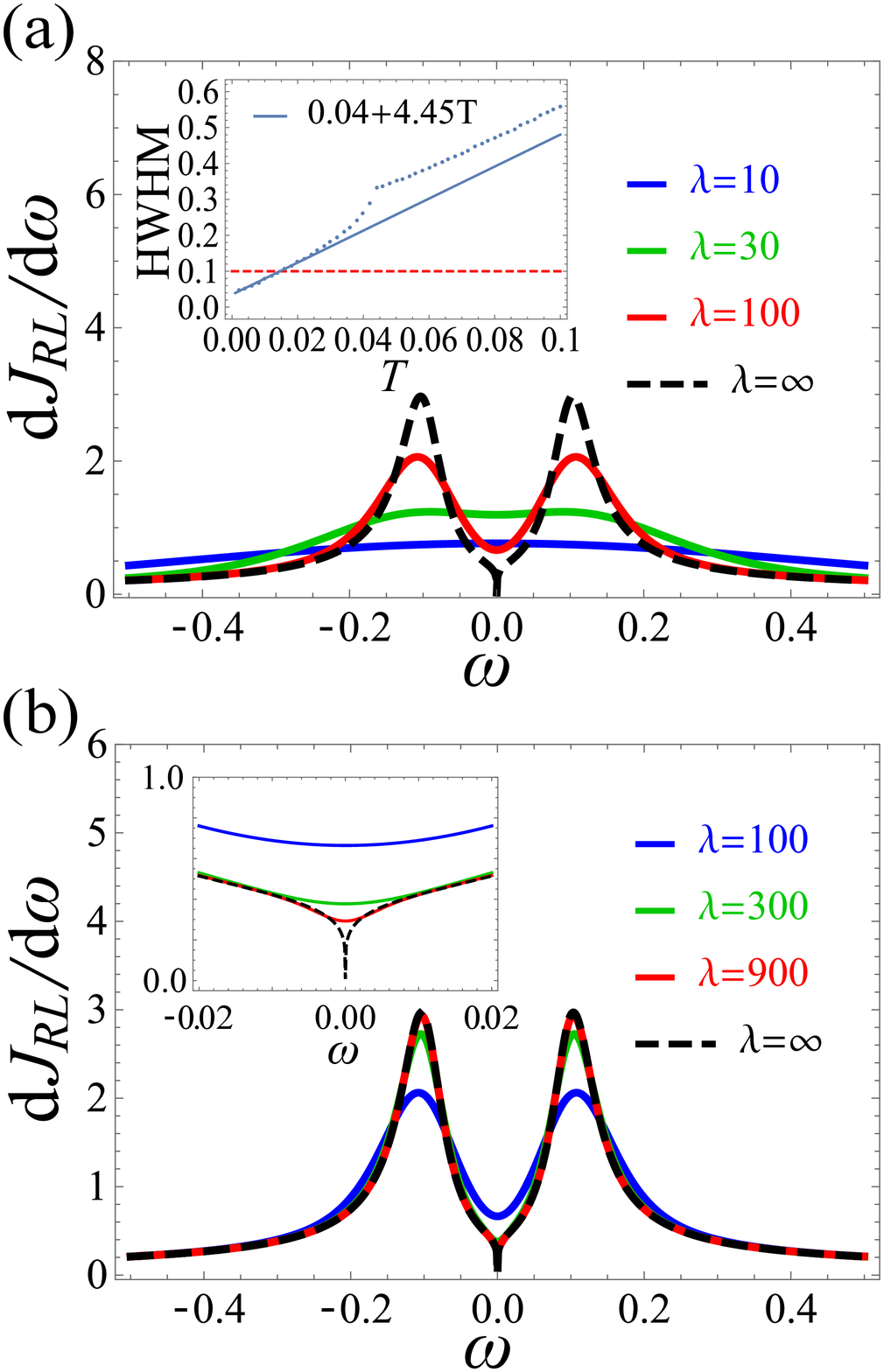}
  \caption{$\textit{Finite-temperature}$ differential tunneling
    conductance with \textit{forward-scattering disorders} for
    \textit{identical} edge velocities. The momentum is taken to be
    $vq=-0.1$.  The intraedge interaction parameter is set to
    $\gamma=0.05$ and disorder length $\xi=30$. (a) Thermal broadening
    of the tunneling peak. Inset: Half width at half maximum depends
    linearly on T for $\text{HWHM}<|vq|$ (red dashed line) (b) Thermal
    rounding of zero-bias anomaly. The frequency and the 
    length are in units of $v\alpha^{-1}$ and $\alpha$ respectively.}
  \label{Fig:JRL_finiteT_dis}
\end{figure}
	
As a complementary experimental setup, we consider horizontal geometry
illustrated in Fig.~\ref{Fig:horizontal setup}, where the right/left
movers of the two edges have opposite spins. In this case, the
low-energy expressions of the fermionic fields are given by
\begin{align}
  &c_{1\uparrow}\approx
  e^{ik_{F,1}x}R_1(x),\quad\,c_{1\downarrow}\approx
  e^{-ik_{F,1}x}L_1(x)
  \nonumber\\
  &c_{2\uparrow}\approx
  e^{-ik_{F,2}x}L_2(x),\quad\,c_{2\downarrow}\approx
  e^{ik_{F,2}x}R_2(x).
\end{align}
The imaginary-time correlator is now given by
\begin{align}
  \mathcal{J}_{\text{QSH},\parallel}(\tau,x)=\mathcal{J}_{RL}(\tau,x)+\mathcal{J}_{LR}(\tau,x),
\end{align}
where $t^{RR}=t^{LL}=0$, by $S_z$ conserving $U(1)$ spin rotational
symmetry. The momentum-resolved tunneling current is given by
\begin{align}
  &J(\omega,Q)\approx J_{RL}(\omega,Q+k_{F,T})+J_{LR}(\omega,Q-k_{F,T}).
\end{align}
Below we will focus on $J_{RL}(\omega,q)$ because the reverse current
contribution can be obtained via the relation
$J_{LR}(\omega,q)=J_{RL}(\omega,-q)$, if we assume that TR symmetry holds
independently on each edge (i.e., \textit{small} Zeeman field).

The space imaginary-time correlator can be calculated with both the
intraedge and interedge LL interactions (see
Appendix~\ref{App:Bosonization_two_edge}), given by
\begin{align}
  &\left\langle\hat{T}_{\tau}L_1^{\dag}R_2(\tau,x)R_2^{\dag}L_1(0,0)\right\rangle=\frac{1}{(2\pi\alpha)^2}\prod_{b=\pm}
  \nonumber\\
  &\!\times\!\!\frac{(\frac{\pi\alpha}{\beta
      v_b})^{2\gamma_{b}+1}}{\left[\sinh\left(\!\frac{\pi(x+iv_b\tau)}{\beta
          v_b}\!\right)\right]^{\frac{1+2\gamma_b+b\bar{\gamma}_{12}}{2}}\left[\sinh\left(\!\frac{\pi(x-iv_b\tau)}{\beta
          v_b}\!\right)\right]^{\frac{1+2\gamma_b-b\bar{\gamma}_{12}}{2}}},
  \label{Eq:JRL_xt}
\end{align}
where $\bar{\gamma}_{12}$ are the anomalous exponents. Note that
$\bar{\gamma}_{12}$ is different from $\gamma_{12}$; the explicit
expression is given in Appendix~\ref{App:Bosonization_two_edge}. For
$U_{12}=0$, $\mathcal{J}_{RL}$ simply reduces to a product of two
single-particle Green functions with the parameters given by
$\bar{\gamma}_{12}=1$, $\gamma_{\pm}=\gamma_{1,2}$ and
$v_{\pm}=v_{1,2}$.
	
\paragraph{Zero-temperature, clean case: }
In the zero temperature, non-interacting and clean limit, the
tunneling current is simply given by a step function
\begin{align}
  J^0_{RL}(\omega,q)=\frac{et_0^2\sgn(\omega)}{|v_1+v_2|}\Theta\left[(\omega-v_2q)(\omega+v_1q)\right].
\end{align}
As shown in Fig.~\ref{Fig:JRL_zeroT_cl}, the presence of interaction
smears out the steps and generate finite tunneling weights at
opposite momentum, i.e. $\omega>-v_2q$ ($\omega<v_1q$) for
$\gamma_2>0$ ($\gamma_1>0$). We also plot the differential tunneling
conductance in the inset of Fig.~\ref{Fig:JRL_zeroT_cl}.
	
	\paragraph{Finite-temperature, clean case: }
	For the special case $v_1=v_2$ and $U_{12}=0$, we can derive the finite-temperature clean tunneling current [using Eq.~(\ref{Eq:F_wn_q})], given by
	\begin{align}
	J^{\text{cl}}_{RL}(\omega,q)&=-2et_0^2\left(\frac{2\pi\alpha}{\beta v}\right)^{4\gamma}\frac{1}{4\pi^2v}\sin(2\pi\gamma)
	\nonumber\\
	&\quad\,\times \text{Im}\left\lbrace B\left[\frac{\beta(-i\omega+ivq)}{4\pi}+\gamma+\frac{1}{2},-2\gamma\right]\right.
	\nonumber\\
	&\quad\quad\,\,\times \left. B\left[\frac{\beta(-i\omega-ivq)}{4\pi}+\gamma+\frac{1}{2},-2\gamma\right]\right\rbrace.\label{Eq:J_RL_equalv}
	\end{align}
	In this case, the tunneling current (differential tunneling conductance) is an odd (even) function in $\omega$. With increasing temperature, the two peaks of the differential tunneling conductance at $\omega=\pm vq$ are broadened, move toward the center and merge into a single peak at $\omega=0$ [see Fig.~\ref{Fig:JRL_finiteT_cl}(a)]. The thermal broadening of the peaks is quantified by the half width at half maximum (HWHM). Specifically, we calculate the distance between the positions of the right peak and its right half maximum. The peak width is proportional to the temperature until the two (left and right) peaks start to merge [see Fig.~\ref{Fig:JRL_finiteT_cl}(b)]. Although the magnitude of the two edge velocities are identical, the kinematic constraint on the tunneling current is weaker than that in the left-to-left tunneling discussed previously. As a result, there is a strong thermal broadening even in the non-interacting limit [see the inset in Fig.~\ref{Fig:JRL_finiteT_cl}(b)]. 
	
\paragraph{Disordered case: }	
Now we discuss the effects of forward-scattering disorder. At zero
temperature, the increasing strength of forward-scattering disorder smears out the power-law peak but the position of the peaks do not
move much (comparing to the thermal effect) as shown in
Fig.~\ref{Fig:JRL_zeroT_dis}(a). Again, a ZBA appears with an exponent
$2\gamma_1+2\gamma_2$ independent of the disorder strength. The
disorder-induced peak broadening is proportional to the strength
$\xi^{-1}$ for $\text{HWHM}<|vq|$. For $\text{HWHM}>|vq|$, the linear
dependence on $\xi^{-1}$ still roughly holds since the two peak do not
merge [see Fig.~\ref{Fig:JRL_zeroT_dis}(b)]. At finite temperatures,
there is a crossover between the zero-temperature disordered and the
finite-temperature clean behaviors [see
Fig.~\ref{Fig:JRL_finiteT_dis}(a)] with the peak width increased
linearly with temperature for
$\text{HWHM}<|vq|$. Figure~\ref{Fig:JRL_finiteT_dis}(b) shows that ZBA
gets rounded at finite temperatures. The effect due to thermal
rounding is similar to Eq.~(\ref{Eq:G_dis_low_T}).

\paragraph{Distinct velocity: }
For distinct edge velocities, the zero-temperature clean tunneling
current is qualitatively modified from the case of identical
velocities, as shown in Fig.~\ref{Fig:JRL_zeroT_cl}. However, in the
presence of forward-scattering disorders, the power-law peaks become
rounded and a ZBA appears characterized by a modified exponent
$2\gamma_++2\gamma_-$. The linear dependence of the peak width still
holds and can be used for quantifying the disorder and interaction
strengths.
		
\subsubsection{Misaligned spin quantization axes}
As discussed above, for the ideal cases where the two spin
quantization axes are parallel, some of the tunneling processes vanish
identically in the TR symmetric limit. However, when the two 2D TI
layers are misaligned such that the quantization axes differ by an
angle $\phi_{12}\in [0,\pi/2]$, all the tunneling amplitudes in
Eq.~(\ref{Eq:S_z_conserved_tunneling_matrix_elemtents}) are expected
to be nonzero. To $O(t_0^2)$ order, the tunneling constants obey the
sum rule
\begin{align}
  \sum_{\alpha'=R,L}t^{\alpha\alpha'}=1,
\end{align}
and the ratio $t^{RL}/t^{RR}=\tan^2\phi_{12}$ $(=\cot^2\phi_{12})$ for the vertical
(co-planar) setup. Also, $t^{RR}=t^{LL}$ and $t^{RL}=t^{LR}$ due to
the time-reversal symmetry on the edges, which will be broken if we
consider Zeeman effect discussed in the next section.
	
\subsection{Other subleading corrections}
\begin{figure}[t!]
  \includegraphics[width=0.4\textwidth]{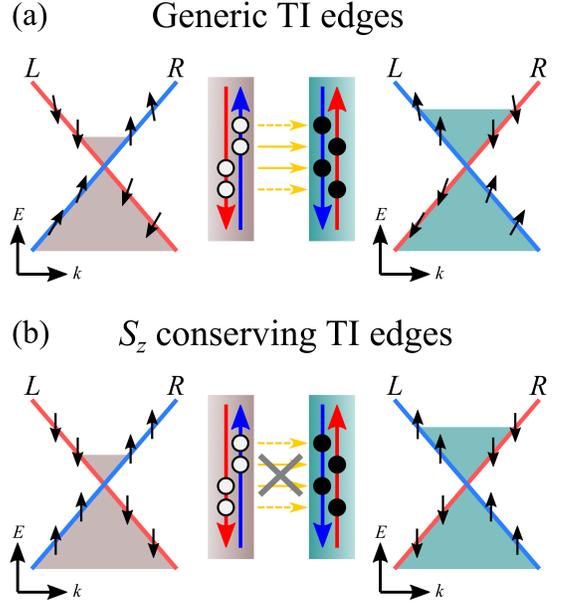}
  \caption{Bands, spin textures, and single-particle tunnelings
    between TI edges, illustrated with the horizontal setup in Fig.~\ref{Fig:horizontal setup}. (a) 
    Generic TI edges include Rashba spin-orbit
    coupling, leading to the illustrated momentum-dependent spin
    texture. Despite a non-conservation of $S_z$, TR symmetry protects
    a degenerate pair of Kramer states at every energy. The tunnelings
    can be classified by small momentum transfer (yellow solid arrows)
    and large momentum transfer (yellow dashed arrows.) (b) $S_z$
    conserving TI edges. Due to the additional conservation of $S_z$,
    the small momentum transfer tunneling is forbidden.}
  \label{Fig:ghLL_bands}
\end{figure}

\subsubsection{Zeeman effect}\label{Subsec:B_field}
With Zeeman effect, for vertical geometry $\vec{B}\perp\hat{z}$, we
note there will be finite tunneling current between right and left
Fermi points and a gap will open at the charge neutral point. In
contrast, for the co-planar geometry $\vec{B}\parallel\hat{z}$, the
spin quantization axis will remain along the z-axis in the presence of
the magnetic field and tunneling current contribution between the two
right/left Fermi points will remain zero. The charge neutral point in
this case remains gapless but moves away from the time-reversal point
in the Brillouin zone.
	
\subsubsection{Rashba spin-orbit coupling}
	
The effects of Rashba spin-orbit coupling on MRTS is a bit more
complicated, but as we discuss below, is sub-leading for a large bare
(without RSOC) tunneling amplitude. We expect the RSOC effects to be
manifest for the right-to-right (right-to-left) tunneling process for
the perfectly-aligned horizontal (vertical) geometry, where bare
tunneling vanishes otherwise. For concreteness, here we briefly
discuss the perfectly-aligned horizontal geometry with identical edges [see Fig.~\ref{Fig:ghLL_bands}(a)], focusing on tunneling current between two right Fermi points. The analysis for the right-to-left tunneling current and for the vertical geometry are quite similar. 

Using chiral decomposition in Eqs.~(\ref{Eq:chiral_dcomp_up}) and (\ref{Eq:chiral_dcomp_down}), we
express the ``hopping term'' as follows:
\begin{align}
  &\sum_{s=\uparrow\downarrow}c^{\dag}_{1s}c_{2s}
  \nonumber\\
  &\approx e^{-i\delta k_Fx}\left\lbrace\frac{\delta k_F
      k_{F,T}}{k_0^2}R^{\dag}_{1}R_{2}+i\frac{k_{F,T}}{k_0^2}\partial_x(R^{\dag}_{1}R_{2})\right.
  \nonumber\\
  &\quad +\left. i\frac{\delta
      k_F}{k_0^2}\left[\partial_x(R^{\dag}_{1})R_{2}-R^{\dag}_{1}\partial_x(R_{2})\right]\right\rbrace
  \nonumber\\
  &\quad +\text{(hopping between other Fermi points)}
      ,\label{Eq:tunn_RSOC}
\end{align}
where $\delta k_F=k_{F,1}-k_{F,2}$,
$k_{F,T}=k_{F,1}+k_{F,2}$. We assume that $k_{F,T}\gg\delta k_F$, so
only the first two terms in Eq.~(\ref{Eq:tunn_RSOC}) are
considered. The imaginary-time correlator for the right-to-right
tunneling current is given by
\begin{align}
  \mathcal{J}_{RR}^{\text{RSOC}}(\tau,x)=&e^{-i\delta
    k_Fx}\left[\alpha_{12}\left\langle\hat{T}_{\tau}R_{1}^{\dag}R_{2}(\tau,x)R_{2}^{\dag}R_{1}(0,0)\right\rangle\right.
  \nonumber\\
  &+\left.\alpha_{12}'\left\langle\hat{T}_{\tau}[\partial_xR_{1}^{\dag}R_{2}](\tau,x)[R_{2}^{\dag}R_{1}](0,0)\right\rangle\right.
  \nonumber\\
  &-\left.\alpha_{12}'\left\langle\hat{T}_{\tau}[R_{1}^{\dag}R_{2}](\tau,x)[\partial_xR_{2}^{\dag}R_{1}](0,0)\right\rangle\right.
  \nonumber\\
  &+\left.\alpha_{12}''\left\langle\hat{T}_{\tau}[\partial_x
      R_{1}^{\dag}R_{2}](\tau,x)[\partial_x
      R_{2}^{\dag}R_{1}](0,0)\right\rangle\right]
  \label{Eq:tun_J_RSOC}
\end{align}
where $\alpha_{12}=(\delta k_{F}k_{F,T})^2/k_{0}^4$,
$\alpha_{12}'=i\delta k_{F}k_{F,T}^2/k_{0}^4$ and
$\alpha_{12}''=k_{F,T}^2/k_{0}^4$.

For distinct edges, we typically expect that $\alpha_{12}\gg\alpha_{12}',\alpha_{12}''$, and thus the momentum-resolved tunneling current is qualitatively the same as that in the quantum spin Hall limit. However, for the identical edges considered here, the $\alpha_{12}$ term becomes less important as $\delta k_{F}\to 0$, for which additional contributions come from the derivative terms in Eq.~(\ref{Eq:tun_J_RSOC}) are manifest. Despite the complicated structures in the
tunneling currents, we argue that there are still universal features
whether RSOC is included or not. Firstly, we expect the
``single-peak'' (``no-peak'') feature of $J_{LL}$ ($J_{RL}$) still remains for the derivatives on the
space-time correlation functions change the exponent by ``-1'',
which, by dimensional analysis make the tunneling current less
divergent. Another observation is that the linear-$T$ thermal
broadening of the tunneling peak should be robust against RSOC, since
the derivatives on the space-time correlation functions does not
change the exponential decay factors at large $v\tau,x$. The
correlation function can in principle be calculated by bosonization,
but we do not pursue this analysis here.	
	
\subsubsection{Interedge backscattering}
Besides the correction in the tunneling current matrix element, the
RSOC also enables backscattering interactions
\cite{Schmidt2012,Kainaris2014,Chou2015}, contributing to the
finite-temperature broadening. The most relevant (in renormalization
group analysis) perturbations involve both edges.  For strong
interaction $K_{\pm}<3/4$ (identical edges), instabilities appear
\cite{Chou2019_2hLL} due to interplay of interedge interactions and
forward-scattering disorder, the tunneling current of the resulting
phase is beyond the scope of present work, but would be of interest to
study in the future in a context of specific experiments.
	
\section{Conclusion}
\label{sec:conclusion}
	
In this manuscript, we developed a finite-temperature spectroscopy of
a hLL as realized on the boundary of the 2D time-reversal symmetric
TI. In our analysis we utilized standard bosonization which enabled
analytical progress in the presence of interactions. Moreover, because
TR symmetry forbids backscattering components of disorder, allowing
only forward scattering nonmagnetic impurities, enabled us to treat
disorder in a hLL exactly. We focused on the \textit{weakly}
interacting regime ($K>3/8$), thereby avoiding edge instability
\cite{Wu2006,Xu2006,Chou2018EdgeGlass}. We thereby analyzed in great
detail various limits of finite-temperature spectral functions and the
interedge tunneling currents in the momentum-resolved tunneling
spectroscopy. For MRTS we explored the vertical and horizontal
geometries with long edges, detailing effects of TR invariant
disorder, interaction, and temperature. We studied how the product
expression for the tunneling current (valid in the noninteracting
limit between edges) is qualitatively modified by the interedge
interaction and distinct edge velocities. Our theory thus provides a
detailed characterization of the emergent hLL, complementary to the
standard transport measurements.
	
Our analysis was limited to the hLL phase, that appears in the weakly
interacting ($K>3/8$) regime of TI edges.  However, as discussed in
\cite{Chou2018EdgeGlass}, TI edge states can become \textit{glassy}
and localized due to an interplay of disorder and interaction for
$K<3/8$ \cite{Wu2006,Xu2006}. This scenario might be relevant to the
earlier InAs/GaSb experiments \cite{Du2015,Li2015}. A detailed
characterization of the finite-temperature spectroscopy in this regime
is beyond present work, but in light of various experiments is of
interest to explore by methods developed here. Here, we only speculate
about some qualitative zero-temperature features inside this glassy
edge states. The localized edges for $K<3/8$ spontaneously break
time-reversal symmetry and exhibit half-charge excitations,
corresponding to domain-walls or equivalently the Luther-Emory
fermions. We expect that this time-reversal breaking eliminates
sensitivity of the response to an applied magnetic field. We thus
expect that the localized nature of the glassy edge will lead to only
weakly momentum-dependent tunneling spectroscopy, contrasting to that
found above for hLL. It might be challenging to distinguish the
single-particle Anderson localization (i.e., trivial edge state) and
the unconventional half-charge localization (i.e. TI edge with
$K<3/8$). Exploring the unique spectroscopic signatures for the
nontrivial half-charge localization is an interesting future
direction.

We conclude by noting that momentum in MRTS setup is tuned by a
magnetic field $B$ that explicitly breaks TR symmetry. Quite
generally, we expect TI phase and the associated hLL edges to be
stable as long as the Zeeman energy associated with this TR breaking
is weak enough, to be below the bulk gap.  Nevertheless, the
bottleneck of our theory is set by the magnetic field induced disorder
backscattering with a localization length $l_{\text{loc}}(B)$.
Although, as we discussed in Sec.~\ref{Subsec:B_field}, the effect of
magnetic field may vary based on the specific setup, we still expect
our theory to be valid in a sufficiently weak magnetic field such that
the length of hLL edge $l_{\text{edge}}\ll l_{\text{loc}}(B)$. As
illustrated in Fig.~\ref{fig:kinematics}, the momentum transfer
$Q=2\pi Bd/\phi_0$ required to access the low-bias tunneling region
between the same ($J_{RR}/J_{LL}$) and the opposite ($J_{RL}/J_{RL}$)
chiral movers are given by the Fermi wavevector difference
$|k_{F,1}-k_{F,2}|$ and the sum $|k_{F,1}+k_{F,2}|$
respectively. Clearly then, typical wavevector range we want to
explore is set by the scale of Fermi wavevector, e.g., for
$Q=|k_{F,1}-k_{F,2}|=0.01 \text{ nm}^{\text{-1}}$ and tunneling
distance $d=15 \text{ nm}$, the corresponding magnetic flux density
$B\sim 1\text{T}$. In principle, the TI materials with larger bulk gap
(e.g., WTe$_2$ \cite{Wu2018}, WSe$_2$
\cite{Chen2018QSH_WSe2,Ugeda2018WSe2}, and BiSiC \cite{Reis2017}) are
best suited for MRTS experiments due to the suppression of
backscattering generated by, e.g., charge puddles \cite{Vayrynen2013}
and Zeeman gap of edge bands \cite{Skolasinski2018}.

\section*{Acknowledgment}
This work is supported by a Simons Investigator Award to Leo
Radzihovsky from the Simons Foundation. Y.-Z.C. is also supported in
part by the Laboratory for Physical Sciences and in part by
JQI-NSF-PFC (supported by NSF grant PHY-1607611).
	
\appendix

\section{chiral decomposition of generic hLL}\
\label{App:chiral decomposition}
	
In the presence of the Rashba spin-orbit coupling (RSOC), the spin is
no longer a good quantum number, and the single particle band develops
a momentum-dependent spin texture. The orientation of the spin
quantization axis at momentum $k$ relative to the one at $k=0$
(denoted by $\uparrow$ and $\downarrow$) is given as follows \cite{Schmidt2012}:
\begin{align}
\left[\begin{array}{cc}
      c_{k\uparrow}\\
      c_{k\downarrow}
    \end{array}\right]=B_{k}\left[\begin{array}{cc}
	c_{k+}\\
	c_{k-}
      \end{array}\right],
\end{align}
where (with the convention $k$ in x-direction and the normal vector of
the 2D TI plane in z-direction)
\begin{align}
  B_{k}=&e^{-i\sigma_2\theta_k} =\left[\begin{array}{cc}
      \cos(\theta_k) & -\sin(\theta_k)\\
      \sin(\theta_k) & \cos(\theta_k)
    \end{array}\right]
  \approx\left[\begin{array}{cc}
      1 & -\frac{k^2}{k_0^2}\\
      \frac{k^2}{k_0^2} & 1
    \end{array}\right].
  \label{eqnAppendix:Bk}
\end{align}
The form of $B_k$, encoding the spin texture is determined by the
unitarity and the time-reversal symmetry (in a particular phase
convention) with spin orientation at momentum $k$ obeying
$\theta_k=\theta_{-k}$. In the last equality in
(\ref{eqnAppendix:Bk}), we use $\theta_k\approx (k/k_0)^2$ for small
$k$, where $k_0$ is a parameter characterizing the scale of spin
rotation across the band. To study the low-energy physics around Fermi
points, we can expand $k\approx \pm k_{F}+q$ for the right (+) and
left (-) movers, respectively. The field operator for spin up is then
given by
\begin{align}
  c_{\uparrow}(x)=&\int_{k}e^{ikx}c_{\uparrow}(k)
  \nonumber\\
  \approx&e^{ik_{F}x}\int_{|q|\ll k_{F}}e^{iqx}c_{\uparrow}(k_{F}+q)
  \nonumber\\
  &+e^{-ik_{F}x}\int_{|q|\ll k_{F}}e^{iqx}c_{\uparrow}(-k_{F}+q)
  \nonumber\\
  \approx&e^{ik_{F}x}\int_{|q|\ll k_{F}}e^{iqx}c_{+}(k_{F}+q)
  \nonumber\\
  &-e^{-ik_{F}x}\int_{|q|\ll
    k_{F}}e^{iqx}\left[\frac{k_F^2}{k_0^2}-\frac{2k_{F}q}{k_0^2}\right]c_{-}(-k_{F}+q)
  \nonumber\\
  =&e^{ik_{F}x}\int_{|q|\ll k_{F}}e^{iqx}c_{+}(k_{F}+q)
  \nonumber\\
  &-e^{-ik_{F}x}\left[\frac{k_F^2}{k_0^2}+i\frac{2k_{F}}{k_0^2}\partial_{x}\right]\int_{|q|\ll
    k_{F}}e^{iqx}c_{-}(-k_{F}+q)
  \nonumber\\
  =&e^{ik_{F}x}R(x)-e^{-ik_{F}x}\left[\frac{k_F^2}{k_0^2}+i\frac{2k_{F}}{k_0^2}\partial_{x}\right]L(x),\label{Eq:chiral_dcomp_up}
\end{align}
Similarly, the field operator for spin down can be expressed as
\begin{align}
  c_{\downarrow}(x)=&\int_{k}e^{ikx}c_{k\downarrow}
  \nonumber\\
  \approx&e^{ik_{F}x}\left[\frac{k_F^2}{k_0^2}-i\frac{2k_{F}}{k_0^2}\partial_{x}\right]R(x)+e^{-ik_{F}x}L(x).\label{Eq:chiral_dcomp_down}
\end{align}

\section{Bosonization convention}
\label{App:bosonization_convention}
To treat interaction and disorder nonperturbatively we utilize
standard bosonization method \cite{Shankar_Book} (with the convention
consistent with Refs.~\cite{Chou2018EdgeGlass,Chou2019_2hLL}), where
left ($L$) and right ($R$) moving fermionic low-energy excitations can
be represented through the bosonic fields $\phi_{R,L}$, according to
\begin{align}
  \label{Eq:bosonization_convention}
  R(x)=&\frac{1}{\sqrt{2\pi\alpha}}e^{i\phi_{R}(x)}=\frac{1}{\sqrt{2\pi\alpha}}e^{i\left[\phi(x)+\theta(x)\right]}
  \nonumber\\
  L(x)=&\frac{1}{\sqrt{2\pi\alpha}}e^{i\phi_{L}(x)}=\frac{1}{\sqrt{2\pi\alpha}}e^{i\left[\phi(x)-\theta(x)\right]},
\end{align}
where $\alpha$ is the ultraviolet cutoff length scale below which the
low-energy description breaks down. The ``phase-like'' ($\phi$) and
the ``phonon-like'' ($\theta$) bosonic fields obey the following
commutation relation
\begin{align}
  [\partial_{x}\theta(x),\phi(x')]=i\pi\delta(x-x').
\end{align}
The commutation relations of the right and left bosons are given by
\begin{align}
  &[\phi_{R}(x),\phi_{R}(x')]=i\pi\sgn(x-x')
  \nonumber\\
  &[\phi_{L}(x),\phi_{L}(x')]=-i\pi\sgn(x-x')
  \nonumber\\
  &[\phi_{R}(x),\phi_{L}(x')]=i\pi.
\end{align}
	
The key characteristic of a helical Luttinger liquid (as
contrasting with superficially similar, spinless fermions) is the
\textit{anomalous} time-reversal operation, $\mathcal{T}$, with $R\rightarrow
L$, $L\rightarrow -R$, and $i\rightarrow -i$, and $\mathcal{T}^2 = -1$, akin to
spin-1/2 fermions.  On the corresponding bosonic operators $\mathcal{T}$ acts
according to, $\phi\rightarrow -\phi+\frac{\pi}{2}$,
$\theta\rightarrow\theta-\frac{\pi}{2}$, and $i\rightarrow-i$. One of
the immediate consequence of the anomalous time reversal symmetry is
the absence of elastic backscattering (i.e., forbidding 
$L^{\dagger}R$ and $R^{\dagger}L$), that clearly breaks it. As discussed in the main
text, the {\em forward}-scattering nonmagnetic disorder (allowed by
$\mathcal{T}$) alone cannot result in Anderson localization, and thus TI hLL
edge is stable to nonmagnetic impurities in the absence of strong
interactions.
	
	
\section{Derivation of clean imaginary time-ordered Green function}
\label{App:Der_space_time_G}
	
In this appendix, we provide a step-by-step derivation of imaginary
time-ordered single fermion Green function in $\tau,x$ domain at
finite temperature using the bosonization formalism. The
generalization to multi-particle Green function for a harmonic
bosonized model is straightforward utilizing Wick's theorem. With the helical edge Hamiltonian $H_{\text{hLL}}=H_{\text{0}}+H_{\text{int}}$ given by Eqs.~(\ref{Eq:H_hLL_kinetic})
and~(\ref{Eq:H_hLL_int}), the bosonized imaginary-time action reads
\begin{align}
\label{Eq:S_hLL_cl}
S_{\text{hLL}}=\int\limits_{\tau,x}\left\{\frac{i}{\pi}\left(\partial_x\theta\right)\left(\partial_{\tau}\phi\right)+\frac{v}{2\pi}\left[K\left(\partial_x\phi\right)^2+\frac{1}{K}\left(\partial_x\theta\right)^2\right]\right\}.
\end{align}

Using the chiral decomposition $\Psi(x)=e^{ik_Fx}R(x)+e^{-ik_Fx}L(x)$ of fermionic
field at low energy, the imaginary time-ordered single fermion
correlation function is given by
\begin{align}
  \mathcal{G}(\tau,x)=&\left\langle\Psi(\tau,x)\Psi^{\dag}(0,0)\right\rangle_{\tau}
  \nonumber\\
  =&e^{ik_Fx}\langle R(\tau,x)R^{\dag}(0,0)\rangle_{\tau}
  +e^{-ik_Fx}\langle L(\tau,x)L^{\dag}(0,0)\rangle_{\tau},
\end{align}
where the subscript $\tau$ denotes imaginary time-ordered average, and
for forward scattering only, appropriate to the hLL studied in this
manuscript, the cross term vanishes. We calculate the left mover
contribution and then deduce the right mover component using the
time-reversal operation according to the relation $\langle
R(\tau,x)R^{\dag}(0,0)\rangle_{\tau}=\langle
L(\tau,x)L^{\dag}(0,0)\rangle^{*}_{\tau}$. Using the bosonization
representation Eq.~(\ref{Eq:bosonization_convention}) and Wick's
theorem for the Gaussian bosonic phase fields, the left-moving part is
given by
\begin{align}
  \langle L(\tau,x)L^{\dag}(0,0)\rangle_{\tau}
  &=\frac{1}{2\pi\alpha}e^{-\frac{1}{2}\langle[\phi(\tau,x)-\theta(\tau,x)-\phi(0,0)+\theta(0,0)]^2\rangle_{\tau}}
  \nonumber\\
  &=\frac{1}{2\pi\alpha}e^{-\frac{1}{2}(K+K^{-1})F_1(\tau,x)+F_2(\tau,x)},
  \label{Eq:<L(tau,x)L(0,0)>}
\end{align}
where
\begin{align}
  F_1(\tau,x)=&2K[\langle\phi(0,0)\phi(0,0)\rangle_{\tau}-\langle\phi(\tau,x)\phi(0,0)\rangle_{\tau}]
  \nonumber\\
  =&2K^{-1}[\langle\theta(0,0)\theta(0,0)\rangle_{\tau}-\langle\theta(\tau,x)\theta(0,0)\rangle_{\tau}]
  \nonumber\\
  F_2(\tau,x)=&2\langle\theta(\tau,x)\phi(0,0)\rangle_{\tau}=2\langle\phi(\tau,x)\theta(0,0)\rangle_{\tau}
  \label{Eq:F1_and_F2_definition}
\end{align}
and $\langle\phi(0,0)\theta(0,0)\rangle_{\tau}=0$.  The correlators
$F_1(\tau,x)$ and $F_2(\tau,x)$ are easily computed with a quadratic
imaginary-time bosonic action, (\ref{Eq:S_hLL_cl}), that in Fourier
domain is given by
\begin{align}
  S_{\text{hLL}}&=\frac{1}{2}\int_{\omega_n,k} \left(\begin{array}{cc}
      \phi^{*}_{\omega_n,k} & \theta^{*}_{\omega_n,k}
    \end{array}\right)
  M^{-1} \left(\begin{array}{cc}
      \phi_{\omega_n,k} \\
      \theta_{\omega_n,k}
    \end{array}\right),
\end{align}
where
\begin{equation}
  M=\frac{\pi}{k^2(v^2k^2+\omega_n^2)}\left(\begin{array}{cc}
      \frac{vk^2}{K} & ik\omega_n\\
      ik\omega_n & vk^2K
    \end{array}\right).
\end{equation}
By rewriting the bosonic fields of Eq.~(\ref{Eq:F1_and_F2_definition})
in Fourier space and performing standard Gaussian integral, we obtain
the following integral expressions
\begin{align}
  F_1(\tau,x)&=\frac{1}{\beta}\sum_{n=-\infty}^{\infty}\int_{0}^{\infty}dk\frac{2v[1-\cos(kx)e^{-i\omega_n\tau}]}{v^2k^2+\omega_n^2}
  \nonumber\\
  F_2(\tau,x)&=-\frac{1}{\beta}\sum_{n=-\infty}^{\infty}\int_{0}^{\infty}dk\frac{2\omega_n\sin(kx)e^{-i\omega_n\tau}}{k(v^2k^2+\omega_n^2)}.
\end{align}
The Matsubara sum can be carried out by using Poisson summation
formula
$\sum_{n=-\infty}^{\infty}\delta(x-nT)=T^{-1}\sum_{m=-\infty}^{\infty}e^{i2\pi
  mx/T}$:
	
\begin{widetext}
  \begin{align}
    F_1(\tau,x)&=\frac{1}{\beta}\int_{-\infty}^{\infty}d\omega\sum_{n=-\infty}^{\infty}\delta(\omega-\omega_n)\int_{0}^{\infty}dk\frac{2v[1-\cos(kx)e^{-i\omega\tau}]}{v^2k^2+\omega^2}=\frac{1}{\beta}\int_{-\infty}^{\infty}d\omega\sum_{m=-\infty}^{\infty}\frac{\beta}{2\pi}e^{im\beta\omega}\int_{0}^{\infty}dk\frac{2v[1-\cos(kx)e^{-i\omega\tau}]}{v^2k^2+\omega^2}
    \nonumber\\
    &=\frac{v}{\pi}\int_{0}^{\infty}dk\sum_{m=-\infty}^{\infty}\frac{\pi}{vk}\left[e^{-|m\beta|vk}-\cos(kx)e^{-|m\beta-\tau|vk}\right]
    \nonumber\\
    &=\int_{0}^{\infty}dk\frac{2n_B(\beta
      vk)}{k}\left[1-\cos(kx)\cosh(\tilde{\tau}vk)\right]+\int_{0}^{\infty}dk\frac{1}{k}\left[1-\cos(kx)e^{-\tilde{\tau}vk}\right],
  \end{align}
\end{widetext}
where $\tilde{\tau}\equiv\text{mod}(\tau,\beta)\in [0,\beta)$.
	
Similarly,
\begin{widetext}
  \begin{align}
    F_2(\tau,x)&=-\frac{1}{\beta}\int_{-\infty}^{\infty}d\omega\sum_{n=-\infty}^{\infty}\delta(\omega-\omega_n)\int_{0}^{\infty}dk\frac{2\omega\sin(kx)e^{-i\omega\tau}}{k(v^2k^2+\omega^2)}=-\frac{1}{\beta}\int_{-\infty}^{\infty}d\omega\sum_{m=-\infty}^{\infty}\frac{\beta}{2\pi}e^{im\beta\omega}\int_{0}^{\infty}dk\frac{-2i\omega\sin(kx)e^{-i\omega\tau}}{k(v^2k^2+\omega^2)}
    \nonumber\\
    &=-i\int_{0}^{\infty}dk\frac{\sin(kx)}{k}\sum_{m=-\infty}^{\infty}\sgn(m\beta-\tau)e^{-|m\beta-\tau|vk}
    =i\int_{0}^{\infty}\frac{dk}{k}\sin(kx)[e^{-\tilde{\tau}vk}-2n_B(\beta
    vk)\sinh(\tilde{\tau}vk)]
  \end{align}
\end{widetext}
	
The integrals are over $k$, with the convergence factor
$e^{-\alpha|k|}$ then gives,
	
\begin{widetext}
  \begin{align}
   \label{Eq:F_1}
    F_1(\tau,x)&=\int_{0}^{\infty}dke^{-\alpha k}\frac{2n_B(\beta
      vk)}{k}\left[1-\cos(kx)\cosh(\tilde{\tau}vk)\right]+\int_{0}^{\infty}dke^{-\alpha
      k}\frac{1}{k}\left[1-\cos(kx)e^{-\tilde{\tau}vk}\right]
    \nonumber\\
    &=-\int_{0}^{\infty}dk\frac{e^{-\left(\alpha+\frac{\beta
            v}{2}\right)k}}{k\sinh(\frac{\beta
        vk}{2})}\left[\sinh^{2}\left[\frac{(v\tilde{\tau}-ix)k}{2}\right]+\sinh^{2}\left[\frac{(v\tilde{\tau}+ix)k}{2}\right]\right]
    -\int_{0}^{\infty}dk\frac{e^{-\alpha
        k}}{k}\left[\frac{e^{i(i\tilde{\tau}v-x)k}}{2}+\frac{e^{i(i\tilde{\tau}v+x)k}}{2}-1\right]
    \nonumber\\
    &\approx\frac{1}{2}\ln\left[\frac{\beta^2v^2}{\pi^2\alpha^2}\sinh\left(\frac{\pi(x+iv\tau)}{\beta
          v}\right)\sinh\left(\frac{\pi(x-iv\tau)}{\beta
          v}\right)\right],
  \end{align}
\end{widetext}
where we have assumed $\alpha\ll x,v\tau,\beta v$ and used the
following integral identities:
\begin{align}
  &\int_{0}^{\infty}dx\frac{e^{-x}\sinh^{2}(\lambda x)}{x\sinh(x)}=\frac{1}{2}\ln\left[\frac{\lambda\pi}{\sin(\lambda\pi)}\right],\text{ for $\text{Re}(\lambda)<1$}\\
  &\int_{0}^{\infty}dx\frac{e^{-\alpha x}(e^{i\lambda
      x}-1)}{x}=\ln\left[\frac{\alpha}{\alpha-i\lambda}\right].
\end{align}
	
In the last line of Eq.~(\ref{Eq:F_1}), we make a replacement $\tilde{\tau}\to\tau$ using the identity $\sin(x+n\pi)\sin(y-n\pi)=\sin x\sin y$ for $n\in \mathbb{Z}$. Similarly,
	
\begin{widetext}
  \begin{align}
    \label{Eq:F_2_naive}
    F_2(\tau,x)&=i\int_{0}^{\infty}\frac{dk}{k}e^{-\alpha
      k}\sin(kx)[e^{-\tilde{\tau}vk}-2n_B(\beta
    vk)\sinh(\tilde{\tau}vk)]
    \nonumber\\
    &=i\int_{0}^{\infty}\frac{dk}{k}e^{-\alpha
      k}\left\lbrace\left[\frac{e^{i(i\tilde{\tau}v+x)k}-e^{i(i\tilde{\tau}v-x)k}}{2i}\right]-i\frac{e^{-\frac{\beta
            vk}{2}}\left[\sinh^{2}\left[\frac{(v\tilde{\tau}-ix)k}{2}\right]-\sinh^{2}\left[\frac{(v\tilde{\tau}+ix)k}{2}\right]\right]}{\sinh\left(\frac{\beta
            vk}{2}\right)}\right\rbrace
    \nonumber\\
    &\approx-i\text{Arg}\left[-i\sinh\left(\frac{\pi(x+iv\tilde{\tau})}{\beta
          v}\right)\right],
  \end{align}
\end{widetext}
where we have used $i\text{Arg}(z)=[\ln(z)-\ln(z^{*})]/2$ and assumed
$\alpha\ll x,v\tau$. As discussed in Ref.~\cite{Giamarchi_Book}, the
expression of $F_2(\tau,x)$ above is not quite correct since it is
bosonic time-ordered. To calculate fermionic correlation function, we
need to add an additional minus sign for $\tau<0$, which can be taken
into account by replacing $\tilde{\tau}\to\tau$ in the last line of
Eq.~(\ref{Eq:F_2_naive}). Following similar procedure, the zero
temperature results are given by
\begin{align}
  &F^{T=0}_1(\tau,x)=\frac{1}{2}\ln\left[\frac{x^2+(v|\tau|+\alpha)^2}{\alpha^2}\right]
  \nonumber\\
  &F^{T=0}_2(\tau,x)=i\text{Arg}\left[v\tau+\alpha\sgn(\tau)+ix\right],
\end{align}
where we have taken replacements $\tilde{\tau}\to|\tau|$ for
$F^{T=0}_1(\tau,x)$ since $\beta=\infty$ and
$v\tilde{\tau}+\alpha\to v\tau+\alpha\sgn(\tau)$ for $F^{T=0}_2(\tau,x)$
for the reason of restoring fermionic time ordering.
	
Plugging $F_1(\tau,x)$ and $F_2(\tau,x)$ into
Eq.~(\ref{Eq:<L(tau,x)L(0,0)>}), we find a standard result,
\begin{align}
  \langle R&(\tau,x)R^{\dag}(0,0)\rangle_{\tau}
  \nonumber\\
  &=\frac{i}{2\pi\alpha}\frac{(\frac{\pi\alpha}{\beta
      v})^{2\gamma+1}}{\left[\sinh\left(\frac{\pi(x+iv\tau)}{\beta
          v}\right)\right]^{\gamma+1}\left[\sinh\left(\frac{\pi(x-iv\tau)}{\beta
          v}\right)\right]^{\gamma}}
  \nonumber\\
  \langle L&(\tau,x)L^{\dag}(0,0)\rangle_{\tau}
  \nonumber\\
  &=-\frac{i}{2\pi\alpha}\frac{(\frac{\pi\alpha}{\beta
      v})^{2\gamma+1}}{\left[\sinh\left(\frac{\pi(x+iv\tau)}{\beta
          v}\right)\right]^{\gamma}\left[\sinh\left(\frac{\pi(x-iv\tau)}{\beta
          v}\right)\right]^{\gamma+1}}
\end{align}
	
\section{Derivation of clean retarded Green function in Fourier space}
\label{App:Der_ret_G}
	
Here, we provide a detailed derivation of the retarded Green function
given by Eq.~(\ref{Eq:G_clean}) in the main text. A similar derivation
for density-density correlation function was discussed in
Ref.~\cite{Chou2016thesis}. We first compute the Green function in the
Matsubara frequency-momentum domain and then perform analytic
continuation to the retarded Green function at real frequency. Below
we compute the left-mover Greens function, with the extension to
right-mover one is straightforward.
	
We first rewrite the above imaginary time-ordered Green function in a
more convenient form:
\begin{align}
  \mathcal{G}_{L}(\tau,x)=&-\langle
  L(\tau,x)L^{\dag}(0,0)\rangle_{\tau}
  \nonumber\\
  =&\frac{i}{2\beta
    v}\frac{(\frac{\sqrt{2}\pi\alpha}{\beta
      v})^{2\gamma}}{\left[\cosh\left(\frac{2\pi x}{\beta
          v}\right)-\cos\left(\frac{2\pi\tau}{\beta}\right)\right]^{\gamma+1}}
  \nonumber\\
  &\times\left(e^{\frac{\pi x}{\beta v}}e^{i\frac{\pi
        \tau}{\beta}}-e^{-\frac{\pi x}{\beta v}}e^{-i\frac{\pi
        \tau}{\beta}}\right).
\end{align}
By using the identity,
$z^{-\nu}=\Gamma(\nu)^{-1}\int_{0}^{\infty}d\lambda\exp(-z\lambda)\lambda^{\nu-1}$
(for $\text{Re}[z]>0$ and $\text{Re}[\nu]>0$), Fourier transform of
the Green function can be written as
	
\begin{widetext}
  \begin{align}
    \mathcal{G}_{L}(i\omega_{n},q)=&\int_{\tau,x}e^{-i(qx-\omega_{n}\tau)}\mathcal{G}_{L}(\tau,x)
    \nonumber\\
    =&-i\frac{(\frac{\sqrt{2}\pi\alpha}{\beta v})^{2\gamma}}{2\beta
      v\Gamma\left(\gamma+1\right)}\int_{\tau,x}\int_{0}^{\infty}d\lambda
    e^{-\lambda\left[\cosh\left(\frac{2\pi x}{\beta
            v}\right)-\cos\left(\frac{2\pi\tau}{\beta}\right)\right]}\lambda^{\gamma}e^{-i(qx-\omega_{n}\tau)}\left[e^{\frac{\pi
          x}{\beta v}}e^{i\frac{\pi \tau}{\beta}}-e^{-\frac{\pi
          x}{\beta v}}e^{-i\frac{\pi \tau}{\beta}}\right]
    \nonumber\\
    =&i\frac{(\frac{\sqrt{2}\pi\alpha}{\beta v})^{2\gamma}}{2\beta
      v\Gamma\left(\gamma+1\right)}\frac{\beta^2
      v}{4\pi^2}\int_{0}^{\infty}d\lambda\lambda^{\gamma}\left\lbrace\int_{-\infty}^{\infty}dx'
      e^{[\frac{1}{2}-u]x'}e^{-\lambda\cosh(x')}
      \int_{0}^{2\pi}d\theta
      e^{i(n+1)\theta}e^{\lambda\cos\left(\theta\right)}\right.
    \nonumber\\
    &\left.\int_{-\infty}^{\infty}dx'
      e^{[-\frac{1}{2}-u]x'}e^{-\lambda\cosh(x')}
      \int_{0}^{2\pi}d\theta
      e^{in\theta}e^{\lambda\cos\left(\theta\right)}\right\rbrace,
  \end{align}
\end{widetext}
where $\int_{x}\equiv \int_{-\infty}^{\infty}dx$,
$\int_{\tau}\equiv\int_{0}^{\beta}d\tau$, $\int_{k}\equiv
\int_{-\infty}^{\infty}\frac{dk}{2\pi}$,
$\omega_{n}=2\pi(n+1/2)/\beta$ because of the boundary condition
$\langle L(\tau+\beta,x)L^{\dag}(0,0)\rangle_{\tau}=-\langle
L(\tau,x)L^{\dag}(0,0)\rangle_{\tau}$ and $u=i\beta
vq/2\pi$. $\Gamma$ denotes the Gamma function. We can use the following identities to carry out the
integrals:
	
\begin{widetext}
  \begin{align}
    &\int_{-\infty}^{\infty}dx\exp\left[-b
      x-a\cosh(x)\right]=2K_{b}(a),\text{ for
      $|\text{Arg}(a)|<\frac{\pi}{2}$}
    \nonumber\\
    &\int_{0}^{2\pi}d\theta\exp\left[in\theta+a\cos(\theta)\right]=2\pi
    I_{n}(a)
    \nonumber\\
    &\int_{0}^{\infty}dxJ_{a+b}\left[2\lambda\sinh(x)\right]e^{(-a+b)x}=I_{a}(\lambda)K_{b}(\lambda),\text{
      for $\lambda>0$, $\text{Re}(a-b)>-\frac{1}{2}$,
      $\text{Re}(a+b)>-1$}
    \nonumber\\
    &\int_{0}^{\infty}dxx^{a}J_{b}(x)=2^{a}\frac{\Gamma\left[\frac{1}{2}(b+a+1)\right]}{\Gamma\left[\frac{1}{2}(b-a+1)\right]},\text{
      for $\text{Re}(a+b)>-1$, $\text{Re}(a)<\frac{1}{2}$}
    \nonumber\\
    &\int_{0}^{\infty}dx\frac{e^{-ax}}{\left[2\sinh(x)\right]^{b}}=\frac{1}{2}B\left(\frac{a}{2}+\frac{b}{2},1-b\right),
    \text{ for $\text{Re}(a+b)>0$, $\text{Re}(b)<1$}
  \end{align}
\end{widetext}
where $\lambda\in\mathbb{R}$, $n\in\mathbb{Z}$, $a,b\in\mathbb{C}$ and
the integral variables $x,\theta$ are along the real axis. $I_b(x)$ and $K_b(x)$ are the modified Bessel function of the first kind and the second kind respectively. (Not to confuse with the Luttinger parameter $K$.)
	
The Green function becomes
\begin{widetext}
  \begin{align}
    \mathcal{G}_{L}(i\omega_{n},q)&=i\frac{(\frac{\sqrt{2}\pi\alpha}{\beta
        v})^{2\gamma}}{2\beta
      v\Gamma\left(\gamma+1\right)}\frac{\beta^2
      v}{\pi}\int_{0}^{\infty}d\lambda\lambda^{\gamma}\left[K_{-\frac{1}{2}+u}(\lambda)I_{n+1}(\lambda)-K_{\frac{1}{2}+u}(\lambda)I_{n}(\lambda)\right]
    \nonumber\\
    &=i\frac{\beta(\frac{\sqrt{2}\pi\alpha}{\beta
        v})^{2\gamma}}{2\pi\Gamma\left(\gamma+1\right)}\int_{0}^{\infty}d\lambda\lambda^{\gamma}\int_{0}^{\infty}dz
    J_{n+\frac{1}{2}+u}\left(2\lambda\sinh(z)\right)\left[e^{-(n+\frac{3}{2}-u)z}-e^{-(n-\frac{1}{2}-u)z}\right]
    \nonumber\\
    &=i\frac{\beta(\frac{\sqrt{2}\pi\alpha}{\beta
        v})^{2\gamma}}{2\pi\Gamma\left(\gamma+1\right)}\int_{0}^{\infty}dz
    \frac{e^{-(n+\frac{3}{2}-u)z}-e^{-(n-\frac{1}{2}-u)z}}{\left[2\sinh(z)\right]^{\gamma+1}}\int_{0}^{\infty}d\lambda'\lambda'^{\gamma}J_{n+\frac{1}{2}+u}\left(\lambda'\right)
    \nonumber\\
    &=i\frac{\beta(\frac{\sqrt{2}\pi\alpha}{\beta
        v})^{2\gamma}}{2\pi\Gamma\left(\gamma+1\right)}\frac{1}{2}\left\lbrace
      B\left(\frac{n}{2}+\frac{3}{4}-\frac{u}{2}+\frac{\gamma+1}{2},-\gamma\right)-B\left(\frac{n}{2}-\frac{1}{4}-\frac{u}{2}+\frac{\gamma+1}{2},-\gamma\right)\right\rbrace
    \nonumber\\
    &\ \ \ \ \times
    2^{\gamma}\frac{\Gamma\left(\frac{n}{2}+\frac{3}{4}+\frac{u}{2}+\frac{\gamma}{2}\right)}{\Gamma\left(\frac{n}{2}+\frac{3}{4}+\frac{u}{2}-\frac{\gamma}{2}\right)}.
  \end{align}
\end{widetext}
We note that the individual terms in the z-dependent integrands are
individually divergent at $z=0$. However, the full integrand is
convergent for $\gamma<1$ by a Taylor expansion.
	
Now using the properties of Gamma and Beta functions:
$\Gamma(z)\Gamma(1-z)=\pi/\sin(\pi z)$,
$B(a,b)=\Gamma(a)\Gamma(b)/\Gamma(a+b)$ and
$B(x,y)=B(x+1,y)+B(x,y+1)$, the above expression simplifies to:
\begin{widetext}
  \begin{align}
    \mathcal{G}_{L}(i\omega_{n},q)
    &=i\frac{\beta(\frac{2\pi\alpha}{\beta
        v})^{2\gamma}}{4\pi^2}\sin\left(\pi\gamma\right)B\left[\frac{\beta(\omega_n-ivq)}{4\pi}+\frac{\gamma}{2},1-\gamma\right]B\left[\frac{\beta(\omega_n+ivq)}{4\pi}+\frac{\gamma+1}{2},-\gamma\right].
  \end{align}
\end{widetext}
Now performing the analytical continuation
$i\omega_n\to\omega_{\eta}\equiv\omega+i\eta$ ($\eta\rightarrow 0^+$) to get the retarded
Green function for the left movers:
\begin{align}
\label{Eq:G_L_clean}
  G^{\text{ret}}_{L}(\omega,q)=&i\frac{\beta(\frac{2\pi\alpha}{\beta
      v})^{2\gamma}}{4\pi^2}\sin\left(\pi\gamma\right)
  \nonumber\\
  &\times
  B\left[-i\frac{\beta(\omega_{\eta}+vq)}{4\pi}+\frac{\gamma}{2},1-\gamma\right]
  \nonumber\\
  &\times
  B\left[-i\frac{\beta(\omega_{\eta}-vq)}{4\pi}+\frac{\gamma+1}{2},-\gamma\right].
\end{align}
Similarly, the retarded Green function for the right movers is given
by
\begin{align}
  G^{\text{ret}}_{R}(\omega,q)=&i\frac{\beta(\frac{2\pi\alpha}{\beta
      v})^{2\gamma}}{4\pi^2}\sin\left(\pi\gamma\right)
  \nonumber\\
  &\times
  B\left[-i\frac{\beta(\omega_{\eta}-vq)}{4\pi}+\frac{\gamma}{2},1-\gamma\right]
  \nonumber\\
  &\times
  B\left[-i\frac{\beta(\omega_{\eta}+vq)}{4\pi}+\frac{\gamma+1}{2},-\gamma\right].
\end{align}
The retarded Green function above is consistent with the finite
temperature results in Ref.~ \cite{Orgad2001} (imaginary part) and the
zero temperature results in Ref.~\cite{Meden1992} (both real and
imaginary parts) for $\gamma<0.5$ at low energy
$\omega/v,q<1/\alpha$. To the best of our knowledge, the full
expression of $G^{\text{ret}}_{R/L}(\omega,q)$ has not appeared in the
literature.
	
Following similar procedure in this appendix, we are able to perform
the Fourier transform for a generalized Euclidean function:
\begin{widetext}
  \begin{align}
    \label{Eq:F_wn_q}
    \mathcal{F}(i\omega_{n},q)=&\int_{\tau,x}e^{-i(qx-\omega_{n}\tau)}\mathcal{F}(\tau,x)
    \nonumber\\
    =&\frac{\beta^2v}{4\pi^2}\left(\frac{i}{\beta
        v}\right)^{n+m}\left(\frac{2\pi\alpha}{\beta
        v}\right)^{2\gamma}\sin\left(\pi\gamma\right)B\left[\frac{\beta(\omega-ivq)}{4\pi}+\frac{\gamma+m}{2},1-\gamma-m\right]B\left[\frac{\beta(\omega+ivq)}{4\pi}+\frac{\gamma+n}{2},1-\gamma-n\right],
  \end{align}
where
\begin{align}
  \mathcal{F}(\tau,x)=\left(\frac{i}{2\pi\alpha}\right)^{n}\left(\frac{-i}{2\pi\alpha}\right)^{m}\frac{(\frac{\pi\alpha}{\beta
      v})^{2\gamma+n+m}}{\sinh\left[\frac{\pi}{\beta}\left(\frac{x}{v}+i\tau\right)\right]^{\gamma+n}\sinh\left[\frac{\pi}{\beta}\left(\frac{x}{v}-i\tau\right)\right]^{\gamma+m}}.
\end{align}
\end{widetext}

\section{Derivation of disorder-averaged retarded Green function in Fourier space}
\label{App:Der_ret_G_dis}
At low temperature, by using Stirling's approximation on the Beta function, $B(x,y)\sim\Gamma(y)x^{-y}$, for a fixed $y$ and $|x|\gg 1$, $\text{Re}(x)>0$, the \textit{clean} Green function in Eq.~(\ref{Eq:G_L_clean}) can be written in the following asymptotic form
\begin{align}
  G^{\text{ret}}_{L}(\omega,q)&\sim
  -i\left(\frac{\alpha}{2v}\right)^{2\gamma}\frac{\Gamma\left(1-\gamma\right)}{\Gamma\left(1+\gamma\right)}\left[-i(\omega+vq)+2\pi\gamma
    T\right]^{\gamma-1}
  \nonumber\\
  &\left[-i(\omega-vq)+2\pi(\gamma+1)T\right]^{\gamma}.
\end{align}
	
The \textit{disordered} Green function, as discussed in the main text,
can be calculated via a convolution with a Lorenzian [see
Eq.~(\ref{Eq:G_dis})]. With the asymptotic approximation in
Eq.~(\ref{Eq:G_clean_asym}), the \textit{disordered} Green function
can be evaluated by residue theorem and is given by
\begin{align}
  G^{\text{ret}}_{\text{dis},L}(\omega,q)=G^{\text{ret}}_{L}(\omega,q+i\xi^{-1})+G^{\text{ret}}_{2,L}(\omega,q).
\end{align}
where $G^{\text{ret}}_{\text{th},L}$ is given by the following
integral
\begin{align}
  G^{\text{ret}}_{2,L}(\omega,q)=&-i\frac{2}{v}\sin\left(\pi\gamma\right)\left(\frac{\alpha}{2}\right)^{2\gamma}\frac{\Gamma\left(1-\gamma\right)}{\Gamma\left(1+\gamma\right)}
  \nonumber\\
  &\times\int_0^{\infty}dk\frac{\xi^{-1}/\pi}{\left[k+\frac{2\pi(\gamma+1)}{\beta
        v}-i\left(\frac{\omega}{v}-q\right)\right]^2-\xi^{-2}}
  \nonumber\\
  &\times\left[k+\frac{2\pi(2\gamma+1)}{\beta
      v}-i\frac{2\omega}{v}\right]^{\gamma-1}k^{\gamma}.
\end{align}
	
Using the following identity
\begin{align}
  &\int_0^{\infty}dx x^\gamma
  (x+a)^{\gamma-1}(x+b)^{-1}=\left(1-\frac{a}{b}\right)^{\gamma-1}b^{2\gamma-1}\frac{\pi}{\sin\left(2\pi\gamma\right)}
  \nonumber\\
  &\quad\,+a^{2\gamma}b^{-1}B(1+\gamma,-2\gamma)
  _2F_1\left(1,1+\gamma,1+2\gamma,\frac{a}{b}\right),
\end{align}
we derive the following expression
\begin{widetext}
  \begin{align}
    \label{Eq:G_dis_2}
    G^{\text{ret}}_{2,L}(\omega,q)=&\sum_{s=\pm}s\frac{i}{\pi}\sin\left(\pi\gamma\right)\left(\frac{\alpha}{2v}\right)^{2\gamma}\frac{\Gamma\left(1-\gamma\right)}{\Gamma\left(1+\gamma\right)}\left\lbrace\frac{\pi}{\sin\left(2\pi\gamma\right)}\left[\frac{i\left(\omega+vq\right)-\frac{2\pi\gamma}{\beta}+sv\xi^{-1}}{-i\left(\omega-vq\right)+\frac{2\pi(\gamma+1)}{\beta}+sv\xi^{-1}}\right]^{\gamma-1}\right.
    \nonumber\\
    &\left.\times\left[-i\left(\omega-vq\right)+\frac{2\pi(\gamma+1)}{\beta}+sv\xi^{-1}\right]^{2\gamma-1}+\left[-i2\omega+\frac{2\pi(2\gamma+1)}{\beta}\right]^{2\gamma}\left[-i\left(\omega-vq\right)+\frac{2\pi(\gamma+1)}{\beta}+sv\xi^{-1}\right]^{-1}\right.
    \nonumber\\
    &\left.\times
      B(1+\gamma,-2\gamma)_2F_1\left(1,1+\gamma,1+2\gamma,\frac{-i2\omega+\frac{2\pi(2\gamma+1)}{\beta}}{-i\left(\omega-vq\right)+\frac{2\pi(\gamma+1)}{\beta}+sv\xi^{-1}}\right)\right\rbrace,
  \end{align}
\end{widetext}
where $_2F_1$ is the ordinary hypergeometric function.
	
\section{Derivation of the tunneling current $J$}
\label{App:Der_J}
	
In this appendix, we provide the derivation of
Eq.~(\ref{Eq:J_xt_formalism2}) in the main text. Working in the
interaction representation, the expectation value of the tunneling
current density $J$, Eq.~(\ref{Eq:J_xt_formalism}), is given by
\begin{align}
  J =\frac{1}{Z}\text{Tr}\left[e^{-\beta H_{12}}
    \hat{U}^{\dagger}(t)\hat{J}(x)\hat{U}(t) \right],
\end{align}
where $\hat{U}(t)=\hat{U}_{12}(t)\hat{U}_I(t)$,
$\hat{U}_{12}(t)=e^{-iH_{12}t}$,
$\hat{U}_I(t)=\hat{T}\exp\left[-i\int_{-\infty}^{t}dt'
  H^I_{\text{tun}}(t')\right]$ ($\hat{T}$ the time-ordering operator),
$H^I_{\text{tun}}(t)\equiv
e^{iH_{12}t}H^{Q}_{\text{tun}}e^{-iH_{12}t}$, and $\beta$ is the
inverse temperature. Expanding in the weak tunneling matrix element
$t_0$, we find the leading contribution to $J\approx J^{(2)}$ is at
$O(t_0^2)$ and is given by
\begin{widetext}
  \begin{align}
    J^{(2)}(t,x)=&\text{Tr}\left\{\frac{e^{-\beta
          H_{12}}}{Z}(-i)(-t_0)
      \sum_{s'=\uparrow\downarrow}\int_{-\infty}^tdt'\int\limits_{x'}
      \left[\hat{U}^{\dag}_{12}(t)\hat{J}(x)\hat{U}_{12}(t),c^{\dagger}_{2s'}(t',x')c_{1s'}(t',x')e^{iQx'}+\text{H.c.}\right]
    \right\}\\
    \nonumber=&-et_0^2\sum_{s,s'=\uparrow\downarrow}\int_{-\infty}^tdt'\int\limits_{x'}\text{Tr}\left\{
      \frac{e^{-\beta H_{12}}}{Z}
      \left[c^{\dagger}_{2s}c_{1s}(t,x)e^{iQx}-c^{\dagger}_{1s}c_{2s}(t,x)e^{-iQx}
        ,c^{\dagger}_{2s'}c_{1s'}(t',x')e^{iQx'}+\text{H.c.}\right]
    \right\}.
  \end{align}
\end{widetext}
	
In the interaction picture, the fermionic creation and annihilation
operators have time dependence controlled by the zero-tunneling
Hamiltonian, $H_{12}$. In the weak tunneling setup, we use a
source-drain bias to control the electro-chemical potential
difference, $eV$ (with electron density fixed) between the two
edges. We take the two edges to be in thermal equilibrium at a common
temperature $T$, at densities controlled by $k_{F1}$ and $k_{F2}$, and
at the fixed electro-chemical potential imbalance, that drives a
steady-state tunneling current. Accordingly, the effect of the
source-drain bias can be included by the substitution,
$c_{as}(t,x)\rightarrow c_{as}(t,x)e^{-i\mu_at/\hbar}$, where $\mu_1=eV$,
$\mu_2=0$. With straightforward algebraic manipulations, at time long
since the tunneling was turned on, we arrive at the steady-state
current
\begin{widetext}
  \begin{align}
    J^{(2)}=&
    -et_0^2\sum_{s,s'=\uparrow\downarrow}\int_{-\infty}^0dt'\int\limits_{x'}
    \left[\begin{array}{r}
        \left\langle c^{\dagger}_{2s}c_{1s}(0,0)c^{\dagger}_{1s'}c_{2s'}(t',x') \right\rangle e^{i\omega t'}e^{-iQx'}\\[2mm]
        -\left\langle c^{\dagger}_{1s'}c_{2s'}(t',x')c^{\dagger}_{2s}c_{1s}(0,0) \right\rangle e^{i\omega t'}e^{-iQx'}\\[2mm]
        -\left\langle c^{\dagger}_{1s}c_{2s}(0,0)c^{\dagger}_{2s'}c_{1s'}(t',x') \right\rangle e^{-i\omega t'}e^{iQx'}\\[2mm]
        +\left\langle
          c^{\dagger}_{2s'}c_{1s'}(t',x')c^{\dagger}_{1s}c_{2s}(0,0)
        \right\rangle e^{-i\omega t'}e^{iQx'}
      \end{array}
    \right]\\
    \nonumber=&et_0^2
    \sum_{s,s'=\uparrow\downarrow}\int_{-\infty}^{\infty}dt'\int_{-\infty}^{\infty}dx'\,e^{i\omega t'}e^{-iQx'}
    \left[ \left\langle
        c^{\dagger}_{1s'}c_{2s'}(t',x')c^{\dagger}_{2s}c_{1s}(0,0)
      \right\rangle -\left\langle
        c^{\dagger}_{2s}c_{1s}(0,0)c^{\dagger}_{1s'}c_{2s'}(t',x')
      \right\rangle \right],
  \end{align}
\end{widetext}
where $\omega=eV/\hbar$ and $\langle\mathcal{O}\rangle$ denotes the expectation value with
respect to $H_{12}$ under thermal density matrix $e^{-\beta
  H_{12}}/Z$ with $H_{12}$ including the interedge interaction but
not the interedge tunneling. We have used translational invariance in
the first equality.  The derived expression is
Eq.~(\ref{Eq:J_xt_formalism}) of the main text and coincides with the
result in Ref.~\onlinecite{Carpentier2002}. We note that this current
expression is quite general, not relying on the linearized band or
chiral decomposition.

\section{analytic continuation of correlation function}
\label{App:analytic_con_J}
	
For notation simplicity, we define
$O=\sum_{s=\uparrow\downarrow}c_{1,s}^{\dag}c_{2,s}$. We will also
drop the spatial argument since the discussion here is only related to
the analytical properties in time. The response function of interest
is given by
\begin{align}
  &J^{(2)}_{1\to 2}(\omega)=\int_{-\infty}^{\infty}dt e^{i\omega t}\left\langle
    O(t)O^{\dag}(0)\right\rangle
  \nonumber\\
  &=\int_{-\infty}^{\infty}dt e^{i\omega t}\sum_{n,m}\left |\left\langle
      n\right |O(0)\left |m\right\rangle \right
  |^2e^{i(E_{n}-E_{m})t}e^{-\beta E_{n}}
  \nonumber\\
  &=2\pi\sum_{n,m}\left |\left\langle n\right |O(0)\left
      |m\right\rangle \right |^2 e^{-\beta
    E_{n}}\delta(\omega+E_{n}-E_{m}).
\end{align}
Similarly, the tunneling current from edge 2 to 1 can be written as
\begin{align}
  &J^{(2)}_{2\to 1}(\omega)=\int_{-\infty}^{\infty}dt e^{i\omega t}\left\langle
    O^{\dag}(0)O(t)\right\rangle
  \nonumber\\
  &=2\pi\sum_{n,m}\left |\left\langle n\right |O(0)\left
      |m\right\rangle \right |^2 e^{-\beta E_{m}}\delta(\omega+E_{n}-E_{m})
  \nonumber\\
  &=2\pi e^{-\beta \omega}\sum_{n,m}\left |\left\langle n\right |O(0)\left
      |m\right\rangle \right |^2 e^{-\beta E_{n}}\delta(\omega+E_{n}-E_{m}).
\end{align}
	
The corresponding Matsubara correlation function is given by
\begin{align}
  &\mathcal{J}(i\omega_n)=\int_{0}^{\beta}d\tau
  e^{i\omega_n\tau}\left\langle\hat{T}_{\tau}O(\tau)O^{\dag}(0)\right\rangle
  \nonumber\\
  &=\int_{0}^{\beta}d\tau e^{i\omega_n\tau}\sum_{n}e^{-\beta E_{n}}
  \nonumber\\
  &\quad\times\left\langle n\right
  |u(\tau)O(\tau)O^{\dag}(0)+u(-\tau)O^{\dag}(0)O(\tau)\left
    |n\right\rangle
  \nonumber\\
  &=\int_{0}^{\beta}d\tau
  e^{i\omega_n\tau}\sum_{n,m}e^{(E_{n}-E_{m})\tau}
  \nonumber\\
  &\quad\times\left[u(\tau)e^{-\beta E_{n}}+u(-\tau)e^{-\beta
      E_{m}}\right]\left |\left\langle n\right |O(0)\left
      |m\right\rangle\right |^2
  \nonumber\\
  &=\sum_{n,m}\left |\left\langle n\right |O(0)\left
      |m\right\rangle\right
  |^2\frac{\left[e^{\beta(E_{n}-E_{m})}-1\right]e^{-\beta
      E_{n}}}{i\omega_n+E_{n}-E_{m}}.
\end{align}
	
By taking the imaginary part of the Matsubara correlator, and do
analytic continuation $i\omega_n\to \omega+i\eta$, we obtain the following
fluctuation-dissipation relation:
\begin{align}
  &2\text{Im}\left[\mathcal{J}(\omega+i\eta)\right]
  \nonumber\\
  &=\left[1-e^{-\beta \omega}\right]J^{(2)}_{1\to 2}(\omega)=\left[e^{\beta
      \omega}-1\right]J^{(2)}_{2\to 1}(\omega)
  \nonumber\\
  &=J^{(2)}_{1\to 2}(\omega)-J^{(2)}_{2\to 1}(\omega).
\end{align}	
	
\section{Bosonization and derivation of $\mathcal{J}(\tau,x)$}
\label{App:Bosonization_two_edge}
The action $S_{12}$ in Eq.~\ref{Eq:S_12} can be generally written as
two decoupled spinless LLs via a basis transformation shown in
Ref.~\cite{Orignac2011}. Here we briefly summarize the result. By the
transformation $\phi_a=\sum_{b=\pm}P_{ab}\phi_b$ and
$\theta_a=\sum_{b=\pm}Q_{ab}\theta_b$ ($a=1,2$), the action can be
written as
\begin{align}
  S_{12}=&\sum_{b=\pm}\,\int\limits_{\tau,x}\left\{\frac{v_{b}}{2\pi}\left[(\partial_x
      \phi_{b})^{2}+(\partial_x \theta_{b})^{2}\right]\right.
  \nonumber\\
  &+\left.\frac{i}{\pi}\left(\partial_x\theta_b\right)\left(\partial_{\tau}\phi_b\right)\right\},
\end{align}
where
\begin{align}
  v_{\pm}^2&=\frac{v_1^2+v_2^2}{2}\pm\sqrt{\left(\frac{v_1^2-v_2^2}{2}\right)^2+\left(\frac{U_{12}}{\pi}\right)^2v_1K_1v_2K_2}
  \nonumber\\
  P&=\left(\begin{array}{cc}
      \sqrt{\frac{v_+}{v_1K_1}}\cos\frac{\theta_{12}}{2} & -\sqrt{\frac{v_-}{v_1K_1}}\sin\frac{\theta_{12}}{2}\\
      \sqrt{\frac{v_+}{v_2K_2}}\sin\frac{\theta_{12}}{2} &
      \sqrt{\frac{v_-}{v_2K_2}}\cos\frac{\theta_{12}}{2}
    \end{array}\right)
  \nonumber\\
  Q&=\left(\begin{array}{cc}
      \sqrt{\frac{v_1K_1}{v_+}}\cos\frac{\theta_{12}}{2} & -\sqrt{\frac{v_1K_1}{v_-}}\sin\frac{\theta_{12}}{2}\\
      \sqrt{\frac{v_2K_2}{v_+}}\sin\frac{\theta_{12}}{2} &
      \sqrt{\frac{v_2K_2}{v_-}}\cos\frac{\theta_{12}}{2}
    \end{array}\right),
\end{align}
with $\tan\theta_{12}=2(U_{12}/\pi)\sqrt{v_1K_1v_2K_2}/(v_1^2-v_2^2)$
($v_1>v_2$ is assumed without loss of generality). The matrices $P$
and $Q$ are chosen to decouple the bosonic fields in the edge basis in the presence of interedge interactions, but under constraint to maintain their canonical commutation relations, which corresponds to requiring $PQ^{T}=1$ or keeping the Berry phase term diagonal. The four-point correlation function for tunneling
current can be calculated as follows
\begin{widetext}
  \begin{align}
    \left\langle
      L_1^{\dag}L_2(\tau,x)L_2^{\dag}L_1(0,0)\right\rangle_{\tau}&=\frac{1}{(2\pi\alpha)^2}e^{-\frac{1}{2}\left\langle\left[\phi_1(\tau,x)-\theta_1(\tau,x)-\phi_2(\tau,x)+\theta_2(\tau,x)-\phi_1(0,0)+\theta_1(0,0)+\phi_2(0,0)-\theta_2(0,0)\right]^2\right\rangle}
    \nonumber\\
    &=\frac{1}{(2\pi\alpha)^2}e^{-\frac{1}{2}\sum_{b=\pm}\left[(P_{1b}-P_{2b})^2+(Q_{1b}-Q_{2b})^2\right]F_{1b}(\tau,x)-\sum_{b=\pm}(P_{1b}-P_{2b})(Q_{1b}-Q_{2b})F_{2b}(\tau,x)}
    \nonumber\\
    &=\frac{1}{(2\pi\alpha)^2}\prod_{b=\pm}e^{-\frac{1}{2}\left[(P_{1b}-P_{2b})^2+(Q_{1b}-Q_{2b})^2\right]F_{1b}(\tau,x)-\left(1-P_{1b}Q_{2b}-P_{2b}Q_{1b}\right)F_{2b}(\tau,x)}
    \nonumber\\
    &=-\frac{1}{(2\pi\alpha)^2}\prod_{b=\pm}\frac{(\frac{\pi\alpha}{\beta
        v_b})^{2\gamma_{b}+1}}{\left[\sinh\left(\frac{\pi(x+iv_b\tau)}{\beta
            v_b}\right)\right]^{\gamma_b-\frac{b}{2}\gamma_{12}}\left[\sinh\left(\frac{\pi(x-iv_b\tau)}{\beta
            v_b}\right)\right]^{\gamma_b+\frac{b}{2}\gamma_{12}+1}}
    \nonumber\\
    \left\langle
      R_1^{\dag}R_2(\tau,x)R_2^{\dag}R_1(0,0)\right\rangle_{\tau}&=-\frac{1}{(2\pi\alpha)^2}\prod_{b=\pm}\frac{(\frac{\pi\alpha}{\beta
        v_b})^{2\gamma_{b}+1}}{\left[\sinh\left(\frac{\pi(x+iv_b\tau)}{\beta
            v_b}\right)\right]^{\gamma_b+\frac{b}{2}\gamma_{12}+1}\left[\sinh\left(\frac{\pi(x-iv_b\tau)}{\beta
            v_b}\right)\right]^{\gamma_b-\frac{b}{2}\gamma_{12}}},
  \end{align}
\end{widetext}
where $F_{1b}$ and $F_{2b}$ are given in
Appendix~\ref{App:Der_space_time_G} with $v\to v_b$ and the
interaction parameters are given by
\begin{align}
  \label{Eq:gamma_12}\gamma_{12}&=-\frac{1}{2}\left(\sqrt{\frac{v_2K_2}{v_1K_1}}+\sqrt{\frac{v_1K_1}{v_2K_2}}\right)\sin\theta_{12},
  \\
  \gamma_b&=\frac{1}{4}(P_{1b}-P_{2b})^2+\frac{1}{4}(Q_{1b}-Q_{2b})^2-\frac{1}{2}
\end{align}
The tunneling current between right and left Fermi points can also be
calculated by similar way
\begin{widetext}
  \begin{align}
    \left\langle
      L_1^{\dag}R_2(\tau,x)R_2^{\dag}L_1(0,0)\right\rangle_{\tau}
    &=\frac{1}{(2\pi\alpha)^2}e^{-\frac{1}{2}\sum_{b=\pm}\left[(P_{1b}-P_{2b})^2+(Q_{1b}-Q_{2b})^2\right]F_{1b}(\tau,x)-\sum_{b=\pm}(P_{1b}-P_{2b})(Q_{1b}+Q_{2b})F_{2b}(\tau,x)}
    \nonumber\\
    &=\frac{1}{(2\pi\alpha)^2}\prod_{b=\pm}e^{-\frac{1}{2}\left[(P_{1b}-P_{2b})^2+(Q_{1b}-Q_{2b})^2\right]F_{1b}(\tau,x)-b\bar{\gamma}_{12}F_{2b}(\tau,x)}
    \nonumber\\
    &=\frac{1}{(2\pi\alpha)^2}\prod_{b=\pm}\frac{(\frac{\pi\alpha}{\beta
        v_b})^{2\gamma_{b}+1}}{\left[\sinh\left(\frac{\pi(x+iv_b\tau)}{\beta
            v_b}\right)\right]^{\gamma_b+\frac{1}{2}-\frac{b}{2}\bar{\gamma}_{12}}\left[\sinh\left(\frac{\pi(x-iv_b\tau)}{\beta
            v_b}\right)\right]^{\gamma_b+\frac{1}{2}+\frac{b}{2}\bar{\gamma}_{12}}}
    \nonumber\\
    \left\langle
      R_1^{\dag}L_2(\tau,x)L_2^{\dag}R_1(0,0)\right\rangle_{\tau}&=\frac{1}{(2\pi\alpha)^2}\prod_{b=\pm}\frac{(\frac{\pi\alpha}{\beta
        v_b})^{2\gamma_{b}+1}}{\left[\sinh\left(\frac{\pi(x+iv_b\tau)}{\beta
            v_b}\right)\right]^{\gamma_b+\frac{1}{2}+\frac{b}{2}\bar{\gamma}_{12}}\left[\sinh\left(\frac{\pi(x-iv_b\tau)}{\beta
            v_b}\right)\right]^{\gamma_b+\frac{1}{2}-\frac{b}{2}\bar{\gamma}_{12}}},
  \end{align}
\end{widetext}
where
\begin{align}
  \bar{\gamma}_{12}&=\frac{1}{2}\left(\sqrt{\frac{v_2K_2}{v_1K_1}}-\sqrt{\frac{v_1K_1}{v_2K_2}}\right)\sin\theta_{12}+\cos\theta_{12}.
\end{align}
	
\subsubsection{Identical edge limit}
We discuss the simplified special case of $K_1=K_2\equiv K$ and
$v_1=v_2\equiv v$. It is convenient in this case to work with the
symmetric and anti-symmetric fields (denoted by subscripts $+$ and
$-$, respectively, not to be confused with the chiral band index for
generic hLL), which are given by
\begin{align}
  \phi_{\pm}=\frac{\phi_1\pm\phi_2}{\sqrt{2}},\quad\theta_{\pm}=\frac{\theta_1\pm\theta_2}{\sqrt{2}}.
\end{align}
In terms of these the Hamiltonian decouples and is given by
\begin{align}\label{Eq:S_12_identical}
  &H_{12}=\sum_{b=\pm}v_{b}\int_{x}\left[K_{b}(\partial_x
    \phi_{b})^{2}+\frac{1}{K_{b}}(\partial_x \theta_{b})^{2}\right]
  \nonumber\\
  &\mathcal{S}_{\text{dis}}=\int\limits_{\tau,x}\left[V_+(x)\frac{1}{\pi}\partial_x\theta_++V_-(x)\frac{1}{\pi}\partial_x\theta_-\right],
\end{align}
where $v_{\pm}=v\sqrt{1\pm UK/2\pi^2v}$ and $K_{\pm}=K/\sqrt{1\pm
  UK/2\pi^2v}$, $V_{\pm}=\left(V_1\pm V_2\right)/\sqrt{2}$, satisfying
$\overline{V_b(x)V_{b'}(y)}=\Delta_{bb'}\delta(x-y)$,
$\Delta_{++}=\Delta_{--}=\left(\Delta_{1}+\Delta_{2}\right)/2$,
$\Delta_{+-}=\Delta_{-+}=\left(\Delta_{1}-\Delta_{2}\right)/2$. For
repulsive interedge density-density interaction $U>0$, we have
$v_-<v_+$ and $K_+<K_-$. As for a single edge, we can eliminate the
disorder potentials via a linear transformation on the $\theta_a$
fields,
$\mathcal{S}_{12}[\theta_b,\phi_b]+\mathcal{S}_{\text{dis}}[\theta_b]\rightarrow\mathcal{S}_{12}[\tilde{\theta}_b,\phi_b]+\text{constant}$,
where
\begin{align}
  \tilde{\theta}_b(\tau,x)=\theta_b(\tau,x)+\frac{K_b}{v_b}\int_{-\infty}^{x}V_b(y)dy.
\end{align}
With this transformation, the disorder-averaged correlator can then be
straightforwardly calculated using Eq.~(\ref{Eq:dis_ave_corr}).
	
\section{Four-point correlation function for tunneling current
  computation}
The calculation of tunneling current requires a Fourier transformation
of a four-point correlation function with two different velocities,
which is generally quite difficult, even numerically. In this
Appendix, we consider the following two complimentary cases, which
cover a broad spectrum of situations with significantly-simplified
calculations: (i) Finite temperature in the absence of interedge
interaction and (ii) Zero temperature in the presence of interedge
interaction. We will also discuss the special case of identical edges,
where analytical expressions are derived.
\subsection{Finite temperature, no interedge interaction}
In the absence of interedge interaction, we consider a space-imaginary
time correlation function of the following form
\begin{align}
  \mathcal{C}(\tau,x)=\mathcal{G}_1(\tau,x)\mathcal{G}_2(\tau,x)
\end{align}
that in Fourier space is a convolution
\begin{align}
  \mathcal{C}(i\omega_n,q)=&\frac{1}{\beta}\sum_{\omega_m}\int\limits_k\mathcal{G}_1(i\omega_n-i\omega_m,q-k)\mathcal{G}_2(i\omega_m,k),
\end{align}
where $i\omega_n=2\pi T n$ ($i\omega_m=2\pi T(m+1/2)$) is bosonic
(fermionic) Matsubara frequency. To this end, it is convenient to
first trade the Matsubara summation for an integration. Using the
standard Lehmann spectral representation
\begin{align}
	\mathcal{G}(i\omega_n,q)=\frac{1}{\pi}\int_{-\infty}^{\infty}dz\frac{\text{Im}\left[G^{\text{ret}}(z,q)\right]}{z-i\omega_n},
\end{align}
we can express $\mathcal{C}(i\omega_n,q)$ in terms of the retarded
Green functions $G^{\text{ret}}$ as follows
\begin{align}
  \mathcal{C}(i\omega_n,q)=&\frac{4}{\beta}\sum_{\omega_m}\int\limits_{k}\int\limits_{z,z'}\frac{\text{Im}\left[G^{\text{ret}}_1(z,q-k)\right]}{z-i\omega_n+i\omega_m}\frac{\text{Im}\left[G^{\text{ret}}_2(z',k)\right]}{z'-i\omega_m}
  \nonumber\\
  =&4\int\limits_{k}\int\limits_{z,z'}\frac{\text{Im}\left[G^{\text{ret}}_1(z,q-k)\right]\text{Im}\left[G^{\text{ret}}_2(z',k)\right]}{z+z'-i\omega_n}
  \nonumber\\
  &\times\,\left[n_F(-z)-n_F(z')\right],
\end{align}
where $\int_z=\int_{-\infty}^{\infty}dz/2\pi$ and the Matsubara summation was done by, e.g., the Poisson summation formula. After the
analytic continuation $i\omega_n\to \omega+i\eta$, the imaginary part
of the correlation function is given by ($z'$ replaced by $\Omega$)
\begin{align}
  \text{Im}\left[\mathcal{C}(\omega+i\eta,q)\right]=&2\int\limits_{\Omega,k}\text{Im}\left[G^{\text{ret}}_1(\omega-\Omega,q-k)\right]
  \nonumber\\
  &\times\text{Im}\left[G^{\text{ret}}_2(\Omega,k)\right]\left[n_F(\Omega-\omega)-n_F(\Omega)\right].
\end{align}
Considering a special case that one of the edge is non-interacting,
e.g. $K_2=1$, the spectral function of edge 2 becomes a delta function
\begin{align}
  \text{Im}\left[G^{\text{ret}}_{R(L)}(\Omega,k)\right]=-\pi\delta\left(\Omega\mp
    vk\right).
\end{align}
After integrating over $k$, we obtained the following integral
expression for the correlation function:
\begin{align}
  \text{Im}\left[\mathcal{C}(\omega+i\eta,q)\right]=&-\frac{1}{v}\int\limits_{\Omega}\text{Im}\left[G^{\text{ret}}_1(\omega-\Omega,q\mp\frac{\Omega}{v})\right]
  \nonumber\\
  &\times\left[n_F(\Omega-\omega)-n_F(\Omega)\right].
\end{align}
	
\subsection{Zero temperature, with interedge interaction}
The tunneling current can also be directly calculated by Fourier
transforming Eq.~(\ref{Eq:JLL_xt}) and ~(\ref{Eq:JRL_xt}) following
the approach in Ref.~\cite{Carpentier2002}. Specifically, we rewrite
Eq.~(\ref{Eq:J_xt_formalism2}) as an integral from $t=0$ to $\infty$
with the integrand (space-time correlator) obtained by an analytic
continuation $\tau=it+\epsilon\sgn(t)$ from the Euclidean correlation
function. Below we will discuss the calculation of $J_{LL}$ and
$J_{RL}$. The tunneling current $J_{LL}$ at zero temperature is given
by the following integral
\begin{widetext}
  \begin{align}
    \label{Eq:J_LL_U12}
    J_{LL}(V,q)=&-\frac{1}{2\pi^2}\text{Re}\int_{-\infty}^{\infty}dx\int_{0}^{\infty}dte^{i(V+i\eta)t}e^{-iqx}\prod_{b=\pm}\frac{\alpha^{2\gamma_{b}}}{\left(x-v_bt+i\epsilon\right)^{\gamma_b-\frac{b}{2}\gamma_{12}}\left(x+v_bt-i\epsilon\right)^{\gamma_b+\frac{b}{2}\gamma_{12}+1}}
    \nonumber\\
    =&-\frac{\Gamma(-2\gamma_+-2\gamma_-)}{2\pi^2}\text{Re}\int_{-\infty}^{\infty}du\left[\eta-i(V-qu)\right]^{2\gamma_++2\gamma_-}\prod_{b=\pm}\frac{\alpha^{2\gamma_{b}}}{\left(u-v_b+i\epsilon\right)^{\gamma_b-\frac{b}{2}\gamma_{12}}\left(u+v_b-i\epsilon\right)^{\gamma_b+\frac{b}{2}\gamma_{12}+1}}
    ,
  \end{align}
\end{widetext}
where in the second equality we change the variable $x=ut$ and
integrate over $t$ using the gamma function identity. The tunneling current $J_{RL}$
can also be calculated with the same procedure, and is given by
\begin{widetext}
  \begin{align}
    \label{Eq:J_RL_U12}
    J_{RL}(V,q)=\frac{\Gamma(-2\gamma_+-2\gamma_-)}{2\pi^2}\text{Re}\int_{-\infty}^{\infty}du\left[\eta-i(V-qu)\right]^{2\gamma_++2\gamma_-}\prod_{b=\pm}\frac{\alpha^{2\gamma_{b}}}{\left(u-v_b+i\epsilon\right)^{\gamma_b+\frac{b}{2}\bar{\gamma}_{12}+\frac{1}{2}}\left(u+v_b-i\epsilon\right)^{\gamma_b+\frac{b}{2}\bar{\gamma}_{12}+\frac{1}{2}}}.
  \end{align}
\end{widetext}
In evaluation of the integrals in Eq.~(\ref{Eq:J_LL_U12})
and~(\ref{Eq:J_RL_U12}), one can detour the integration
contour~\cite{Carpentier2002} to yield accurate numerical results. The
effects of forward-scattering disorders can be included by replacing
$q$ in the integrand with $q-i\sgn(u)/\xi$.
	
\subsection{identical edges}
For identical edges $v_1=v_2$, $K_1=K_2$ and in the absence of
interedge interaction, we can derive the exact \textit{disordered
  zero-temperature} expression using similar procedure as in
Appendix~\ref{App:Der_ret_G}. We were not able to derive a
low-temperature asymptotic expression since Stirling approximation
gives a qualitatively wrong answer in low temperature in this
case. The exact zero-temperature \textit{clean} tunneling current is
given by
\begin{align}
  J^{\text{T=0}}_{LL}(\omega,q)=&-\frac{et_0^2}{2\pi
    v}\left(\frac{\alpha}{2v}\right)^{4\gamma}\frac{\Gamma\left(1-2\gamma\right)}{\Gamma\left(2+2\gamma\right)}\text{Im}\left\lbrace\left[-i(\omega+vq)+\eta\right]^{2\gamma-1}\right.
  \nonumber\\
  &\left.\left[-i(\omega-vq)+\eta\right]^{2\gamma+1}\right\rbrace
  \nonumber\\
  J^{\text{T=0}}_{RL}(\omega,q)=&\frac{et_0^2}{2\pi
    v}\left(\frac{\alpha}{2v}\right)^{4\gamma}\frac{\Gamma\left(1-2\gamma\right)}{\Gamma\left(2+2\gamma\right)}\text{Im}\left\lbrace\left[-i(\omega+vq)+\eta\right]^{2\gamma}\right.
  \nonumber\\
  &\left.\left[-i(\omega-vq)+\eta\right]^{2\gamma}\right\rbrace
\end{align}
	
The \textit{disordered} tunneling current can be calculated by Residue
theorem and is given by
\begin{align}
  \label{Eq:J_dis_T0}
  J^{\text{T=0}}_{\text{dis},RL/LL}(\omega,q)=J^{\text{T=0}}_{1,RL/LL}(\omega,q+i\xi^{-1})+J^{\text{T=0}}_{2,RL/LL}(\omega,q),
\end{align}
where
\begin{widetext}
  \begin{align}
    J^{\text{T=0}}_{2,LL}(\omega,q)=&\frac{et_0^2}{2\pi
      v}\sum_{s=\pm}s\frac{1}{\pi}\sin\left(2\pi\gamma\right)\left(\frac{\alpha}{2v}\right)^{4\gamma}\frac{\Gamma\left(1-2\gamma\right)}{\Gamma\left(2+2\gamma\right)}\left\lbrace\frac{\pi}{\sin\left(4\pi\gamma\right)}\left[\frac{i\left(\omega+vq\right)+sv\xi^{-1}}{-i\left(\omega-vq\right)+sv\xi^{-1}}\right]^{2\gamma-1}\right.
    \nonumber\\
    &\left.\times\left[-i\left(\omega-vq\right)+sv\xi^{-1}\right]^{4\gamma}-\left[-i2\omega+\eta\right]^{4\gamma+1}\left[-i\left(\omega-vq\right)+sv\xi^{-1}\right]^{-1}\right.
    \nonumber\\
    &\left.\times
      B(2+2\gamma,-1-4\gamma)_2F_1\left(1,2+2\gamma,2+4\gamma,\frac{-i2\omega+\eta}{-i\left(\omega-vq\right)+sv\xi^{-1}}\right)\right\rbrace
    \nonumber\\
    J^{\text{T=0}}_{2,RL}(\omega,q)=&\frac{et_0^2}{2\pi
      v}\sum_{s=\pm}s\frac{1}{\pi}\sin\left(2\pi\gamma\right)\left(\frac{\alpha}{2v}\right)^{4\gamma}\frac{\Gamma\left(-2\gamma\right)}{\Gamma\left(1+2\gamma\right)}\left\lbrace\frac{\pi}{\sin\left(4\pi\gamma\right)}\left[\frac{i\left(\omega+vq\right)+sv\xi^{-1}}{-i\left(\omega-vq\right)+sv\xi^{-1}}\right]^{2\gamma}\right.
    \nonumber\\
    &\left.\times\left[-i\left(\omega-vq\right)+sv\xi^{-1}\right]^{4\gamma}-2^{-2-4\gamma}\left[-i2\omega+\eta\right]^{4\gamma+1}\left[-i\left(\omega-vq\right)+sv\xi^{-1}\right]^{-1}\right.
    \nonumber\\
    &\left.\times
      B(1+2\gamma,-1/2-2\gamma)_2F_1\left(1,1+2\gamma,2+4\gamma,\frac{-i2\omega+\eta}{-i\left(\omega-vq\right)+sv\xi^{-1}}\right)\right\rbrace.
  \end{align}
\end{widetext}

\bibliography{TI_ref}
	
\end{document}